\documentclass[longauth]{aa}

\usepackage{graphicx}
\usepackage{natbib}
\usepackage{scalerel}

\usepackage[table]{xcolor}

\usepackage{txfonts}
\usepackage[pdfencoding=auto,psdextra]{hyperref}
\hypersetup{
    colorlinks=true,
    linkcolor=blue,
    filecolor=magenta,      
    urlcolor=blue,
    citecolor=blue
}
\makeatletter
\renewcommand*\aa@pageof{, page \thepage{} of \pageref*{LastPage}}
\makeatother

\usepackage[utf8]{inputenc}

\usepackage[switch, modulo]{lineno}
\usepackage{placeins}
\usepackage{euclid}

\begin{document} 

\title{\Euclid preparation}
\subtitle{XLIII. Measuring detailed galaxy morphologies for \Euclid with machine learning} 
	   
\newcommand{\orcid}[1]{} 
\author{Euclid Collaboration: B.~Aussel\orcid{0000-0003-2592-6806}\thanks{\email{ben.aussel@uni-muenster.de}}\inst{\ref{aff1}}
\and S.~Kruk\orcid{0000-0001-8010-8879}\inst{\ref{aff2}}
\and M.~Walmsley\orcid{0000-0002-6408-4181}\inst{\ref{aff3}}
\and M.~Huertas-Company\orcid{0000-0002-1416-8483}\inst{\ref{aff4},\ref{aff5},\ref{aff6},\ref{aff7}}
\and M.~Castellano\orcid{0000-0001-9875-8263}\inst{\ref{aff8}}
\and C.~J.~Conselice\inst{\ref{aff3}}
\and M.~Delli~Veneri\orcid{0000-0002-8178-2942}\inst{\ref{aff9}}
\and H.~Dom\'{i}nguez~S\'{a}nchez\orcid{0000-0002-9013-1316}\inst{\ref{aff10}}
\and P.-A.~Duc\inst{\ref{aff11}}
\and J.~H.~Knapen\orcid{0000-0003-1643-0024}\inst{\ref{aff4},\ref{aff5}}
\and U.~Kuchner\orcid{0000-0002-0035-5202}\inst{\ref{aff12}}
\and A.~La~Marca\orcid{0000-0002-7217-5120}\inst{\ref{aff13},\ref{aff14}}
\and B.~Margalef-Bentabol\inst{\ref{aff13}}
\and F.~R.~Marleau\inst{\ref{aff15}}
\and G.~Stevens\orcid{0000-0002-8885-4443}\inst{\ref{aff16}}
\and Y.~Toba\orcid{0000-0002-3531-7863}\inst{\ref{aff17}}
\and C.~Tortora\orcid{0000-0001-7958-6531}\inst{\ref{aff18}}
\and L.~Wang\orcid{0000-0002-6736-9158}\inst{\ref{aff13},\ref{aff14}}
\and N.~Aghanim\inst{\ref{aff19}}
\and B.~Altieri\orcid{0000-0003-3936-0284}\inst{\ref{aff2}}
\and A.~Amara\inst{\ref{aff20}}
\and S.~Andreon\orcid{0000-0002-2041-8784}\inst{\ref{aff21}}
\and N.~Auricchio\inst{\ref{aff22}}
\and M.~Baldi\orcid{0000-0003-4145-1943}\inst{\ref{aff23},\ref{aff22},\ref{aff24}}
\and S.~Bardelli\orcid{0000-0002-8900-0298}\inst{\ref{aff22}}
\and R.~Bender\orcid{0000-0001-7179-0626}\inst{\ref{aff25},\ref{aff26}}
\and C.~Bodendorf\inst{\ref{aff25}}
\and D.~Bonino\inst{\ref{aff27}}
\and E.~Branchini\orcid{0000-0002-0808-6908}\inst{\ref{aff28},\ref{aff29},\ref{aff21}}
\and M.~Brescia\orcid{0000-0001-9506-5680}\inst{\ref{aff30},\ref{aff18},\ref{aff9}}
\and J.~Brinchmann\inst{\ref{aff31}}
\and S.~Camera\orcid{0000-0003-3399-3574}\inst{\ref{aff32},\ref{aff33},\ref{aff27}}
\and V.~Capobianco\orcid{0000-0002-3309-7692}\inst{\ref{aff27}}
\and C.~Carbone\orcid{0000-0003-0125-3563}\inst{\ref{aff34}}
\and J.~Carretero\orcid{0000-0002-3130-0204}\inst{\ref{aff35},\ref{aff36}}
\and S.~Casas\inst{\ref{aff37}}
\and S.~Cavuoti\orcid{0000-0002-3787-4196}\inst{\ref{aff18},\ref{aff9}}
\and A.~Cimatti\inst{\ref{aff38}}
\and G.~Congedo\orcid{0000-0003-2508-0046}\inst{\ref{aff39}}
\and L.~Conversi\orcid{0000-0002-6710-8476}\inst{\ref{aff40},\ref{aff2}}
\and Y.~Copin\orcid{0000-0002-5317-7518}\inst{\ref{aff41}}
\and F.~Courbin\orcid{0000-0003-0758-6510}\inst{\ref{aff42}}
\and H.~M.~Courtois\orcid{0000-0003-0509-1776}\inst{\ref{aff43}}
\and M.~Cropper\inst{\ref{aff44}}
\and A.~Da~Silva\orcid{0000-0002-6385-1609}\inst{\ref{aff45},\ref{aff46}}
\and H.~Degaudenzi\orcid{0000-0002-5887-6799}\inst{\ref{aff47}}
\and A.~M.~Di~Giorgio\orcid{0000-0002-4767-2360}\inst{\ref{aff48}}
\and J.~Dinis\inst{\ref{aff46},\ref{aff45}}
\and F.~Dubath\orcid{0000-0002-6533-2810}\inst{\ref{aff47}}
\and X.~Dupac\inst{\ref{aff2}}
\and S.~Dusini\orcid{0000-0002-1128-0664}\inst{\ref{aff49}}
\and M.~Farina\inst{\ref{aff48}}
\and S.~Farrens\orcid{0000-0002-9594-9387}\inst{\ref{aff50}}
\and S.~Ferriol\inst{\ref{aff41}}
\and S.~Fotopoulou\orcid{0000-0002-9686-254X}\inst{\ref{aff51}}
\and M.~Frailis\orcid{0000-0002-7400-2135}\inst{\ref{aff52}}
\and E.~Franceschi\orcid{0000-0002-0585-6591}\inst{\ref{aff22}}
\and P.~Franzetti\inst{\ref{aff34}}
\and M.~Fumana\inst{\ref{aff34}}
\and S.~Galeotta\orcid{0000-0002-3748-5115}\inst{\ref{aff52}}
\and B.~Garilli\orcid{0000-0001-7455-8750}\inst{\ref{aff34}}
\and B.~Gillis\orcid{0000-0002-4478-1270}\inst{\ref{aff39}}
\and C.~Giocoli\orcid{0000-0002-9590-7961}\inst{\ref{aff22},\ref{aff53}}
\and A.~Grazian\orcid{0000-0002-5688-0663}\inst{\ref{aff54}}
\and F.~Grupp\inst{\ref{aff25},\ref{aff26}}
\and S.~V.~H.~Haugan\orcid{0000-0001-9648-7260}\inst{\ref{aff55}}
\and W.~Holmes\inst{\ref{aff56}}
\and I.~Hook\orcid{0000-0002-2960-978X}\inst{\ref{aff57}}
\and F.~Hormuth\inst{\ref{aff58}}
\and A.~Hornstrup\orcid{0000-0002-3363-0936}\inst{\ref{aff59},\ref{aff60}}
\and P.~Hudelot\inst{\ref{aff61}}
\and K.~Jahnke\orcid{0000-0003-3804-2137}\inst{\ref{aff62}}
\and E.~Keih\"anen\orcid{0000-0003-1804-7715}\inst{\ref{aff63}}
\and S.~Kermiche\orcid{0000-0002-0302-5735}\inst{\ref{aff64}}
\and A.~Kiessling\orcid{0000-0002-2590-1273}\inst{\ref{aff56}}
\and M.~Kilbinger\orcid{0000-0001-9513-7138}\inst{\ref{aff65}}
\and B.~Kubik\inst{\ref{aff41}}
\and M.~K\"ummel\orcid{0000-0003-2791-2117}\inst{\ref{aff26}}
\and M.~Kunz\orcid{0000-0002-3052-7394}\inst{\ref{aff66}}
\and H.~Kurki-Suonio\orcid{0000-0002-4618-3063}\inst{\ref{aff67},\ref{aff68}}
\and R.~Laureijs\inst{\ref{aff69}}
\and S.~Ligori\orcid{0000-0003-4172-4606}\inst{\ref{aff27}}
\and P.~B.~Lilje\orcid{0000-0003-4324-7794}\inst{\ref{aff55}}
\and V.~Lindholm\orcid{0000-0003-2317-5471}\inst{\ref{aff67},\ref{aff68}}
\and I.~Lloro\inst{\ref{aff70}}
\and E.~Maiorano\orcid{0000-0003-2593-4355}\inst{\ref{aff22}}
\and O.~Mansutti\orcid{0000-0001-5758-4658}\inst{\ref{aff52}}
\and O.~Marggraf\orcid{0000-0001-7242-3852}\inst{\ref{aff71}}
\and K.~Markovic\orcid{0000-0001-6764-073X}\inst{\ref{aff56}}
\and N.~Martinet\orcid{0000-0003-2786-7790}\inst{\ref{aff72}}
\and F.~Marulli\orcid{0000-0002-8850-0303}\inst{\ref{aff73},\ref{aff22},\ref{aff24}}
\and R.~Massey\orcid{0000-0002-6085-3780}\inst{\ref{aff74}}
\and S.~Maurogordato\inst{\ref{aff75}}
\and E.~Medinaceli\orcid{0000-0002-4040-7783}\inst{\ref{aff22}}
\and S.~Mei\orcid{0000-0002-2849-559X}\inst{\ref{aff76}}
\and Y.~Mellier\inst{\ref{aff77},\ref{aff61}}
\and M.~Meneghetti\orcid{0000-0003-1225-7084}\inst{\ref{aff22},\ref{aff24}}
\and E.~Merlin\orcid{0000-0001-6870-8900}\inst{\ref{aff8}}
\and G.~Meylan\inst{\ref{aff42}}
\and M.~Moresco\orcid{0000-0002-7616-7136}\inst{\ref{aff73},\ref{aff22}}
\and L.~Moscardini\orcid{0000-0002-3473-6716}\inst{\ref{aff73},\ref{aff22},\ref{aff24}}
\and E.~Munari\orcid{0000-0002-1751-5946}\inst{\ref{aff52}}
\and S.-M.~Niemi\inst{\ref{aff69}}
\and C.~Padilla\orcid{0000-0001-7951-0166}\inst{\ref{aff35}}
\and S.~Paltani\inst{\ref{aff47}}
\and F.~Pasian\inst{\ref{aff52}}
\and K.~Pedersen\inst{\ref{aff78}}
\and W.~J.~Percival\orcid{0000-0002-0644-5727}\inst{\ref{aff79},\ref{aff80},\ref{aff81}}
\and V.~Pettorino\inst{\ref{aff82}}
\and S.~Pires\orcid{0000-0002-0249-2104}\inst{\ref{aff50}}
\and G.~Polenta\orcid{0000-0003-4067-9196}\inst{\ref{aff83}}
\and M.~Poncet\inst{\ref{aff84}}
\and L.~A.~Popa\inst{\ref{aff85}}
\and L.~Pozzetti\orcid{0000-0001-7085-0412}\inst{\ref{aff22}}
\and F.~Raison\orcid{0000-0002-7819-6918}\inst{\ref{aff25}}
\and R.~Rebolo\inst{\ref{aff4},\ref{aff5}}
\and A.~Renzi\orcid{0000-0001-9856-1970}\inst{\ref{aff86},\ref{aff49}}
\and J.~Rhodes\inst{\ref{aff56}}
\and G.~Riccio\inst{\ref{aff18}}
\and E.~Romelli\orcid{0000-0003-3069-9222}\inst{\ref{aff52}}
\and M.~Roncarelli\orcid{0000-0001-9587-7822}\inst{\ref{aff22}}
\and E.~Rossetti\inst{\ref{aff23}}
\and R.~Saglia\orcid{0000-0003-0378-7032}\inst{\ref{aff26},\ref{aff25}}
\and D.~Sapone\orcid{0000-0001-7089-4503}\inst{\ref{aff87}}
\and B.~Sartoris\inst{\ref{aff26},\ref{aff52}}
\and M.~Schirmer\orcid{0000-0003-2568-9994}\inst{\ref{aff62}}
\and P.~Schneider\orcid{0000-0001-8561-2679}\inst{\ref{aff71}}
\and A.~Secroun\orcid{0000-0003-0505-3710}\inst{\ref{aff64}}
\and G.~Seidel\orcid{0000-0003-2907-353X}\inst{\ref{aff62}}
\and S.~Serrano\orcid{0000-0002-0211-2861}\inst{\ref{aff88},\ref{aff89},\ref{aff90}}
\and C.~Sirignano\orcid{0000-0002-0995-7146}\inst{\ref{aff86},\ref{aff49}}
\and G.~Sirri\orcid{0000-0003-2626-2853}\inst{\ref{aff24}}
\and L.~Stanco\orcid{0000-0002-9706-5104}\inst{\ref{aff49}}
\and J.-L.~Starck\orcid{0000-0003-2177-7794}\inst{\ref{aff65}}
\and P.~Tallada-Cresp\'{i}\orcid{0000-0002-1336-8328}\inst{\ref{aff91},\ref{aff36}}
\and A.~N.~Taylor\inst{\ref{aff39}}
\and H.~I.~Teplitz\orcid{0000-0002-7064-5424}\inst{\ref{aff92}}
\and I.~Tereno\inst{\ref{aff45},\ref{aff93}}
\and R.~Toledo-Moreo\orcid{0000-0002-2997-4859}\inst{\ref{aff94}}
\and F.~Torradeflot\orcid{0000-0003-1160-1517}\inst{\ref{aff36},\ref{aff91}}
\and I.~Tutusaus\orcid{0000-0002-3199-0399}\inst{\ref{aff95}}
\and E.~A.~Valentijn\inst{\ref{aff14}}
\and L.~Valenziano\orcid{0000-0002-1170-0104}\inst{\ref{aff22},\ref{aff96}}
\and T.~Vassallo\orcid{0000-0001-6512-6358}\inst{\ref{aff26},\ref{aff52}}
\and A.~Veropalumbo\orcid{0000-0003-2387-1194}\inst{\ref{aff21},\ref{aff29}}
\and Y.~Wang\orcid{0000-0002-4749-2984}\inst{\ref{aff92}}
\and J.~Weller\orcid{0000-0002-8282-2010}\inst{\ref{aff26},\ref{aff25}}
\and A.~Zacchei\orcid{0000-0003-0396-1192}\inst{\ref{aff52},\ref{aff97}}
\and G.~Zamorani\orcid{0000-0002-2318-301X}\inst{\ref{aff22}}
\and J.~Zoubian\inst{\ref{aff64}}
\and E.~Zucca\orcid{0000-0002-5845-8132}\inst{\ref{aff22}}
\and A.~Biviano\orcid{0000-0002-0857-0732}\inst{\ref{aff52},\ref{aff97}}
\and M.~Bolzonella\orcid{0000-0003-3278-4607}\inst{\ref{aff22}}
\and A.~Boucaud\orcid{0000-0001-7387-2633}\inst{\ref{aff76}}
\and E.~Bozzo\orcid{0000-0002-8201-1525}\inst{\ref{aff47}}
\and C.~Burigana\orcid{0000-0002-3005-5796}\inst{\ref{aff98},\ref{aff96}}
\and C.~Colodro-Conde\inst{\ref{aff4}}
\and D.~Di~Ferdinando\inst{\ref{aff24}}
\and R.~Farinelli\inst{\ref{aff22}}
\and J.~Graci\'{a}-Carpio\inst{\ref{aff25}}
\and G.~Mainetti\inst{\ref{aff99}}
\and S.~Marcin\inst{\ref{aff100}}
\and N.~Mauri\orcid{0000-0001-8196-1548}\inst{\ref{aff38},\ref{aff24}}
\and C.~Neissner\inst{\ref{aff35},\ref{aff36}}
\and A.~A.~Nucita\inst{\ref{aff101},\ref{aff102},\ref{aff103}}
\and Z.~Sakr\orcid{0000-0002-4823-3757}\inst{\ref{aff104},\ref{aff95},\ref{aff105}}
\and V.~Scottez\inst{\ref{aff77},\ref{aff106}}
\and M.~Tenti\orcid{0000-0002-4254-5901}\inst{\ref{aff24}}
\and M.~Viel\orcid{0000-0002-2642-5707}\inst{\ref{aff97},\ref{aff52},\ref{aff107},\ref{aff108},\ref{aff109}}
\and M.~Wiesmann\inst{\ref{aff55}}
\and Y.~Akrami\orcid{0000-0002-2407-7956}\inst{\ref{aff110},\ref{aff111}}
\and V.~Allevato\orcid{0000-0001-7232-5152}\inst{\ref{aff18}}
\and S.~Anselmi\orcid{0000-0002-3579-9583}\inst{\ref{aff86},\ref{aff49},\ref{aff112}}
\and C.~Baccigalupi\orcid{0000-0002-8211-1630}\inst{\ref{aff107},\ref{aff52},\ref{aff108},\ref{aff97}}
\and M.~Ballardini\orcid{0000-0003-4481-3559}\inst{\ref{aff113},\ref{aff114},\ref{aff22}}
\and S.~Borgani\orcid{0000-0001-6151-6439}\inst{\ref{aff115},\ref{aff97},\ref{aff52},\ref{aff108}}
\and A.~S.~Borlaff\orcid{0000-0003-3249-4431}\inst{\ref{aff116},\ref{aff117},\ref{aff118}}
\and H.~Bretonni\`{e}re\inst{\ref{aff119}}
\and S.~Bruton\orcid{0000-0002-6503-5218}\inst{\ref{aff120}}
\and R.~Cabanac\orcid{0000-0001-6679-2600}\inst{\ref{aff95}}
\and A.~Calabro\orcid{0000-0003-2536-1614}\inst{\ref{aff8}}
\and A.~Cappi\inst{\ref{aff22},\ref{aff75}}
\and C.~S.~Carvalho\inst{\ref{aff93}}
\and G.~Castignani\orcid{0000-0001-6831-0687}\inst{\ref{aff73},\ref{aff22}}
\and T.~Castro\orcid{0000-0002-6292-3228}\inst{\ref{aff52},\ref{aff108},\ref{aff97},\ref{aff109}}
\and G.~Ca\~{n}as-Herrera\orcid{0000-0003-2796-2149}\inst{\ref{aff69},\ref{aff121}}
\and K.~C.~Chambers\orcid{0000-0001-6965-7789}\inst{\ref{aff122}}
\and J.~Coupon\inst{\ref{aff47}}
\and O.~Cucciati\orcid{0000-0002-9336-7551}\inst{\ref{aff22}}
\and S.~Davini\inst{\ref{aff29}}
\and G.~De~Lucia\orcid{0000-0002-6220-9104}\inst{\ref{aff52}}
\and G.~Desprez\inst{\ref{aff123}}
\and S.~Di~Domizio\orcid{0000-0003-2863-5895}\inst{\ref{aff28},\ref{aff29}}
\and H.~Dole\orcid{0000-0002-9767-3839}\inst{\ref{aff19}}
\and A.~D\'{i}az-S\'{a}nchez\orcid{0000-0003-0748-4768}\inst{\ref{aff124}}
\and J.~A.~Escartin~Vigo\inst{\ref{aff25}}
\and S.~Escoffier\orcid{0000-0002-2847-7498}\inst{\ref{aff64}}
\and I.~Ferrero\orcid{0000-0002-1295-1132}\inst{\ref{aff55}}
\and F.~Finelli\orcid{0000-0002-6694-3269}\inst{\ref{aff22},\ref{aff96}}
\and L.~Gabarra\inst{\ref{aff86},\ref{aff49}}
\and K.~Ganga\orcid{0000-0001-8159-8208}\inst{\ref{aff76}}
\and J.~Garc\'ia-Bellido\orcid{0000-0002-9370-8360}\inst{\ref{aff110}}
\and E.~Gaztanaga\orcid{0000-0001-9632-0815}\inst{\ref{aff89},\ref{aff88},\ref{aff20}}
\and K.~George\inst{\ref{aff26}}
\and F.~Giacomini\orcid{0000-0002-3129-2814}\inst{\ref{aff24}}
\and G.~Gozaliasl\orcid{0000-0002-0236-919X}\inst{\ref{aff125},\ref{aff67}}
\and A.~Gregorio\orcid{0000-0003-4028-8785}\inst{\ref{aff115},\ref{aff52},\ref{aff108}}
\and D.~Guinet\orcid{0000-0002-8132-6509}\inst{\ref{aff41}}
\and A.~Hall\orcid{0000-0002-3139-8651}\inst{\ref{aff39}}
\and H.~Hildebrandt\orcid{0000-0002-9814-3338}\inst{\ref{aff126}}
\and A.~Jimenez~Mu\~noz\orcid{0009-0004-5252-185X}\inst{\ref{aff127}}
\and J.~J.~E.~Kajava\orcid{0000-0002-3010-8333}\inst{\ref{aff128},\ref{aff129}}
\and V.~Kansal\inst{\ref{aff130},\ref{aff131},\ref{aff132}}
\and D.~Karagiannis\orcid{0000-0002-4927-0816}\inst{\ref{aff133}}
\and C.~C.~Kirkpatrick\inst{\ref{aff63}}
\and L.~Legrand\orcid{0000-0003-0610-5252}\inst{\ref{aff66}}
\and A.~Loureiro\orcid{0000-0002-4371-0876}\inst{\ref{aff134},\ref{aff135}}
\and J.~Macias-Perez\inst{\ref{aff127}}
\and M.~Magliocchetti\orcid{0000-0001-9158-4838}\inst{\ref{aff48}}
\and R.~Maoli\orcid{0000-0002-6065-3025}\inst{\ref{aff136},\ref{aff8}}
\and M.~Martinelli\orcid{0000-0002-6943-7732}\inst{\ref{aff8},\ref{aff137}}
\and C.~J.~A.~P.~Martins\orcid{0000-0002-4886-9261}\inst{\ref{aff138},\ref{aff31}}
\and S.~Matthew\inst{\ref{aff39}}
\and M.~Maturi\orcid{0000-0002-3517-2422}\inst{\ref{aff104},\ref{aff139}}
\and L.~Maurin\orcid{0000-0002-8406-0857}\inst{\ref{aff19}}
\and R.~B.~Metcalf\orcid{0000-0003-3167-2574}\inst{\ref{aff73},\ref{aff22}}
\and M.~Migliaccio\inst{\ref{aff140},\ref{aff141}}
\and P.~Monaco\inst{\ref{aff115},\ref{aff52},\ref{aff108},\ref{aff97}}
\and G.~Morgante\inst{\ref{aff22}}
\and S.~Nadathur\orcid{0000-0001-9070-3102}\inst{\ref{aff20}}
\and Nicholas~A.~Walton\orcid{0000-0003-3983-8778}\inst{\ref{aff142}}
\and A.~Peel\orcid{0000-0003-0488-8978}\inst{\ref{aff42}}
\and A.~Pezzotta\inst{\ref{aff25}}
\and V.~Popa\inst{\ref{aff85}}
\and C.~Porciani\orcid{0000-0002-7797-2508}\inst{\ref{aff71}}
\and D.~Potter\orcid{0000-0002-0757-5195}\inst{\ref{aff143}}
\and M.~P\"{o}ntinen\orcid{0000-0001-5442-2530}\inst{\ref{aff67}}
\and P.~Reimberg\orcid{0000-0003-3410-0280}\inst{\ref{aff77}}
\and P.-F.~Rocci\inst{\ref{aff19}}
\and A.~G.~S\'anchez\orcid{0000-0003-1198-831X}\inst{\ref{aff25}}
\and A.~Schneider\orcid{0000-0001-7055-8104}\inst{\ref{aff143}}
\and E.~Sefusatti\orcid{0000-0003-0473-1567}\inst{\ref{aff52},\ref{aff108},\ref{aff97}}
\and M.~Sereno\orcid{0000-0003-0302-0325}\inst{\ref{aff22},\ref{aff24}}
\and P.~Simon\inst{\ref{aff71}}
\and A.~Spurio~Mancini\orcid{0000-0001-5698-0990}\inst{\ref{aff44}}
\and S.~A.~Stanford\inst{\ref{aff144}}
\and J.~Steinwagner\inst{\ref{aff25}}
\and G.~Testera\inst{\ref{aff29}}
\and M.~Tewes\orcid{0000-0002-1155-8689}\inst{\ref{aff71}}
\and R.~Teyssier\orcid{0000-0001-7689-0933}\inst{\ref{aff145}}
\and S.~Toft\orcid{0000-0003-3631-7176}\inst{\ref{aff60},\ref{aff146},\ref{aff147}}
\and S.~Tosi\orcid{0000-0002-7275-9193}\inst{\ref{aff28},\ref{aff29},\ref{aff21}}
\and A.~Troja\orcid{0000-0003-0239-4595}\inst{\ref{aff86},\ref{aff49}}
\and M.~Tucci\inst{\ref{aff47}}
\and C.~Valieri\inst{\ref{aff24}}
\and J.~Valiviita\orcid{0000-0001-6225-3693}\inst{\ref{aff67},\ref{aff68}}
\and D.~Vergani\orcid{0000-0003-0898-2216}\inst{\ref{aff22}}
\and I.~A.~Zinchenko\inst{\ref{aff26}}}
										   
\institute{Institut f\"{u}r Planetologie, Universit\"at M\"unster, Wilhelm-Klemm-Str. 10, 48149 M\"unster, Germany\label{aff1}
\and
ESAC/ESA, Camino Bajo del Castillo, s/n., Urb. Villafranca del Castillo, 28692 Villanueva de la Ca\~nada, Madrid, Spain\label{aff2}
\and
Jodrell Bank Centre for Astrophysics, Department of Physics and Astronomy, University of Manchester, Oxford Road, Manchester M13 9PL, UK\label{aff3}
\and
Instituto de Astrof\'isica de Canarias, Calle V\'ia L\'actea s/n, 38204, San Crist\'obal de La Laguna, Tenerife, Spain\label{aff4}
\and
Departamento de Astrof\'isica, Universidad de La Laguna, 38206, La Laguna, Tenerife, Spain\label{aff5}
\and
Universit\'e PSL, Observatoire de Paris, Sorbonne Universit\'e, CNRS, LERMA, 75014, Paris, France\label{aff6}
\and
Universit\'e Paris-Cit\'e, 5 Rue Thomas Mann, 75013, Paris, France\label{aff7}
\and
INAF-Osservatorio Astronomico di Roma, Via Frascati 33, 00078 Monteporzio Catone, Italy\label{aff8}
\and
INFN section of Naples, Via Cinthia 6, 80126, Napoli, Italy\label{aff9}
\and
Centro de Estudios de F\'isica del Cosmos de Arag\'on (CEFCA), Plaza San Juan, 1, planta 2, 44001, Teruel, Spain\label{aff10}
\and
Universit\'e de Strasbourg, CNRS, Observatoire astronomique de Strasbourg, UMR 7550, 67000 Strasbourg, France\label{aff11}
\and
University of Nottingham, University Park, Nottingham NG7 2RD, UK\label{aff12}
\and
SRON Netherlands Institute for Space Research, Landleven 12, 9747 AD, Groningen, The Netherlands\label{aff13}
\and
Kapteyn Astronomical Institute, University of Groningen, PO Box 800, 9700 AV Groningen, The Netherlands\label{aff14}
\and
Universit\"at Innsbruck, Institut f\"ur Astro- und Teilchenphysik, Technikerstr. 25/8, 6020 Innsbruck, Austria\label{aff15}
\and
School of Computer Science, Merchant Venturers Building, University of Bristol, Woodland Road, Bristol, BS8 1UB, UK\label{aff16}
\and
National Astronomical Observatory of Japan, 2-21-1 Osawa, Mitaka, Tokyo 181-8588, Japan\label{aff17}
\and
INAF-Osservatorio Astronomico di Capodimonte, Via Moiariello 16, 80131 Napoli, Italy\label{aff18}
\and
Universit\'e Paris-Saclay, CNRS, Institut d'astrophysique spatiale, 91405, Orsay, France\label{aff19}
\and
Institute of Cosmology and Gravitation, University of Portsmouth, Portsmouth PO1 3FX, UK\label{aff20}
\and
INAF-Osservatorio Astronomico di Brera, Via Brera 28, 20122 Milano, Italy\label{aff21}
\and
INAF-Osservatorio di Astrofisica e Scienza dello Spazio di Bologna, Via Piero Gobetti 93/3, 40129 Bologna, Italy\label{aff22}
\and
Dipartimento di Fisica e Astronomia, Universit\`a di Bologna, Via Gobetti 93/2, 40129 Bologna, Italy\label{aff23}
\and
INFN-Sezione di Bologna, Viale Berti Pichat 6/2, 40127 Bologna, Italy\label{aff24}
\and
Max Planck Institute for Extraterrestrial Physics, Giessenbachstr. 1, 85748 Garching, Germany\label{aff25}
\and
Universit\"ats-Sternwarte M\"unchen, Fakult\"at f\"ur Physik, Ludwig-Maximilians-Universit\"at M\"unchen, Scheinerstrasse 1, 81679 M\"unchen, Germany\label{aff26}
\and
INAF-Osservatorio Astrofisico di Torino, Via Osservatorio 20, 10025 Pino Torinese (TO), Italy\label{aff27}
\and
Dipartimento di Fisica, Universit\`a di Genova, Via Dodecaneso 33, 16146, Genova, Italy\label{aff28}
\and
INFN-Sezione di Genova, Via Dodecaneso 33, 16146, Genova, Italy\label{aff29}
\and
Department of Physics "E. Pancini", University Federico II, Via Cinthia 6, 80126, Napoli, Italy\label{aff30}
\and
Instituto de Astrof\'isica e Ci\^encias do Espa\c{c}o, Universidade do Porto, CAUP, Rua das Estrelas, PT4150-762 Porto, Portugal\label{aff31}
\and
Dipartimento di Fisica, Universit\`a degli Studi di Torino, Via P. Giuria 1, 10125 Torino, Italy\label{aff32}
\and
INFN-Sezione di Torino, Via P. Giuria 1, 10125 Torino, Italy\label{aff33}
\and
INAF-IASF Milano, Via Alfonso Corti 12, 20133 Milano, Italy\label{aff34}
\and
Institut de F\'{i}sica d'Altes Energies (IFAE), The Barcelona Institute of Science and Technology, Campus UAB, 08193 Bellaterra (Barcelona), Spain\label{aff35}
\and
Port d'Informaci\'{o} Cient\'{i}fica, Campus UAB, C. Albareda s/n, 08193 Bellaterra (Barcelona), Spain\label{aff36}
\and
Institute for Theoretical Particle Physics and Cosmology (TTK), RWTH Aachen University, 52056 Aachen, Germany\label{aff37}
\and
Dipartimento di Fisica e Astronomia "Augusto Righi" - Alma Mater Studiorum Universit\`a di Bologna, Viale Berti Pichat 6/2, 40127 Bologna, Italy\label{aff38}
\and
Institute for Astronomy, University of Edinburgh, Royal Observatory, Blackford Hill, Edinburgh EH9 3HJ, UK\label{aff39}
\and
European Space Agency/ESRIN, Largo Galileo Galilei 1, 00044 Frascati, Roma, Italy\label{aff40}
\and
Universit\'e Claude Bernard Lyon 1, CNRS/IN2P3, IP2I Lyon, UMR 5822, Villeurbanne, F-69100, France\label{aff41}
\and
Institute of Physics, Laboratory of Astrophysics, Ecole Polytechnique F\'ed\'erale de Lausanne (EPFL), Observatoire de Sauverny, 1290 Versoix, Switzerland\label{aff42}
\and
UCB Lyon 1, CNRS/IN2P3, IUF, IP2I Lyon, 4 rue Enrico Fermi, 69622 Villeurbanne, France\label{aff43}
\and
Mullard Space Science Laboratory, University College London, Holmbury St Mary, Dorking, Surrey RH5 6NT, UK\label{aff44}
\and
Departamento de F\'isica, Faculdade de Ci\^encias, Universidade de Lisboa, Edif\'icio C8, Campo Grande, PT1749-016 Lisboa, Portugal\label{aff45}
\and
Instituto de Astrof\'isica e Ci\^encias do Espa\c{c}o, Faculdade de Ci\^encias, Universidade de Lisboa, Campo Grande, 1749-016 Lisboa, Portugal\label{aff46}
\and
Department of Astronomy, University of Geneva, ch. d'Ecogia 16, 1290 Versoix, Switzerland\label{aff47}
\and
INAF-Istituto di Astrofisica e Planetologia Spaziali, via del Fosso del Cavaliere, 100, 00100 Roma, Italy\label{aff48}
\and
INFN-Padova, Via Marzolo 8, 35131 Padova, Italy\label{aff49}
\and
Universit\'e Paris-Saclay, Universit\'e Paris Cit\'e, CEA, CNRS, AIM, 91191, Gif-sur-Yvette, France\label{aff50}
\and
School of Physics, HH Wills Physics Laboratory, University of Bristol, Tyndall Avenue, Bristol, BS8 1TL, UK\label{aff51}
\and
INAF-Osservatorio Astronomico di Trieste, Via G. B. Tiepolo 11, 34143 Trieste, Italy\label{aff52}
\and
Istituto Nazionale di Fisica Nucleare, Sezione di Bologna, Via Irnerio 46, 40126 Bologna, Italy\label{aff53}
\and
INAF-Osservatorio Astronomico di Padova, Via dell'Osservatorio 5, 35122 Padova, Italy\label{aff54}
\and
Institute of Theoretical Astrophysics, University of Oslo, P.O. Box 1029 Blindern, 0315 Oslo, Norway\label{aff55}
\and
Jet Propulsion Laboratory, California Institute of Technology, 4800 Oak Grove Drive, Pasadena, CA, 91109, USA\label{aff56}
\and
Department of Physics, Lancaster University, Lancaster, LA1 4YB, UK\label{aff57}
\and
von Hoerner \& Sulger GmbH, Schlossplatz 8, 68723 Schwetzingen, Germany\label{aff58}
\and
Technical University of Denmark, Elektrovej 327, 2800 Kgs. Lyngby, Denmark\label{aff59}
\and
Cosmic Dawn Center (DAWN), Denmark\label{aff60}
\and
Institut d'Astrophysique de Paris, UMR 7095, CNRS, and Sorbonne Universit\'e, 98 bis boulevard Arago, 75014 Paris, France\label{aff61}
\and
Max-Planck-Institut f\"ur Astronomie, K\"onigstuhl 17, 69117 Heidelberg, Germany\label{aff62}
\and
Department of Physics and Helsinki Institute of Physics, Gustaf H\"allstr\"omin katu 2, 00014 University of Helsinki, Finland\label{aff63}
\and
Aix-Marseille Universit\'e, CNRS/IN2P3, CPPM, Marseille, France\label{aff64}
\and
AIM, CEA, CNRS, Universit\'{e} Paris-Saclay, Universit\'{e} de Paris, 91191 Gif-sur-Yvette, France\label{aff65}
\and
Universit\'e de Gen\`eve, D\'epartement de Physique Th\'eorique and Centre for Astroparticle Physics, 24 quai Ernest-Ansermet, CH-1211 Gen\`eve 4, Switzerland\label{aff66}
\and
Department of Physics, P.O. Box 64, 00014 University of Helsinki, Finland\label{aff67}
\and
Helsinki Institute of Physics, Gustaf H{\"a}llstr{\"o}min katu 2, University of Helsinki, Helsinki, Finland\label{aff68}
\and
European Space Agency/ESTEC, Keplerlaan 1, 2201 AZ Noordwijk, The Netherlands\label{aff69}
\and
NOVA optical infrared instrumentation group at ASTRON, Oude Hoogeveensedijk 4, 7991PD, Dwingeloo, The Netherlands\label{aff70}
\and
Universit\"at Bonn, Argelander-Institut f\"ur Astronomie, Auf dem H\"ugel 71, 53121 Bonn, Germany\label{aff71}
\and
Aix-Marseille Universit\'e, CNRS, CNES, LAM, Marseille, France\label{aff72}
\and
Dipartimento di Fisica e Astronomia "Augusto Righi" - Alma Mater Studiorum Universit\`a di Bologna, via Piero Gobetti 93/2, 40129 Bologna, Italy\label{aff73}
\and
Department of Physics, Institute for Computational Cosmology, Durham University, South Road, DH1 3LE, UK\label{aff74}
\and
Universit\'e C\^{o}te d'Azur, Observatoire de la C\^{o}te d'Azur, CNRS, Laboratoire Lagrange, Bd de l'Observatoire, CS 34229, 06304 Nice cedex 4, France\label{aff75}
\and
Universit\'e Paris Cit\'e, CNRS, Astroparticule et Cosmologie, 75013 Paris, France\label{aff76}
\and
Institut d'Astrophysique de Paris, 98bis Boulevard Arago, 75014, Paris, France\label{aff77}
\and
Department of Physics and Astronomy, University of Aarhus, Ny Munkegade 120, DK-8000 Aarhus C, Denmark\label{aff78}
\and
Waterloo Centre for Astrophysics, University of Waterloo, Waterloo, Ontario N2L 3G1, Canada\label{aff79}
\and
Department of Physics and Astronomy, University of Waterloo, Waterloo, Ontario N2L 3G1, Canada\label{aff80}
\and
Perimeter Institute for Theoretical Physics, Waterloo, Ontario N2L 2Y5, Canada\label{aff81}
\and
Universit\'e Paris-Saclay, Universit\'e Paris Cit\'e, CEA, CNRS, Astrophysique, Instrumentation et Mod\'elisation Paris-Saclay, 91191 Gif-sur-Yvette, France\label{aff82}
\and
Space Science Data Center, Italian Space Agency, via del Politecnico snc, 00133 Roma, Italy\label{aff83}
\and
Centre National d'Etudes Spatiales -- Centre spatial de Toulouse, 18 avenue Edouard Belin, 31401 Toulouse Cedex 9, France\label{aff84}
\and
Institute of Space Science, Str. Atomistilor, nr. 409 M\u{a}gurele, Ilfov, 077125, Romania\label{aff85}
\and
Dipartimento di Fisica e Astronomia "G. Galilei", Universit\`a di Padova, Via Marzolo 8, 35131 Padova, Italy\label{aff86}
\and
Departamento de F\'isica, FCFM, Universidad de Chile, Blanco Encalada 2008, Santiago, Chile\label{aff87}
\and
Institut d'Estudis Espacials de Catalunya (IEEC),  Edifici RDIT, Campus UPC, 08860 Castelldefels, Barcelona, Spain\label{aff88}
\and
Institute of Space Sciences (ICE, CSIC), Campus UAB, Carrer de Can Magrans, s/n, 08193 Barcelona, Spain\label{aff89}
\and
Satlantis, University Science Park, Sede Bld 48940, Leioa-Bilbao, Spain\label{aff90}
\and
Centro de Investigaciones Energ\'eticas, Medioambientales y Tecnol\'ogicas (CIEMAT), Avenida Complutense 40, 28040 Madrid, Spain\label{aff91}
\and
Infrared Processing and Analysis Center, California Institute of Technology, Pasadena, CA 91125, USA\label{aff92}
\and
Instituto de Astrof\'isica e Ci\^encias do Espa\c{c}o, Faculdade de Ci\^encias, Universidade de Lisboa, Tapada da Ajuda, 1349-018 Lisboa, Portugal\label{aff93}
\and
Universidad Polit\'ecnica de Cartagena, Departamento de Electr\'onica y Tecnolog\'ia de Computadoras,  Plaza del Hospital 1, 30202 Cartagena, Spain\label{aff94}
\and
Institut de Recherche en Astrophysique et Plan\'etologie (IRAP), Universit\'e de Toulouse, CNRS, UPS, CNES, 14 Av. Edouard Belin, 31400 Toulouse, France\label{aff95}
\and
INFN-Bologna, Via Irnerio 46, 40126 Bologna, Italy\label{aff96}
\and
IFPU, Institute for Fundamental Physics of the Universe, via Beirut 2, 34151 Trieste, Italy\label{aff97}
\and
INAF, Istituto di Radioastronomia, Via Piero Gobetti 101, 40129 Bologna, Italy\label{aff98}
\and
Centre de Calcul de l'IN2P3/CNRS, 21 avenue Pierre de Coubertin 69627 Villeurbanne Cedex, France\label{aff99}
\and
University of Applied Sciences and Arts of Northwestern Switzerland, School of Engineering, 5210 Windisch, Switzerland\label{aff100}
\and
Department of Mathematics and Physics E. De Giorgi, University of Salento, Via per Arnesano, CP-I93, 73100, Lecce, Italy\label{aff101}
\and
INAF-Sezione di Lecce, c/o Dipartimento Matematica e Fisica, Via per Arnesano, 73100, Lecce, Italy\label{aff102}
\and
INFN, Sezione di Lecce, Via per Arnesano, CP-193, 73100, Lecce, Italy\label{aff103}
\and
Institut f\"ur Theoretische Physik, University of Heidelberg, Philosophenweg 16, 69120 Heidelberg, Germany\label{aff104}
\and
Universit\'e St Joseph; Faculty of Sciences, Beirut, Lebanon\label{aff105}
\and
Junia, EPA department, 41 Bd Vauban, 59800 Lille, France\label{aff106}
\and
SISSA, International School for Advanced Studies, Via Bonomea 265, 34136 Trieste TS, Italy\label{aff107}
\and
INFN, Sezione di Trieste, Via Valerio 2, 34127 Trieste TS, Italy\label{aff108}
\and
ICSC - Centro Nazionale di Ricerca in High Performance Computing, Big Data e Quantum Computing, Via Magnanelli 2, Bologna, Italy\label{aff109}
\and
Instituto de F\'isica Te\'orica UAM-CSIC, Campus de Cantoblanco, 28049 Madrid, Spain\label{aff110}
\and
CERCA/ISO, Department of Physics, Case Western Reserve University, 10900 Euclid Avenue, Cleveland, OH 44106, USA\label{aff111}
\and
Laboratoire Univers et Th\'eorie, Observatoire de Paris, Universit\'e PSL, Universit\'e Paris Cit\'e, CNRS, 92190 Meudon, France\label{aff112}
\and
Dipartimento di Fisica e Scienze della Terra, Universit\`a degli Studi di Ferrara, Via Giuseppe Saragat 1, 44122 Ferrara, Italy\label{aff113}
\and
Istituto Nazionale di Fisica Nucleare, Sezione di Ferrara, Via Giuseppe Saragat 1, 44122 Ferrara, Italy\label{aff114}
\and
Dipartimento di Fisica - Sezione di Astronomia, Universit\`a di Trieste, Via Tiepolo 11, 34131 Trieste, Italy\label{aff115}
\and
NASA Ames Research Center, Moffett Field, CA 94035, USA\label{aff116}
\and
Kavli Institute for Particle Astrophysics \& Cosmology (KIPAC), Stanford University, Stanford, CA 94305, USA\label{aff117}
\and
Bay Area Environmental Research Institute, Moffett Field, California 94035, USA\label{aff118}
\and
Department of Astronomy and Astrophysics, University of California, Santa Cruz, 1156 High Street, Santa Cruz, CA 95064, USA\label{aff119}
\and
Minnesota Institute for Astrophysics, University of Minnesota, 116 Church St SE, Minneapolis, MN 55455, USA\label{aff120}
\and
Institute Lorentz, Leiden University, Niels Bohrweg 2, 2333 CA Leiden, The Netherlands\label{aff121}
\and
Institute for Astronomy, University of Hawaii, 2680 Woodlawn Drive, Honolulu, HI 96822, USA\label{aff122}
\and
Department of Astronomy \& Physics and Institute for Computational Astrophysics, Saint Mary's University, 923 Robie Street, Halifax, Nova Scotia, B3H 3C3, Canada\label{aff123}
\and
Departamento F\'isica Aplicada, Universidad Polit\'ecnica de Cartagena, Campus Muralla del Mar, 30202 Cartagena, Murcia, Spain\label{aff124}
\and
Department of Computer Science, Aalto University, PO Box 15400, Espoo, FI-00 076, Finland\label{aff125}
\and
Ruhr University Bochum, Faculty of Physics and Astronomy, Astronomical Institute (AIRUB), German Centre for Cosmological Lensing (GCCL), 44780 Bochum, Germany\label{aff126}
\and
Univ. Grenoble Alpes, CNRS, Grenoble INP, LPSC-IN2P3, 53, Avenue des Martyrs, 38000, Grenoble, France\label{aff127}
\and
Department of Physics and Astronomy, Vesilinnantie 5, 20014 University of Turku, Finland\label{aff128}
\and
Serco for European Space Agency (ESA), Camino bajo del Castillo, s/n, Urbanizacion Villafranca del Castillo, Villanueva de la Ca\~nada, 28692 Madrid, Spain\label{aff129}
\and
ARC Centre of Excellence for Dark Matter Particle Physics, Melbourne, Australia\label{aff130}
\and
Centre for Astrophysics \& Supercomputing, Swinburne University of Technology, Victoria 3122, Australia\label{aff131}
\and
W.M. Keck Observatory, 65-1120 Mamalahoa Hwy, Kamuela, HI, USA\label{aff132}
\and
Department of Physics and Astronomy, University of the Western Cape, Bellville, Cape Town, 7535, South Africa\label{aff133}
\and
Oskar Klein Centre for Cosmoparticle Physics, Department of Physics, Stockholm University, Stockholm, SE-106 91, Sweden\label{aff134}
\and
Astrophysics Group, Blackett Laboratory, Imperial College London, London SW7 2AZ, UK\label{aff135}
\and
Dipartimento di Fisica, Sapienza Universit\`a di Roma, Piazzale Aldo Moro 2, 00185 Roma, Italy\label{aff136}
\and
INFN-Sezione di Roma, Piazzale Aldo Moro, 2 - c/o Dipartimento di Fisica, Edificio G. Marconi, 00185 Roma, Italy\label{aff137}
\and
Centro de Astrof\'{\i}sica da Universidade do Porto, Rua das Estrelas, 4150-762 Porto, Portugal\label{aff138}
\and
Zentrum f\"ur Astronomie, Universit\"at Heidelberg, Philosophenweg 12, 69120 Heidelberg, Germany\label{aff139}
\and
Dipartimento di Fisica, Universit\`a di Roma Tor Vergata, Via della Ricerca Scientifica 1, Roma, Italy\label{aff140}
\and
INFN, Sezione di Roma 2, Via della Ricerca Scientifica 1, Roma, Italy\label{aff141}
\and
Institute of Astronomy, University of Cambridge, Madingley Road, Cambridge CB3 0HA, UK\label{aff142}
\and
Department of Astrophysics, University of Zurich, Winterthurerstrasse 190, 8057 Zurich, Switzerland\label{aff143}
\and
Department of Physics and Astronomy, University of California, Davis, CA 95616, USA\label{aff144}
\and
Department of Astrophysical Sciences, Peyton Hall, Princeton University, Princeton, NJ 08544, USA\label{aff145}
\and
Niels Bohr Institute, University of Copenhagen, Jagtvej 128, 2200 Copenhagen, Denmark\label{aff146}
\and
Cosmic Dawn Center (DAWN)\label{aff147}}    

\date{Received ---; accepted ---}

  \abstract{The \Euclid mission is expected to image millions of galaxies at high resolution, providing an extensive dataset with which to study galaxy evolution. Because galaxy morphology is both a fundamental parameter and one that is hard to determine for large samples, we investigate the application of deep learning in predicting the detailed morphologies of galaxies in \Euclid using \texttt{Zoobot}, a convolutional neural network pretrained with 450\,000 galaxies from the Galaxy Zoo project. We adapted \texttt{Zoobot} for use with emulated \Euclid images generated based on \HST COSMOS images and with labels provided by volunteers in the Galaxy Zoo: Hubble project. We experimented with different numbers of galaxies and various magnitude cuts during the training process. We demonstrate that the trained \texttt{Zoobot} model successfully measures detailed galaxy morphology in emulated \Euclid images. It effectively predicts whether a galaxy has features and identifies and characterises various features, such as spiral arms, clumps, bars, discs, and central bulges. When compared to volunteer classifications, \texttt{Zoobot} achieves mean vote fraction deviations of less than 12\% and an accuracy of above 91\% for the confident volunteer classifications across most morphology types. However, the performance varies depending on the specific morphological class. For the global classes, such as disc or smooth galaxies, the mean deviations are less than 10\%, with only 1000 training galaxies necessary to reach this performance. On the other hand, for more detailed structures and complex tasks, such as detecting and counting spiral arms or clumps, the deviations are slightly higher, of namely around 12\% with 60\,000 galaxies used for training. In order to enhance the performance on complex morphologies, we anticipate that a larger pool of labelled galaxies is needed, which could be obtained using crowd sourcing. We estimate that, with our model, the detailed morphology of approximately $800$ million galaxies of the Euclid Wide Survey could be reliably measured and that approximately $230$ million of these galaxies would display features. Finally, our findings imply that the model can be effectively adapted to new morphological labels. We demonstrate this adaptability by applying \texttt{Zoobot} to peculiar galaxies. In summary, our trained \texttt{Zoobot} CNN can readily predict morphological catalogues for \Euclid images.}


   \titlerunning{\Euclid\/ preparation. XLIII. Measuring detailed galaxy morphologies for \Euclid with machine learning}
   \authorrunning{B. Aussel et al.}

   \maketitle

\section{Introduction}

\Euclid is a space-based mission of the European Space Agency (ESA) launched in 2023. Operating in the optical and near-infrared, its primary goal is to achieve a better understanding of the accelerated expansion of the Universe and the nature of dark matter (\citealt{laureijs_2011}), and it has a broad range of secondary goals. The Euclid Wide Survey \citep{Scaramella-EP1} will cover approximately 15\,000 deg$^2$ of the extragalactic sky, corresponding to $36\,\%$ of the celestial sphere. The angular resolution of the \Euclid visible imager (VIS, \citealt{Cropper_2016}) of $\ang{;;0.2}$ is comparable to that of the \HST (HST) Advanced Camera for Surveys (ACS), while the field of view of $0.53\,$deg$^2$ is 175 times larger. \Euclid is expected to image billions of galaxies to $z\approx2$ and to a depth of $24.5$\,mag at $10\sigma$ for extended sources (galaxy sizes of $\sim\ang{;;0.3}$) in the VIS band (\citealt{laureijs_2011}). It will therefore resolve the internal morphology of an unprecedented number of galaxies, estimated at approximately 250 million \citep{Bretonniere-EP13}. Many will display complex features, such as clumps, bars, spiral arms, and/or bulges.

Large samples of galaxies with measured detailed morphologies are crucial to understand galaxy evolution and its impact on galaxy structure \citep{Masters_2019}. For example, bars are believed to funnel gas inwards from the spiral arms and may lead to the growth of a central bulge \citep{Sakamoto_1999, Masters_2010, Kruk_2018}. \textit{Euclid} will provide an unprecedentedly large dataset of galaxy images with resolved morphology \citep{Bretonniere-EP13}, which is essential for studies of galaxy evolution. This includes studying the evolution of morphology with redshift and environment, where \Euclid will offer the necessary statistics for analysing trends in stellar mass, colour, and so on, thereby enabling the distinction of complex correlations. However, accurately measuring the morphologies and structures of galaxies will be a challenge. 

Numerous methods for diverse applications have been developed to quantify galaxy morphology from imaging data. These include visual classifications \citep{Hubble_1926, deVaucouleurs_1959,Lintott_2008, Bait_2017}, non-parametric morphologies \citep{Conselice_2003, Lotz_2004}, galaxy profile fitting \citep{Sersic_1968, Peng_2002}, and machine learning techniques \citep{Huertas-Company_2015, Vega-Ferrero_2021}. Many approaches perform measurements in an automated or semi-automated manner, while some facilitate the decomposition of galaxies into multiple constituents, such as bulges and discs, or combine several parameters to scrutinise current models. In a recent study, \citet{Bretonniere-EP26} compared the performance of five modern morphology fitting codes on simulated galaxies mimicking incoming \Euclid images. These galaxies were generated as simplified models with single-S\'ersic and double-S\'ersic profiles and as neural network-generated galaxies with more detailed morphologies. This \Euclid Morphology Challenge was primarily designed to quantify galaxy structures using analytic functions that describe the shape of the surface brightness profile. However, it also highlighted the necessity for additional efforts to fully capture the richness of the detailed morphologies that \Euclid will uncover on a larger scale.

For several decades now, expert visual classifications have proven to be successful in measuring detailed morphology \citep{Hubble_1926, deVaucouleurs_1959, Sandage_1961,vandenBergh_1976,deVaucouleurs_1991,Baillard_2011,Bait_2017}. However, they do not scale well to large surveys and reproducibility is challenging.

The Galaxy Zoo project \citep{Lintott_2008} was set up to harness the collective efforts of thousands of volunteers to classify galaxies from the Sloan Digital Sky Survey (SDSS). With Galaxy Zoo, the number of classified galaxies has significantly increased, with more than 1 million galaxies classified so far. The capability of humans to collectively recognise detailed and faint features in galaxies is unrivalled. However, the number of volunteers on the citizen science platform does not scale well with the sizes of the next generation of surveys, such as those by the Large Synoptic Survey Telescope (LSST, \citealt{Ivezic_2019}) of the Vera Rubin Observatory and by \Euclid. \Euclid will image more than a billion galaxies \citep{laureijs_2011}. It is unfeasible to classify such a large sample with citizen science alone. 

This problem can be solved with machine learning. Machine learning has been shown many times to be a powerful tool for classifying galaxy morphology \citep{Dieleman_2015, Huertas-Company_2015, DominguezSanchez_2018, DominguezSanchez_2019, Cheng_2020, Vega-Ferrero_2021, Walmsley_2022_1}. Supervised approaches using convolutional neural networks (CNNs) have proven to be effective for this task. \citet{Walmsley_2022_1} showed that the Galaxy Zoo volunteer responses can be used to train a deep learning model, called \texttt{Zoobot} \citep{Walmsley_2023}, which is able to automatically predict the volunteer labels and therefore the detailed morphologies of galaxies.

The goal of the present study is to evaluate the feasibility of predicting detailed morphologies for emulated \Euclid galaxy images with \texttt{Zoobot} and to test the performance. For this, we used emulated \Euclid images based on the Cosmic Evolution Survey (COSMOS, \citealt{Scoville_2007}). We trained \texttt{Zoobot} and assessed its performance on these images using morphology labels provided by volunteers in the Galaxy Zoo: Hubble (GZH, \citealt{Willett_2017}) citizen science project. Ultimately, the goal is to apply \texttt{Zoobot} to the future \Euclid galaxy images to generate automated detailed morphology predictions.

This paper is structured as follows: In Sect.\,\ref{sec:Data}, the volunteer morphology classifications from GZH and their corresponding HST COSMOS images are introduced. We explain how these images were converted to emulated \Euclid images. The \texttt{Zoobot} CNN and the process of fine-tuning is presented in Sect.\,\ref{sec:Zoobot}. In Sect.\,\ref{sec:Training}, we describe the training of \texttt{Zoobot} for the GZH labels and emulated \Euclid images. We also describe the different experiments that we conducted in this study. In Sect.\,\ref{sec:Results}, we present and discuss our results. First, we show comparisons of the model trained with different data. We then evaluate the model predictions of the best-performing model in detail. Furthermore, we compare the performance on emulated \Euclid images and on the original \textit{Hubble} images. An example of fine-tuning \texttt{Zoobot} to a new morphology class (finding peculiar galaxies) is presented in Sect.\,\ref{sec:Peculiar}. Finally, we summarise our findings and provide an outlook towards the real \Euclid images in Sect.\,\ref{sec:Conclusions}.


\section{Data}\label{sec:Data}
 
In this study, we aim to generate automated detailed morphology predictions on emulated \Euclid images, test our pipeline, and evaluate its performance to be able to estimate the quality of future predictions.

To emulate the future \Euclid images from existing galaxy images, these need to have at least the same spatial resolution and depth at approximately the same wavelength range as VIS \citep{Cropper_2016}. As we are following a supervised deep learning approach, these existing galaxy images need to have reliable morphology labels to train our model and evaluate our results. All these requirements are fulfilled with the COSMOS \citep{Scoville_2007} galaxy images labelled by volunteers in the GZH \citep{Willett_2017} project. 

\subsection{Images} 

\subsubsection{\HST COSMOS images}

We used COSMOS galaxy images \citep{Scoville_2007}. For the COSMOS survey, an area of $1.64$\,deg$^2$ was observed with the ACS Wide Field Channel of HST in the F814W filter with an angular resolution of $\ang{;;0.09}$ \citep{Scoville_2007a,Koekemoer_2007}. We used the publicly available mosaics in the FITS format with a final drizzle pixel scale of $\ang{;;0.03}$. The limiting point source depth at $5\sigma$ is 27.2\,mag. Therefore, the depth and resolution are better than those estimated for \Euclid (24.5\,mag at $10\sigma$ for sources with $\sim\,\ang{;;0.3}$ extent and $\ang{;;0.2}$, \citealt{Cropper_2016}). The wavelength range of the \Euclid VIS band (550--900\,nm) includes the F814W band of Hubble. While ideally, data from other HST filters, such as F606W, could be combined to emulate the \Euclid VIS observations, the extensive COSMOS survey provides only single-band F814W images. We used the same dataset from COSMOS that was used in GZH \citep{Willett_2017}. For the morphological classifications by the volunteers, \citet{Willett_2017} applied a magnitude restriction of $m_{I814W}<23.5$, yielding a total of 84\,954 galaxies.

\subsubsection{Emulated \Euclid COSMOS images}\label{sec:euclid_images}

We used available emulated \Euclid images generated from the previously described COSMOS images that were created as part of the \Euclid Data Challenge 2, with the goal of testing the steps of the data processing for \Euclid. The area covered by these images is $\ang{1.2;;}\times\ang{1.2;;}$, which is smaller than the original COSMOS field. Therefore, only 76\,176 images from the GZH COSMOS set were available. The images are emulated to be \Euclid VIS-like and are expected to match the properties of \Euclid data, on a reduced scale.

\begin{figure}[htbp!]
\centering
\includegraphics[width=\hsize]{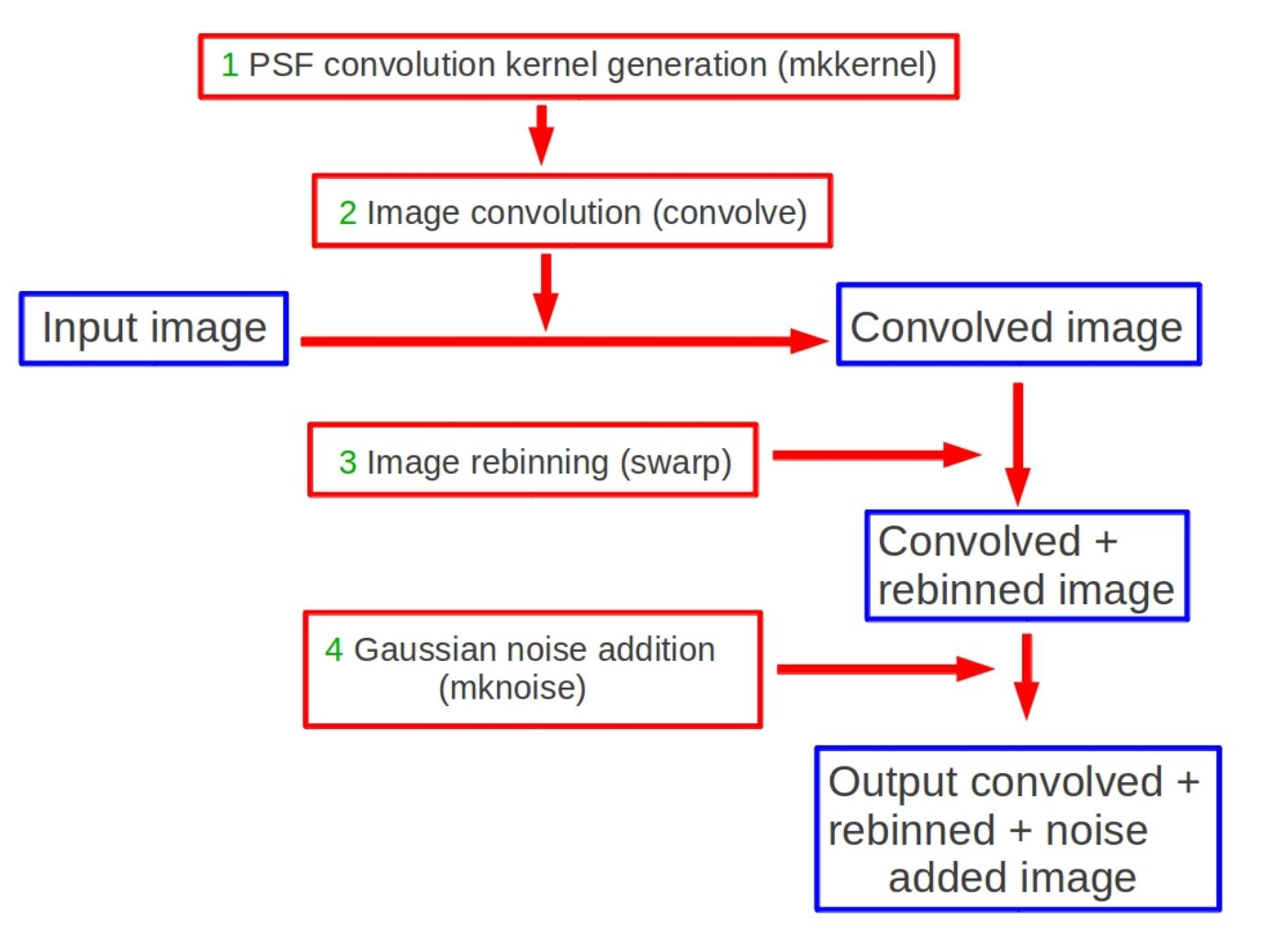}
\caption{Data pipeline scheme for the emulated \Euclid VIS images created as part of the \Euclid Data Challenge 2. The green numbers correspond to the numbers of the description of the pipeline given in the text.}
\label{fig:euclid_pipeline}
\end{figure}

\begin{figure}[htbp!]%
  \includegraphics[width=\linewidth]{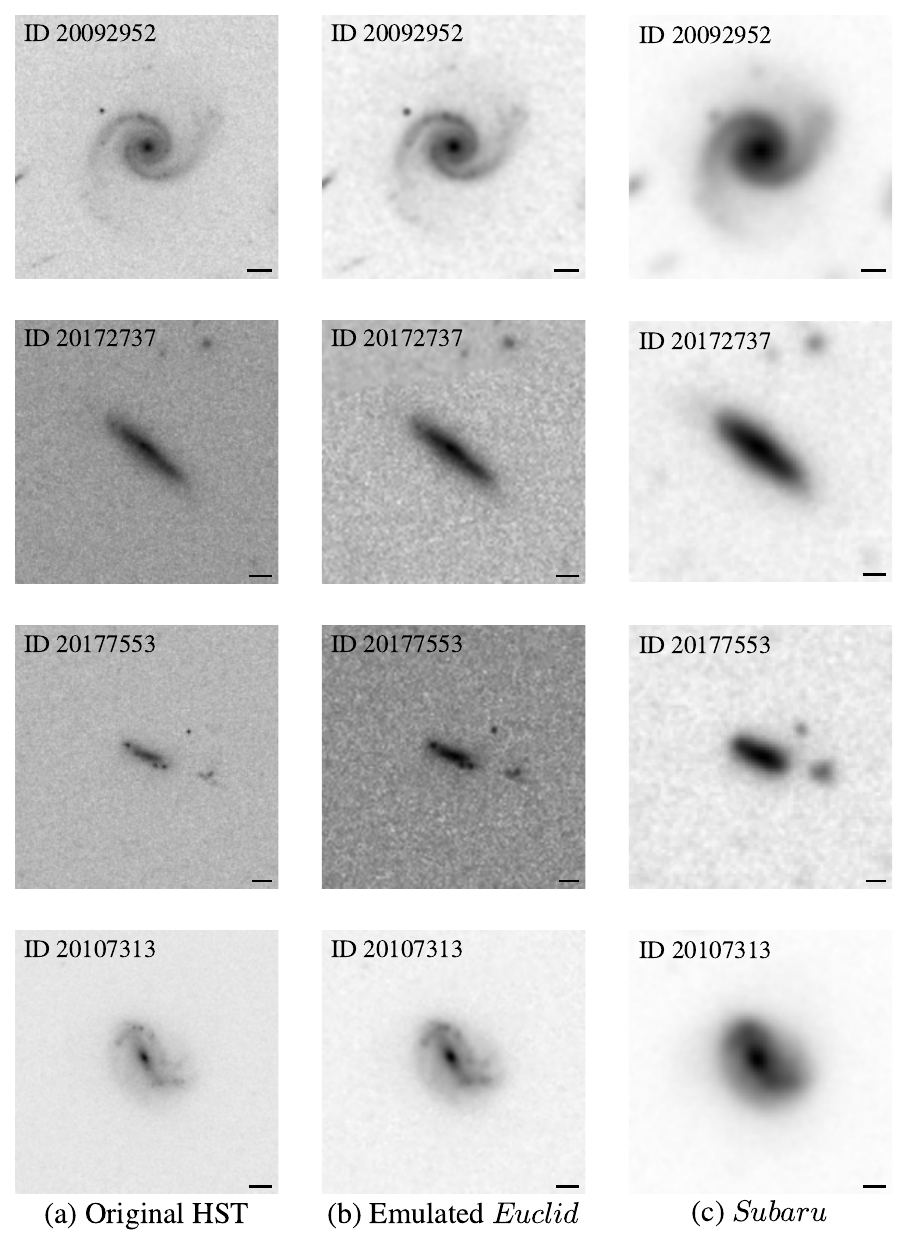}%
  \caption{Examples of galaxy images (inverted greyscale) of different morphological types (image IDs 20092952, 20172737, 20177553, 20107313): (a) from the original HST COSMOS dataset, (b) from the emulated \Euclid VIS dataset, and (c) from the \textit{Subaru} dataset. The images are scaled with galaxy size using three times the Kron radius. The black bars represent a length of $\ang{;;1}$. The image IDs are the unique identifiers for the galaxies of the COSMOS survey \citep{Griffith_2012}.}%
  \label{fig:example_hubble_euclid_images}
\end{figure}

The original HST COSMOS images were rebinned and smoothed to the \Euclid pixel scale ($\ang{;;0.1}$, \citealt{laureijs_2011}), convolved with a kernel of the difference between the HST ACS and \Euclid VIS point spread function (PSF) to emulate the resolution of \Euclid ($\ang{;;0.2}$) and with random Gaussian noise added in order to match the \Euclid VIS depth (24.5\,mag for galaxy sizes of $\sim\,\ang{;;0.3}$, \citealt{Cropper_2016}). The emulation software takes as input a high-resolution image (HST COSMOS image in this case) and processes it to emulate a VIS-like image, taking the following steps (see Fig.\,\ref{fig:euclid_pipeline}):
\begin{enumerate}
    \item First, the software generates an analytical kernel according to the input image PSF of HST ACS and the PSF of the \Euclid VIS instrument.
    \item It then convolves the input image according to the previously generated kernel.
    \item Subsequently, it performs the rebinning of the convolved image to the required pixel scale ($\ang{;;0.1}$).
    \item Finally, Gaussian noise is added to each pixel to reproduce the desired depth in output. 
\end{enumerate}   

For all galaxies of our dataset, we extracted cutouts from the available emulated \Euclid greyscale FITS files with the galaxy in the centre. The sizes of the cutouts were based on the sizes of the galaxies, using three times the Kron radius ($3\times$\texttt{KRON\_RADIUS\_HI} in \citealt{Griffith_2012}) for each galaxy in order to appear large enough to identify features, but not exceeding the image boundaries. We chose the Kron radius as a measure of galaxy size as it is least sensitive to the galaxy type. With this, the influence of relatively smaller galaxy sizes at higher redshifts on the performance of the network was taken out. The size of the images varies between $\ang{;;10.5}$ and $\ang{;;38.3}$, with a median of $\ang{;;12.5}$. As in \citet{Willett_2017}, we applied an arcsinh intensity mapping to the images to avoid a saturation of galaxy centres, while increasing the appearance of faint features. We saved the resulting cutouts as $300\times300$ pixel images in the \texttt{JPG} format to reduce the required memory. To conclude, the images have different pixel scales, but approximately the same relative galaxy size compared to the background.

To measure the impact of the lower resolution and noise of the \Euclid images on the galaxy classifications, we also created $300\times300$ pixel \texttt{JPG} cutouts for the original HST COSMOS images with an arcsinh intensity mapping. Additionally, we created similar cutouts for the same galaxies imaged by the ground-based \textit{Subaru} telescope \citep{Kaifu_2000,Taniguchi_2007}. To illustrate the effect of the emulation, we show in Fig.\,\ref{fig:example_hubble_euclid_images} example galaxy images with different morphologies (a) from the original HST COSMOS dataset, (b) from the emulated \Euclid dataset and (c) from the \textit{Subaru} dataset. These examples demonstrate that although the morphology is still identifiable, in general, the \Euclid images have a lower resolution, potentially leading to different classifications, especially for faint galaxies. 

\subsection{Volunteer labels}

We used the GZH volunteer classifications \citep{Willett_2017} for the same galaxies for which the previously described emulated \Euclid images were created. Volunteers on the citizen science project answered a series of questions about the morphology of a set of galaxy images. GZH used COSMOS images with `pseudo-colour'. The $I_{814W}$ data was used as an illumination map and the colour information was provided from the $B_J$, $r^+$, and $i^+$ filters of the Subaru telescope \citep{Griffith_2012}. Thus, the galaxy images shown to the volunteers had HST's angular resolution for the intensity, but the colour gradients were at ground-based resolution. The size of the cutouts corresponded to the galaxy size. Thus, the galaxies had different resolutions but relatively the same size, similar to our emulated \Euclid images. An arcsinh intensity mapping was applied before the images were shown as $424\times424$ pixels \texttt{PNG}s to the volunteers. 

The series of questions, asked to the volunteers, was structured as a decision tree \citep{Willett_2017} shown in Fig\,\ref{fig:gz_decision_tree}. Some questions were only asked if for the previous question a certain answer was selected. The decision tree was designed similarly to that used in Galaxy Zoo 2 (GZ2, \citealt{Willett_2013}) with some differences, involving questions for clumpiness, as expected for the high-redshift galaxies in the COSMOS dataset. We used the published dataset from \citet{Willett_2017}, which contains for every galaxy and for every classification the number of volunteers that answered the question and the respective vote fractions for each answer. It also provides metadata, such as photometric redshifts and magnitudes. As mentioned before, the publicly available dataset has a restriction of $m_{I814W}<23.5$, meaning that no labels are available for fainter galaxies. We used the GZH volunteer classifications for all available 76\,176 emulated \Euclid galaxy images.

\section{\texttt{Zoobot}}\label{sec:Zoobot}

The newly developed and publicly released Python package \texttt{Zoobot} \citep{Walmsley_2023} is a CNN trained for predicting detailed galaxy morphology, such as bars, spiral arms, and discs. In this section, we describe the \texttt{Zoobot} CNN and how we adapted it to the emulated \Euclid images with the corresponding GZH volunteer labels.

\subsection{Bayesian neural network: \texttt{Zoobot}}

Zoobot was initially developed to automatically predict detailed morphology for Dark Energy Camera Legacy Survey (DECaLS) \citep{Dey_2019} DR5 galaxy images \citep{Walmsley_2022_1}. It was trained on the corresponding volunteer classifications from the Galaxy Zoo: DECaLS (GZD) GZD-5 campaign. The 249\,581 GZD-5 volunteer classifications were used for training \texttt{Zoobot} on the questions in the GZD-5 decision tree. The volunteer responses for the different questions had different uncertainties, depending on how many volunteers answered a question for a specific galaxy image.

The Bayesian \texttt{Zoobot} CNN learns from all volunteer responses while taking the corresponding uncertainty into account \citep{Walmsley_2022_1}. Thus, all GZD-5 galaxies could be included in the training. \texttt{Zoobot} was trained on all classification tasks (all questions of the GZD-5 decision tree) simultaneously, leading to shared representations of the galaxies and to increased performance for all tasks. The base architecture of \texttt{Zoobot} is the EfficientNet B0 model \citep{Tan_2019} with a modified final output layer \citep{Walmsley_2022_1}. The layer consists of one output unit per answer of the decision tree, giving predictions between 1 and 100 using softmax or sigmoid activation. \texttt{Zoobot} does not predict discrete classes, but Dirichlet-Multinomial posteriors that can be transformed into predicted vote fractions. This is achieved by using a Dirichlet-Multinomial loss function for each question $q$
\begin{equation}
    \mathcal{L}_q=\sum_q\int \text{Multinomial}(\Vec{k_q}|\Vec{\rho},N_q)\text{Dirichlet}(\Vec{\rho}|\Vec{\alpha})d\Vec{\rho},\label{eq:dirichlet_dist}
\end{equation}
with the total number of responses $N_q$ to the question $q$, $\Vec{k_q}$ the ground truth number of votes for each answer, and $\Vec{\rho}$ the probabilities of a volunteer giving each answer. The model predicts the Dirichlet parameters $\Vec{\alpha}=\Vec{f_q}$ to the answers measured via the values of the output units of the final layer. Each vector has one element per answer. The integral is analytic as Multinomial and Dirichlet distributions are conjugates. The loss is then applied by summing over all questions of the decision tree
\begin{equation}
    \ln \mathcal{L} = \sum_q \mathcal{L}_q,\label{eq:dirichlet_sum}
\end{equation}
with the assumption that answers to different questions are independent.
The loss naturally handles volunteer votes with different uncertainties (different number of responses), as, for example, questions with no answers do not influence the gradients in training, since $\partial L_q(\Vec{k_q}=0,N_q=0,\Vec{\alpha})/ \partial \alpha=0$. We refer the reader to \citet{Walmsley_2022_1} and \citet{Walmsley_2022_3} for further details.

\texttt{Zoobot} is therefore well suited for our goal of automatically predicting detailed morphology for \Euclid galaxy images. With \texttt{Zoobot}, we can train on all available emulated \Euclid galaxies with their GZH labels, since it takes the uncertainty of the volunteer answers into account. We have to train only one model for all galaxy morphology types, since \texttt{Zoobot} is trained on all questions simultaneously. Rather than just discrete classifications, we generate posteriors.

\subsection{Transfer learning}

The trained \texttt{Zoobot} models can be adapted (`fine-tuned') to solve a new task for galaxy images \citep{Walmsley_2023}. This adaption of a previously trained machine learning model to a new problem is called transfer learning \citep{Lu_2015}. Instead of retraining all model parameters, the original model architecture and the corresponding parameters (weights) learned from the previous training can be reused. Far fewer new labels for the same performance are required using transfer learning compared to training from scratch \citep{DominguezSanchez_2019,Walmsley_2022_2}. In \citet{Walmsley_2022_2} the adaption of \texttt{Zoobot} to the new problem of finding ring galaxies is described. The pretrained \texttt{Zoobot} models outperformed models built from scratch, especially when the number of images involved in the training was limited. Pretraining on all GZD-5 tasks, involving the usage of shared representations, also leads to higher accuracy for finding ring galaxies than pretraining on only a single task.

In \citet{Walmsley_2022_3} the GZ-Evo dataset was introduced, which is a combined dataset from all major Galaxy Zoo campaigns. The included campaigns were Galaxy Zoo 2 (GZ2, \citealt{Willett_2013}) trained on galaxy images from the Sloan Digital Sky Survey (SDSS) Data Release 7, Galaxy Zoo: CANDELS (GZC, \citealt{Simmons_2017}) trained on galaxy images from the Cosmic Assembly Near-infrared Deep Extragalactic Legacy survey (CANDELS) also involving HST images \citep{Grogin_2011}, and the previously described GZD-5 \citep{Walmsley_2022_1} and GZH \citep{Willett_2017}. Additionally, Galaxy Zoo labels from the Mayall \textit{z}-band Legacy Survey (MzLS) and the Beijing-Arizona Sky Survey (BASS, \citealt{Dey_2019}) were used, which are part of Galaxy Zoo DESI \citep{Walmsley_2023_2}. \texttt{Zoobot} was trained on all 206 possible morphology classifications of the different campaigns simultaneously, with the involved Dirichlet loss naturally handling unknown answers from different decision trees \citep{Walmsley_2022_3}. Pretraining with GZ-Evo shows further improvements for the task of finding ring galaxies compared to direct training. With training from different campaigns, \citet{Walmsley_2022_3} hypothesise that because the model was trained on all galaxy images from different campaigns (having different redshifts and magnitudes) and on all possible questions, the model builds a galaxy representation of high generalization. Therefore, we expect this model to be best suited to be adapted to our new tasks. 

We thus used a version of \texttt{Zoobot} pretrained on a modified GZ-Evo catalogue, specifically pretrained on all major Galaxy Zoo campaigns with the exception of GZH in order to not influence our results when training to the GZH decision tree. In total, $450\,000$ galaxy images with volunteer classifications were involved in the pretraining. We also conducted experiments with versions of \texttt{Zoobot} pretrained with different datasets (pretrained on GZD-5 galaxies and without pretraining). The results for these models are presented in Appendix\,\ref{sec:appendix_weights}. We adapted the pretrained \texttt{Zoobot} model to our new problem. This involved two new tasks simultaneously: (i) training on new images, namely the emulated \Euclid VIS images, and (ii) training on a new decision tree.

\section{Training}\label{sec:Training}

In this section, we describe how we used the GZH volunteer labels to train \texttt{Zoobot} (Sect.\,\ref{sec:prep_datasets}). Furthermore, we describe the experiments we conducted for the training, that is, restricting the magnitude and number of examples used for training (Sect.\,\ref{sec:experiments}). Lastly, we present how each model was trained in more detail (Sect.\,\ref{sec:training_zoobot}).

\subsection{Preparing the datasets}\label{sec:prep_datasets}

Unlike the GZD-5 decision tree used in \citet{Walmsley_2022_1}, the GZH decision tree incorporates questions that have multiple possible answers, although not all leading to the same subsequent question (see Fig.\,\ref{fig:gz_decision_tree} and \citealt{Willett_2017}). Since \texttt{Zoobot} does not support this type of structure, we simply excluded the subsequent questions associated with such cases. The remaining questions and their corresponding answers used in this study can be found in Table\,\ref{tab:questions}. Moreover, similar to \citet{Walmsley_2022_1}, we used the raw vote counts as we fine-tuned previously trained \texttt{Zoobot} models that have already been trained on the raw vote counts. Moreover, the used Dirichlet-Multinomial loss (see Eq.\,\ref{eq:dirichlet_dist}) is statistically only valid when using raw vote counts. Assessing \texttt{Zoobot}'s performance when considering votes weighted by user performance or debiased for observational effects is beyond the scope of this research.

Additionally, we provide the average number of volunteer responses for each question in Table\,\ref{tab:questions}. Furthermore, we list the fraction $f_{\textrm{rel}}$ of galaxies for which the question is deemed relevant. We define a galaxy to be relevant for a specific question when at least half of the volunteers answered that question (for example measuring the number of spiral arms is only meaningful if the majority of volunteers classified the galaxy as spiral in the previous question), similar to the approach taken by \citet{Walmsley_2022_1}. Since every volunteer responded to the initial question of `smooth-or-featured', this question has the highest number of responses. However, with the exception of the `how-rounded' question, all subsequent questions were asked only if the answer to the first question was `featured'. Consequently, the number of responses decreases substantially as one progresses in the decision tree, resulting in greater uncertainty. As previously mentioned, \texttt{Zoobot} is able to learn from uncertain volunteer responses. 

\begin{table}[htbp!]
\begin{center}
    \caption{Questions and corresponding answers from GZH used for training  \texttt{Zoobot}. Additionally, we list the mean number of volunteer responses $N$ for every question and the fraction of relevant galaxies $f_{\textrm{rel}}$, i.e. where at least half of the volunteers answered the question.}
    \label{tab:questions}
    \tiny
    \begin{tabular}{p{2.3cm}p{3cm}rr}
    \hline
        Question & Answers & $N$ & $f_{\textrm{rel}}$\\ \hline
        smooth-or-featured & smooth, features, artifact & 46.1 & $100.0\,\%$ \\
        disc-edge-on & yes, no & 8.1 & $6.1\,\%$ \\
        has-spiral-arms & yes, no & 6.5& $4.9\,\%$ \\
        bar & yes, no & 7.1 & $6.1\,\%$ \\
        bulge-size & none, just-noticeable, obvious, dominant & 7.1 & $6.1\,\%$ \\
        how-rounded & completely, in-between, cigar-shaped & 7.1 & $63.4\,\%$\\
        bulge-shape & rounded, boxy, none & 1.6 & $0.4\,\%$ \\
        spiral-winding & tight, medium, loose & 3.6 & $4.8\,\%$\\
        spiral-arm-count & 1, 2, 3, 4, 5-plus, can't-tell & 3.6 & $4.8\,\%$\\
        clumpy-appearance & yes, no & 13.1 & $13.9\,\%$\\
        clump-count & 1, 2, 3, 4, 5-plus, can't-tell & 5.0 & $1.9\,\%$\\
        galaxy-symmetrical & yes, no & 4.4 & $1.2\,\%$\\
        clumps-embedded & yes, no & 4.4 & $1.2\,\%$\\
    \end{tabular}
\end{center}
\end{table}

Our dataset contains 76\,176 greyscale galaxy images with detailed morphology labels. This dataset, referred to as the `complete set', encompasses all available images. It has a magnitude range of $10.5 < m_{I814W} < 23.5$ and a redshift range of $0 < z < 4.1$. In order to ensure an unbiased evaluation of the model, we divided this set into two distinct subsets: one for training and validation, and another independent test set for evaluation purposes. To accomplish this, we performed a random split of 80\,\% for training and validation, and the remaining 20\,\% for the test set. Subsequently, we further split the training and validation set using another random 80/20 percent split. The resulting datasets are listed in Table\,\ref{tab:datasets}.

\renewcommand{\arraystretch}{1.5}
\begin{table}[h]
\begin{center}
    \caption{Datasets of \Euclid images with GZH labels used in this study.}
    \label{tab:datasets}
    \small
    \begin{tabular}{llll}
    \hline
        Dataset & Type of set & Restriction & Number of galaxies\\ \hline
        complete & train / val & - & 60\,940 (48\,752 / 12\,188)\\
        complete & test & - & 15\,236 \\ 
        bright & train / val & $m_{I814W} < 22.5$ & 27\,882 (22\,306 / 5576)\\
    \end{tabular}
\end{center}
\end{table}
\renewcommand{\arraystretch}{1}

\subsection{Experiments}\label{sec:experiments}

The \Euclid mission is anticipated to generate an unparalleled number of galaxy images with approximately 250 million having resolved internal morphology \citep{Bretonniere-EP13}, but humans will only have limited capacity to label them. Consequently, it is important to assess the number of labelled galaxies required to achieve satisfactory performance in morphology predictions (Sect.\,\ref{sec:num_galaxies}). Additionally, we aim to investigate the selection criteria for which galaxies to label (Sect.\,\ref{sec:bright_thresh}). Suppose a person has the capacity to label 1000 galaxy images. An open question is whether the automated predictions will get better if those 1000 galaxies are selected randomly, or if 1000 bright galaxies are used instead.

\subsubsection{Restricting the training set size}\label{sec:num_galaxies}

Our goal is to assess the performance of \texttt{Zoobot} based on a limited number of galaxies used for training. Hence, we randomly chose a specific number $N_{\textrm{train}}$ of galaxy images from the training and validation sets (refer to Table\,\ref{tab:datasets}). These selected images were then used for training. To ensure a fair comparison between all models, we consistently evaluated the performance on the complete test set, without excluding any images.

\subsubsection{Restricting the magnitude}\label{sec:bright_thresh}

Typically, assessing the morphology of brighter galaxies is more straightforward compared to fainter ones. Our goal here is to investigate whether our automated morphology predictions have a better performance when trained on bright galaxies or on randomly selected galaxies from the complete dataset, especially when the number of examples is limited. We therefore created, from our complete training and validation set, a subset which we refer to as the ‘bright set', by applying a magnitude restriction of $m_{I814W} < 22.5$. This resulted in a bright training and validation set comprising 27\,882 images. Similar to the complete set, we then performed an 80/20 percent split for training and validation purposes (see Table\,\ref{tab:datasets}).

\subsection{Training \texttt{Zoobot}}\label{sec:training_zoobot}

We used the TensorFlow \citep{Abadi_2016} implementation of \texttt{Zoobot} \citep{Walmsley_2023}. We trained \texttt{Zoobot} on the datasets shown in Table\,\ref{tab:datasets} by using the fine-tuning procedure described in the code of \citet{Walmsley_2023}. For this, we replaced the original model head with a single dense layer with the number of neurons corresponding to the number of GZH answers used, specifically 40 neurons for 40 answers to 13 questions (see Table\,\ref{tab:questions}). As in \citet{Walmsley_2022_3}, we selected the sigmoid activation function for the final layer to predict scores between 1 and 100 corresponding to the Dirichlet parameters (see Eq.\,\ref{eq:dirichlet_dist}). The \texttt{JPG} images with the applied arcsinh intensity mapping (see Sect.\,\ref{sec:euclid_images}) were normalised to values between 0 and 1 before feeding them into the network. Additionally, we applied similar augmentations as \citet{Walmsley_2022_1} to all images during training, namely a random vertical flip of the image with a probability of 0.5 and a rotation by a random angle. As in the code of \citet{Walmsley_2023}, the training process was divided into two parts: at first, we only trained the new head, and in a second step the entire model, as soon as the validation loss was not decreasing for more than 20 consecutive epochs. Furthermore, we reduced the learning rate by a factor of $0.25$ when the validation loss did not decrease for ten consecutive epochs. The chosen hyperparameters were selected as they lead to the best model performance in comparison to multiple other tested values. We used the Adam optimizer \citep{Kingma_2014} for training. We trained the pretrained model with the bright and complete training sets with different numbers of images ranging between five and all the available images (see Table\,\ref{tab:datasets}).
To evaluate how \Euclid's lower resolution and noise affect the performance of our model, we conducted separate training using the original HST COSMOS images for the same set of galaxies (see Sect.\,\ref{sec:Data}). This approach allows us to analyse the impact independently of training with a new decision tree.

\section{Results: Zoobot for \Euclid images}\label{sec:Results}

We trained \texttt{Zoobot} to emulated \Euclid VIS images with GZH labels. In Sect.\,\ref{sec:comparison_number}, we compare the various models trained in this study, which were trained with different numbers of images from the bright or complete sets. We then evaluate the model with the best performance on \Euclid images in detail in Sect.\,\ref{sec:best_model}.

\subsection{Comparing models – The impact of the number of training galaxies and magnitude restriction}\label{sec:comparison_number}

\begin{figure}[htbp!]
\centering
\includegraphics[width=\hsize]{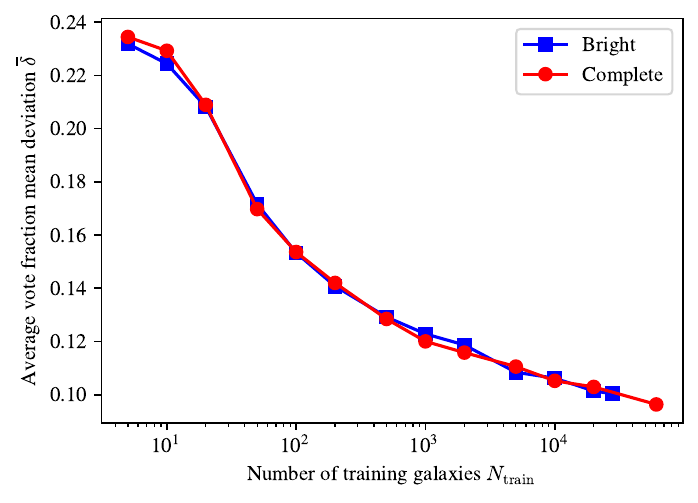}
\caption{Vote fraction mean deviation averaged over all morphology answers $\Bar{\delta}$ as a function of the number of galaxies $N_{\textrm{train}}$ from the bright and complete set used for training. To ensure a consistent comparison, the predictions were done on the complete test set. Lower values indicate better performance.}
\label{fig:num_galaxies_deviations_B}
\end{figure}

\begin{figure*}[htbp!]
\centering
\includegraphics[width=\hsize]{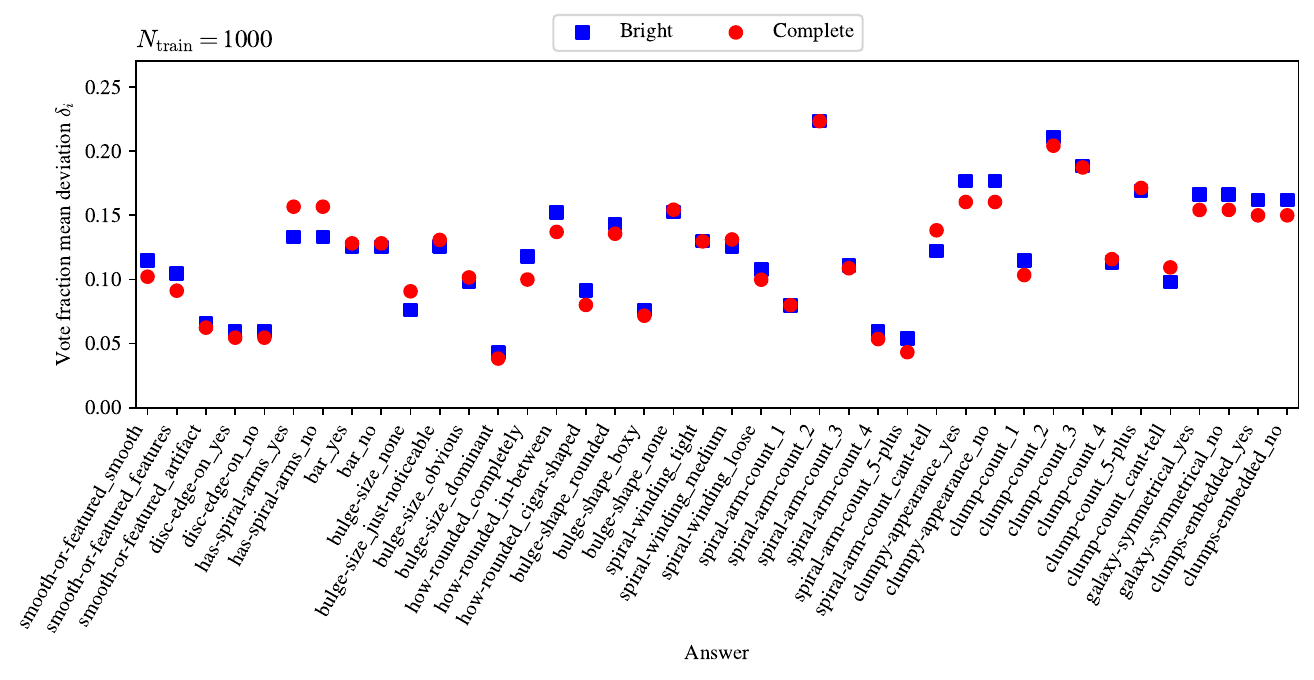}
\caption{Vote fraction mean deviations $\delta_i$ of the model predictions and the volunteer labels for the different morphology answers $i$ (see Eq.\,\ref{eq:vote_frac_mean_dev}), for models trained on 1000 bright or random galaxies from the complete set. Lower $\delta_i$ indicates better performance.}
\label{fig:deviations_datasets_weights_B_1000}
\end{figure*}

\begin{figure}[htbp!]
\centering
\includegraphics[width=\hsize]{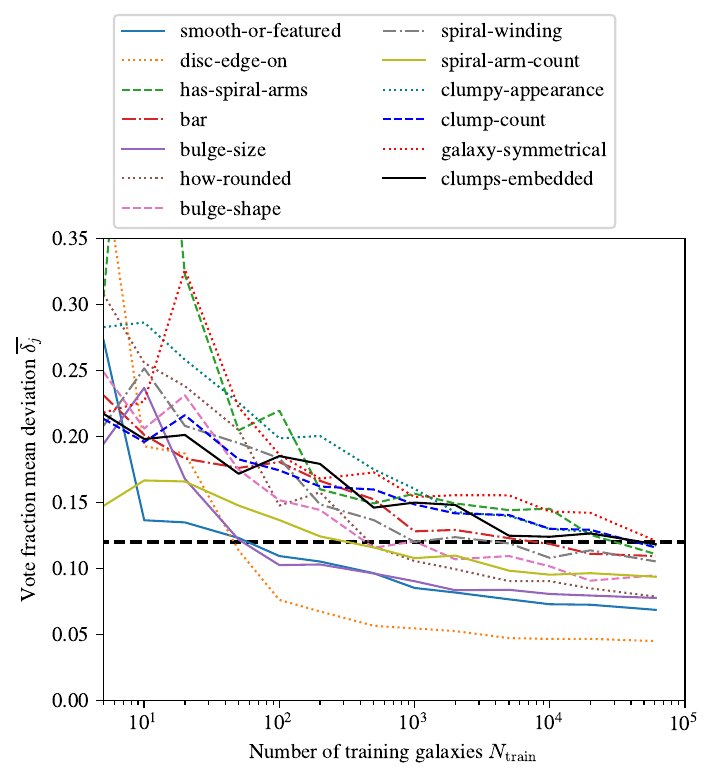}
\caption{Vote fraction mean deviations of the model predictions $\Bar{\delta_j}$ for the different morphology questions $j$ of the decision tree, as a function of number of galaxies included in training $N_{\textrm{train}}$. This is illustrated for the model trained on galaxies from the complete dataset. All questions reach a mean deviation of less than $12\,\%$ (dashed black line) after being trained with all available galaxies.}
\label{fig:deviations_num_galaxies_questions}
\end{figure}

\texttt{Zoobot} is not predicting discrete classes, but rather posteriors that can be converted into vote fractions (values between 0 and 1). This is accomplished by dividing the predicted Dirichlet parameter for a particular answer by the sum of the parameters of all answers to the corresponding question. To evaluate the performance of \texttt{Zoobot}, we used the predicted vote fractions and compared them with the corresponding volunteer vote fractions (considered to be ‘ground truth' vote fractions). This allows for a comprehensive assessment of \texttt{Zoobot}'s performance. To ensure the inclusion of only relevant galaxies for a specific question, we considered galaxies for which at least half of the volunteers provided an answer (see Table\,\ref{tab:questions}). Following the method described in \citet{Walmsley_2022_1}, for a given answer $i$ to a morphology question $j$, we calculated the absolute difference between the predicted vote fraction $f_{\textrm{pred}}$ and the volunteer vote fraction $f_{\textrm{gt}}$ for each relevant galaxy in the test set. We then averaged these differences over all relevant galaxies $n_j$ as
\begin{equation}
    \delta_i := \overline{|f_{\textrm{pred}}-f_{\textrm{gt}}|}.
    \label{eq:vote_frac_mean_dev}
\end{equation}

To allow for easier comparison among different models, while considering the performance on all answers, we calculated the unweighted average of all $\delta_i$ values. This aggregated measure, referred to as the averaged vote fraction mean deviation $\Bar{\delta}$, served as our primary metric for comparison, with lower values indicating better performance. For consistency, we evaluated the models using predictions on the same complete test set consisting of 15\,236 images (see Table\,\ref{tab:datasets}).

\subsubsection{Overview}

We show in Fig.\,\ref{fig:num_galaxies_deviations_B} the model performance (given by the averaged mean vote fraction deviations $\Bar\delta$) depending on the number of training galaxy images used, $N_{\textrm{train}}$, for the models trained on galaxies from the bright and complete set. The figure summarises our experiments with different magnitude restrictions and number of training images. 

As expected, with increasing number of training galaxies, the average mean deviation $\Bar\delta$ is decreasing: the more galaxy examples (of different types) are used for training, the better the model predictions get for all answers. Notably, no substantial discrepancies are observed between training on bright galaxies or randomly selected galaxies from the complete set. The model trained on all available galaxy images from the complete set yields the best performance, characterised by the lowest $\Bar{\delta}$ of approximately $9.5\,\%$ (analysed in Sect.\,\ref{sec:best_model}).

\subsubsection{\texttt{Zoobot} trained on only 1000 galaxy images}\label{sec:zoobot_1000}

Next, we compared the model performance in more detail for the models trained on 1000 galaxies from the bright and complete set. Fig.\,\ref{fig:deviations_datasets_weights_B_1000} shows the vote fraction mean deviations $\delta_i$ for all morphology answers $i$ for both models. We selected 1000 galaxies as a reasonably small quantity that a single expert could potentially label, while still achieving satisfactory performance for most questions.  

All answers reach a mean deviation below $22\,\%$ indicating that training with only 1000 galaxies already leads to high model performance in general. For most answers, there is no substantial difference between training on bright or complete galaxies.   

In particular, for the `disc-edge-on' and `bar' questions, the model shows approximately the same performance when trained on either  1000 bright or  1000 random galaxies. Thus, the relevant features that the model learns do not change qualitatively with different magnitudes. Additionally, the `disc-edge-on' task seems to be easier to learn because the deviations $\delta_i$ are well below $10\,\%$. 

For the `clumpy-appearance', `galaxy-symmetrical' and `clumps-embedded' questions, \texttt{Zoobot} performs slightly better (by about $1\,\%$) when trained on random galaxies from the complete set than when trained on bright galaxies. The better performance for these clump-related questions can thus be explained with the higher number of relevant examples in the complete training set compared to the bright set, as clumpiness is more frequent among fainter galaxies. On the other hand, identifying spiral arms seems to be more effective (by about $2\,\%$) when training on bright galaxies. This suggests that the examples included in the bright training set provide clearer and more reliable labels to learn to identify spiral arms.

\subsubsection{Number of training galaxies for different morphology types}

 Fig.\,\ref{fig:deviations_num_galaxies_questions} shows the dependence of the model performance (vote fraction mean deviation $\Bar{\delta_j}$) on the number of training galaxies $N_{\textrm{train}}$ for the different morphology questions $j$. Here, the vote fraction mean deviation is provided as the average of all answers for a particular morphology question and the models were trained on galaxies randomly selected from the complete set.

An increase in the number of training galaxies generally leads to improved performance, characterised by a decrease in the vote fraction mean deviation. This means that in general for all morphology tasks, performance can be improved with training on more labelled examples. All questions reach an averaged vote fraction mean deviation below $12\,\%$ (highlighted in Fig.\,\ref{fig:deviations_num_galaxies_questions}) when trained with all available galaxies from the complete set. They show different dependencies on the number of training galaxies.

Although in general more training examples increase the quality of the predictions, there are instances where a larger number of galaxies leads to slightly worse performance. These fluctuations in vote fraction mean deviation are particularly noticeable in the low-number regime, for example for the `how-rounded' question with 200 training galaxies. They can be attributed to the model's sensitivity to the specific galaxies randomly selected for training. Nevertheless, these variations do not alter the overall observable trends for the different questions.

When comparing the various questions, the `disc-edge-on' task not only has the lowest mean deviation (as discussed in Sect.\,\ref{sec:best_model}) when trained with the complete set, but it also achieves a deviation below $10\%$ after training with just 100 galaxies. This is even more impressive as only $6.1\,\%$ of the galaxies are relevant (see Table\,\ref{tab:questions}), although \texttt{Zoobot} learns from all galaxies. This further indicates that identifying disc galaxies is easier to learn than other tasks of the decision tree. Similarly for the `bulge-size' question, the model achieves a deviation below $12\%$ after training with only 100 images. Since these tasks were included in all GZ decision trees, this outcome can be interpreted as a demonstration of the effectiveness of fine-tuning. Furthermore, training on only 100 random galaxies leads for the `smooth-or-featured' question to deviations below $12\,\%$. This question was included in all GZ decision trees as the first question and was thus answered by all volunteers, and therefore required fewer new examples compared to other tasks. 

In contrast, for the `has-spiral-arms' question, 60\,000 galaxies are required to achieve deviations below $12\,\%$. Despite the inclusion in all GZ decision trees, a substantial number of examples are still necessary to accurately predict the corresponding vote fractions. This observation suggests that detecting spiral arms might pose a greater challenge for \Euclid images compared to the galaxies in the pretraining datasets. Additionally, questions related to clumps in galaxies exhibit similar patterns, requiring a range of 10\,000 to 60\,000 random galaxies to achieve a deviation below $12\%$. From the campaigns involved in the pretraining of \texttt{Zoobot}, these clump-related questions were exclusively included in the GZC campaign. Consequently, the impact of this pretraining is likely less effective for these tasks. Moreover, given that spiral arms and clumps involve finer structures, the associated tasks are inherently more complex and need a larger number of training examples.

\subsection{Analysis of the best performing model}\label{sec:best_model}

In this section, we analyse the performance of \texttt{Zoobot} for emulated \Euclid VIS images with the lowest averaged vote fraction mean deviation $\Bar{\delta}$, and thus the best performing model, as derived in Sect.\,\ref{sec:comparison_number}. We show examples of \texttt{Zoobot}'s output, then investigate the performance with standard classification metrics after discretizing the vote fractions (Sect.\,\ref{sec:discrete}) and demonstrate how our model can be used to find spiral galaxies in a given dataset (Sect.\,\ref{sec:finding_spiral}). Next, we analyse the predicted vote fractions directly by looking at the mean (Sect.\,\ref{sec:deviations}) and the histograms (Sect.\,\ref{sec:histograms}) of the deviations from their respective volunteer vote fractions, and by investigating their redshift and magnitude dependence (Sect.\,\ref{sec:magnitude_redshift}). Finally, we compare the model performance between HST and \Euclid images (Sect.\,\ref{sec:comparison_hubble}).

\begin{figure}[htbp!]
\centering
\includegraphics[width=\linewidth]{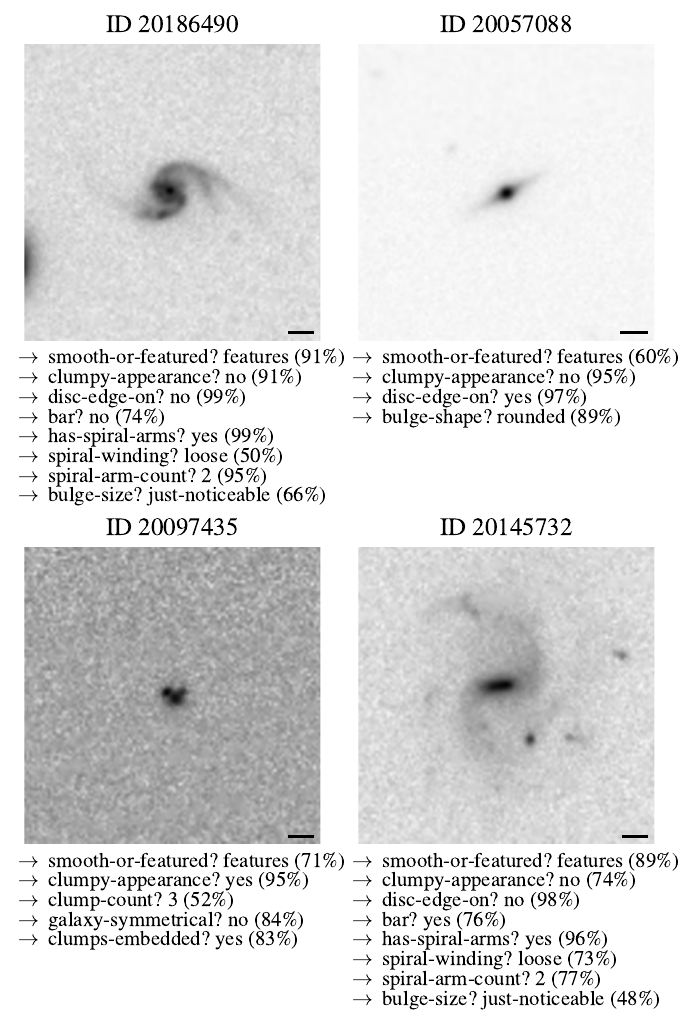}
\caption{Four examples of the predictions of \texttt{Zoobot} following the structure of the GZH decision tree (see Table\,\ref{tab:questions} and Fig.\,\ref{fig:gz_decision_tree}) for galaxies (inverted greyscale, image ID given above each image) from the complete test set. For every question, the answer with the highest predicted vote fraction (denoted in the parenthesis) is selected. The black bars represent a length of $\ang{;;1}$.}
\label{fig:example_detections}
\end{figure}

\begin{figure}[htbp!]%
  \includegraphics[width=\linewidth]{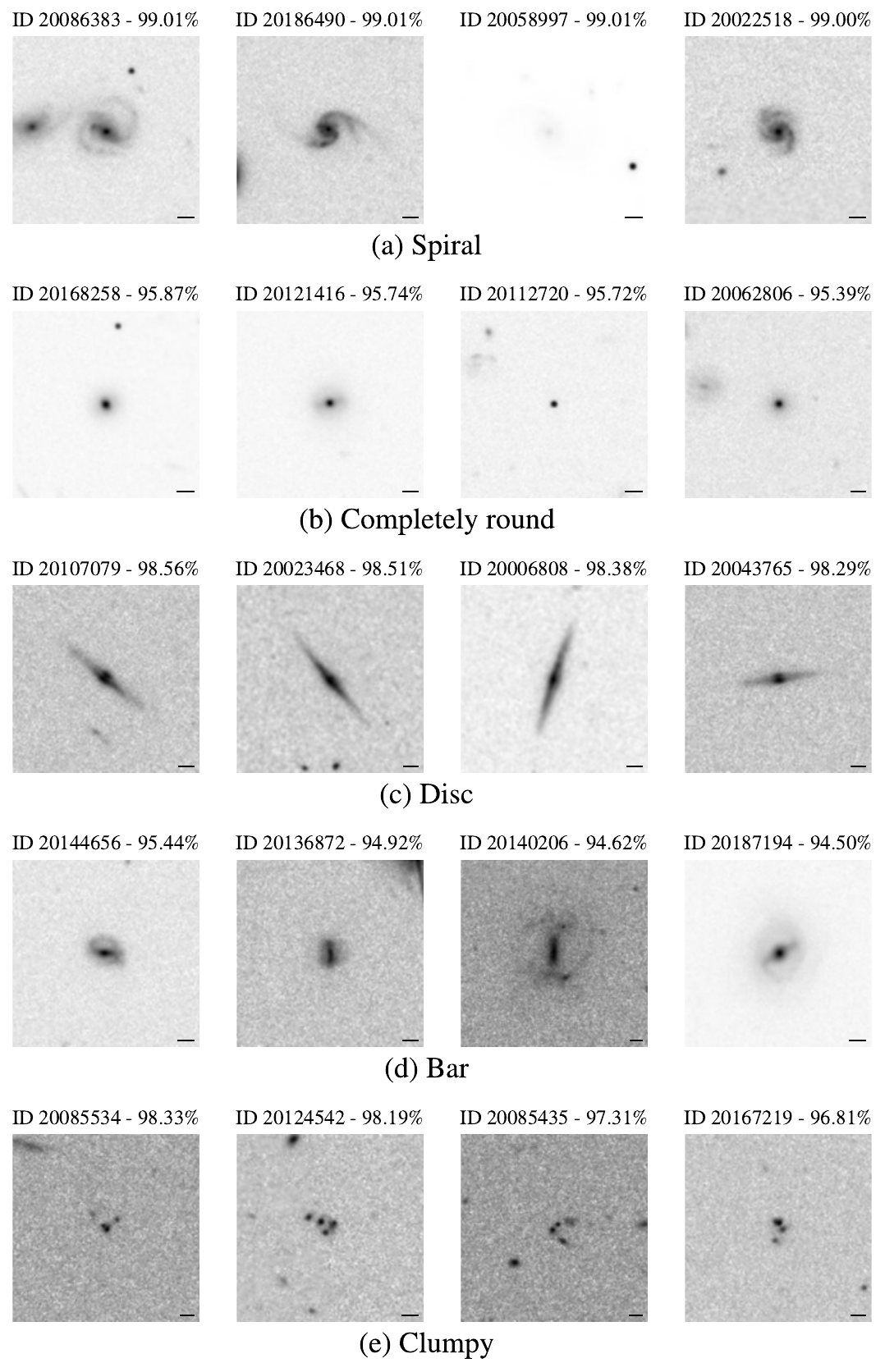}%
  \caption{Examples of galaxies with the highest predicted vote fractions of \texttt{Zoobot} for (a) spiral, (b) completely round, (c) disc, (d) barred, and (e) clumpy galaxies from the complete test set. Above each galaxy image, the corresponding image ID and the predicted vote fraction in percent are given. The black bars represent a length of $\ang{;;1}$.}
  \label{fig:hubble_highest_confidences}
\end{figure}

To verify the quality of the predictions, four examples of \texttt{Zoobot}'s output on different galaxies from the complete test set are shown in Fig.\,\ref{fig:example_detections}. The selected answer for every question is the one with the highest predicted vote fraction, while the asked questions follow the structure of the GZH decision tree (see Table\,\ref{tab:questions} and Fig.\,\ref{fig:gz_decision_tree}). Fig.\,\ref{fig:hubble_highest_confidences} shows four galaxies from the complete test set with the highest predicted vote fractions for five example answers ---(a) spiral, (b) completely rounded, (c) disc, (d) bar, and (e) clumpy--- in order  to demonstrate the quality of \texttt{Zoobot}'s predictions.

\subsubsection{Discrete classifications}\label{sec:discrete}

To get an intuitive sense of \texttt{Zoobot}'s performance for the different morphology tasks, we converted the predicted vote fractions into discrete values by binning them to the class with the highest predicted vote fraction. However, it is important to note that these metrics only provide a basic indication of \texttt{Zoobot}'s performance and do not fully capture its ability to predict morphology, as the information is simplified and reduced. 

We evaluated the discretised predictions with standard classification metrics for the different classes. Accuracy $A$ is the fraction of correct predictions for both the positive and negative class among the total number of galaxy images $N_{\textrm{total}}$. It is calculated as
\begin{equation}
     A=\frac{N_{\textrm{TP}}+N_{\textrm{TN}}}{N_{\textrm{total}}},
\end{equation}
where $N_{\textrm{TP}}$ is the number of true positives and $N_{\textrm{TN}}$ the number of true negatives.

Precision $P$ is the fraction of correct classifications among the galaxies predicted to belong to a particular class. It is calculated as
\begin{equation}
    P = \frac{N_{\textrm{TP}}}{N_{\textrm{TP}}+N_{\textrm{FP}}},    
\end{equation}
where $N_{\textrm{FP}}$ is the number of false positives.

Recall $R$ is defined as the fraction of correct classifications among the galaxies of a certain class and calculated as
\begin{equation}
    R = \frac{N_{\textrm{TP}}}{N_{\textrm{TP}}+N_{\textrm{FN}}},
\end{equation}
where $N_{\textrm{FN}}$ is the number of false negatives.

The $F_1$-score combines precision and recall by taking their harmonic mean. Thus, it is a more general measure for evaluating model performance. It is calculated as
\begin{equation}
     F_1 = 2\,\frac{P\,R}{P+R}.
\end{equation}

All of these metrics have values between $0$ and $1$. Some classification tasks have an imbalanced number of galaxies for the different classes. Moreover, there are some morphology tasks with more than two answers (see Table\,\ref{tab:questions}). Therefore, we calculated the above metrics by treating each class as the positive class and averaging over the results. We also provide the $F_1$-score weighted by the number of galaxies for the different classes, $F_1^\star$, similar to \citet{Walmsley_2022_1}.

The performance of the model for a particular classification task can be summarised by a confusion matrix. The rows of this two-dimensional matrix correspond to the predicted classes, while the columns correspond to the ground truth classes. The diagonal elements are the fraction of correct classifications, while the other elements correspond to false classifications.

\begin{table}[h]
\begin{center}
    \caption{Classification metrics of the model on the complete test set for all galaxies corresponding to Fig.\,\ref{fig:confusion_matrices}a. Precision $P$, recall $R,$ and $F_1$-score are calculated using the unweighted average of all classes. We also show the weighted $F_1$-score in the $F_1^\star$ column. For `has-spiral-arms', we also provide the metrics for finding spiral galaxies in the complete test set (printed in italic), corresponding to the confusion matrix in Fig.\,\ref{fig:confusion_matrices_spiral}a.}
    \label{tab:metrics_all}
    \tiny
    \begin{tabular}{lrlllll}
    \hline
    Question & $N_{\textrm{total}}$ & $A$ & $P$ & $R$ & $F_1$ & $F_1^\star$\\ \hline
    smooth-or-featured & 15\,236 & 0.885 & 0.835 & 0.811 & 0.822 & 0.884 \\
    disc-edge-on & 986 & 0.982 & 0.963 & 0.957 & 0.960 & 0.982 \\
    has-spiral-arms & 764 & 0.916 & 0.584 & 0.725 & 0.614 & 0.935 \\
    & \textit{10\,746} & \textit{0.965} & \textit{0.864} & \textit{0.936} & \textit{0.896} & \textit{0.966} \\
    bar & 974 & 0.878 & 0.822 & 0.744 & 0.774 & 0.869 \\
    bulge-size & 975 & 0.822 & 0.542 & 0.563 & 0.549 & 0.823 \\
    how-rounded & 9915 & 0.874 & 0.872 & 0.868 & 0.869 & 0.874 \\
    bulge-shape & 84 & 0.893 & 0.866 & 0.875 & 0.870 & 0.894 \\
    spiral-winding & 746 & 0.709 & 0.683 & 0.672 & 0.677 & 0.709 \\
    spiral-arm-count & 745 & 0.678 & 0.450 & 0.353 & 0.375 & 0.653 \\
    clumpy-appearance & 2265 & 0.874 & 0.867 & 0.850 & 0.857 & 0.873 \\
    clump-count & 328 & 0.546 & 0.516 & 0.413 & 0.390 & 0.539 \\
    galaxy-symmetrical & 225 & 0.880 & 0.884 & 0.690 & 0.737 & 0.860 \\
    clumps-embedded & 226 & 0.850 & 0.791 & 0.819 & 0.803 & 0.853 \\
    \end{tabular}
\end{center}
\end{table}

\begin{table}[h]
\begin{center}
    \caption{Same classification metrics as in Table\,\ref{tab:metrics_all}, but for galaxies with confident volunteer responses (i.e. one answer has a vote fraction above 0.8) corresponding to Fig.\,\ref{fig:confusion_matrices}b. For `has-spiral-arms', we also provide the metrics for finding spiral galaxies in the complete test set (printed in italics), corresponding to the confusion matrix in Fig.\,\ref{fig:confusion_matrices_spiral}b.}
    \label{tab:metrics_confident}
    \tiny
    \begin{tabular}{lrlllll}
    \hline
    Question & $N_{\textrm{total}}$ & $A$ & $P$ & $R$ & $F_1$ & $F_1^\star$\\ \hline
    smooth-or-featured & 1963 & 0.995 & 0.995 & 0.993 & 0.994 & 0.995 \\
    disc-edge-on & 907 & 0.998 & 0.994 & 0.994 & 0.994 & 0.998 \\
    has-spiral-arms & 666 & 0.950 & 0.542 & 0.975 & 0.564 & 0.971 \\
    & \textit{10\,553} & \textit{0.970} & \textit{0.859} & \textit{0.952} & \textit{0.899} & \textit{0.971} \\
    bar & 511 & 0.977 & 0.968 & 0.907 & 0.935 & 0.976 \\
    bulge-size & 85 & 0.976 & 0.905 & 0.991 & 0.940 & 0.978 \\
    how-rounded & 5119 & 0.979 & 0.979 & 0.977 & 0.978 & 0.979 \\
    bulge-shape & 36 & 0.917 & 0.864 & 0.946 & 0.893 & 0.921 \\
    spiral-winding & 46 & 0.978 & 0.933 & 0.982 & 0.954 & 0.979 \\
    spiral-arm-count & 202 & 0.941 & 0.396 & 0.534 & 0.402 & 0.943 \\
    clumpy-appearance & 1307 & 0.970 & 0.967 & 0.959 & 0.963 & 0.970 \\
    clump-count & 64 & 0.828 & 0.540 & 0.513 & 0.525 & 0.868 \\
    galaxy-symmetrical & 115 & 0.974 & 0.986 & 0.850 & 0.905 & 0.972 \\
    clumps-embedded & 85 & 0.941 & 0.844 & 0.966 & 0.890 & 0.946 \\
    \end{tabular}
\end{center}
\end{table}

\begin{figure}[h]%
  \centering
  \includegraphics[width=0.98\linewidth]{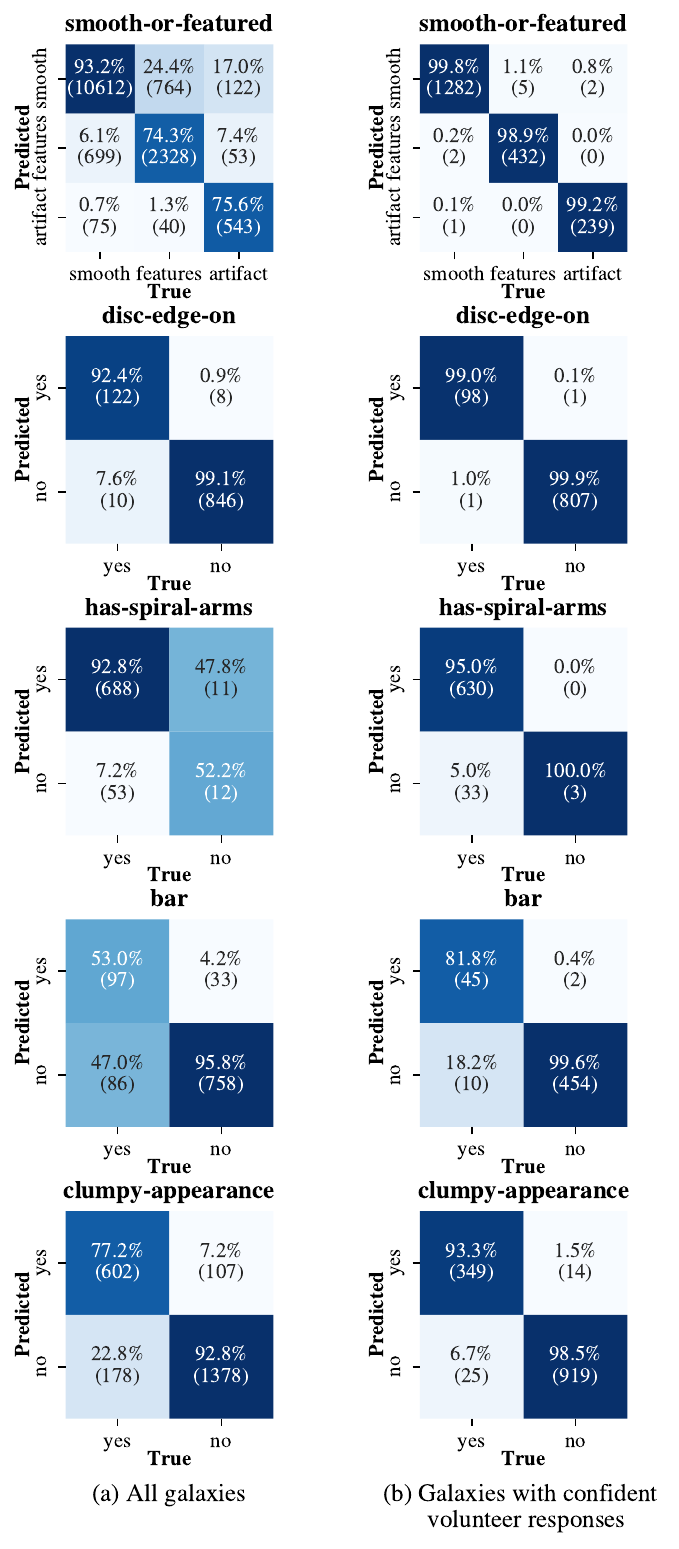}%
  \caption{Confusion matrices for five selected morphology questions after binning to the class with the highest predicted vote fraction. The confusion matrices for the other questions are shown in the Appendix. The colour map corresponds to the fraction of the ground truth values for the different classes (also denoted in the confusion matrices).}%
  \label{fig:confusion_matrices}
\end{figure}

\begin{figure}[h]%
  \centering
  \includegraphics[width=\linewidth]{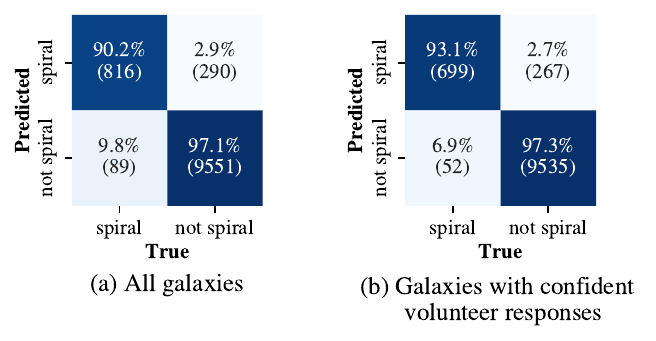}%
  \caption{Confusion matrices for the task of finding spiral galaxies in the complete test set by applying the selection cuts suggested in \citet{Willett_2017}. 
  }%
  \label{fig:confusion_matrices_spiral}
\end{figure}

The resulting metrics are listed in Table\,\ref{tab:metrics_all}. For five selected morphology tasks, we show the corresponding confusion matrices in Fig.\,\ref{fig:confusion_matrices}a. We calculated the same metrics for galaxies from the complete test set where the volunteers are confident, meaning one answer has a vote fraction of higher than $0.8$. Through this procedure, one can analyse the model performance against confident labels \citep{DominguezSanchez_2019, Walmsley_2022_1}. The results are shown in Table\,\ref{tab:metrics_confident}. The corresponding confusion matrices for selected questions are shown in Fig.\,\ref{fig:confusion_matrices}b. We present all confusion matrices for the remaining tasks in Appendix \ref{sec:appendix_conf}.

For the majority of the morphology questions, the accuracy is higher than $97\%$. For all other questions the accuracy is above $91\%$ except for the question of the `clump-count' where it is only $82.8\%$. The $F_1$-scores are all above $89\,\%$ except for the `has-spiral-arms', `spiral-arm-count' and `clump-count' questions. 

The accuracy for all galaxies, as shown in Table\,\ref{tab:metrics_all}, is generally lower compared to confidently classified galaxies, ranging from $54.6\%$ (`clump-count') to $98.2\%$ (`disc-edge-on'). This outcome is expected, since the ground truth labels themselves carry inherent uncertainty. Considering that volunteers may not reach a consensus in these cases, it can be inferred that answering morphology questions for such galaxies could be challenging. Particularly for complex morphologies, such as the number and winding of spiral arms, the size of the bulge and the number of clumps, the performance of the model is lower than for other questions that are less complex, such as determining whether a galaxy is a disc viewed edge-on. This can be attributed to several factors: the limited number of examples for these classes included in the training dataset, making them less represented, and the inherent difficulty associated with accurately identifying these morphological features. 

Furthermore, counting spiral arms and clumps are especially difficult classification tasks, as there are in both cases six classes that can be selected and some arms or clumps might be difficult to identify. Moreover, the distributions of the answers are imbalanced, with classes containing only one (`5-plus' spiral arms) or no examples (one clump) that contribute equally to the averaged metrics. Thus, the $F_1$-scores for confident volunteer responses are substantially lower than compared to other questions. In numerous instances, the predicted count for spiral arms and clumps is off by just one number from the ground truth count. Consequently, the discrete metrics provided do not fully capture the capabilities of \texttt{Zoobot}. Instead, the predicted vote fractions are preferable for assessing the number of spiral arms or clumps.

For the `has-spiral-arms' question, there are only three confident `no' examples, while there are 663 galaxies confidently classified as spiral in the test set (see Fig.\,\ref{fig:confusion_matrices}b). Thus, the test set in this binary case is extremely unbalanced and the derived metrics are therefore not reflecting \texttt{Zoobot}'s overall ability of finding spiral galaxies in a given dataset. We demonstrate this in the following section by not only using the `has-spiral-arms' question, but the whole decision tree to use the full ability of \texttt{Zoobot}.

\subsubsection{Finding spiral galaxies in the test set} \label{sec:finding_spiral}

We investigated how \texttt{Zoobot} can be used to find spiral galaxies in a given dataset. Similar to the approach used for volunteer vote fractions $f$, we applied the suggested criteria from \citet{Willett_2017} for selecting spiral galaxies in the complete test set. These criteria were: $f_{\textrm{edge-on,no}}>0.25$, $f_{\textrm{clumpy,no}}>0.3$, and $f_{\textrm{features}}>0.23$. Additionally, we excluded galaxies where the conditions mentioned above apply, but the number of volunteers was insufficient (as \texttt{Zoobot} is only predicting vote fractions), using the suggested cutoff of $N_{\textrm{spiral}}\ge20$. For the final catalogue, we chose a vote fraction of $f_{\textrm{spiral}}>0.5$ to identify spiral galaxies. Thus, all galaxies for which the conditions were fulfilled were classified as spiral, while all the others were classified as not spiral. For the predicted vote fractions, we applied the same cuts. Once more, we measure the performance for confident labels (volunteer vote fraction for the final answer greater than $0.8$ or smaller than $0.2$).

\texttt{Zoobot} achieves an accuracy of $96.5\,\%$ for finding spiral galaxies in the complete test set, with an $F_1$-score of $89.6\,\%$. The corresponding confusion matrix is shown in Fig.\,\ref{fig:confusion_matrices_spiral}a and the corresponding metrics listed in Table\,\ref{tab:metrics_all}. On confident labels, \texttt{Zoobot} achieves an accuracy of $97.0\,\%$ with an $F_1$-score of $89.9\,\%$ as shown in Fig.\,\ref{fig:confusion_matrices_spiral}b and in Table\,\ref{tab:metrics_confident}. These values demonstrate that \texttt{Zoobot} is indeed well suited for identifying spiral galaxies in a given dataset.

\subsubsection{Vote fraction mean deviations}\label{sec:deviations}

\begin{figure*}[htbp!]
\centering
\includegraphics[width=\hsize]{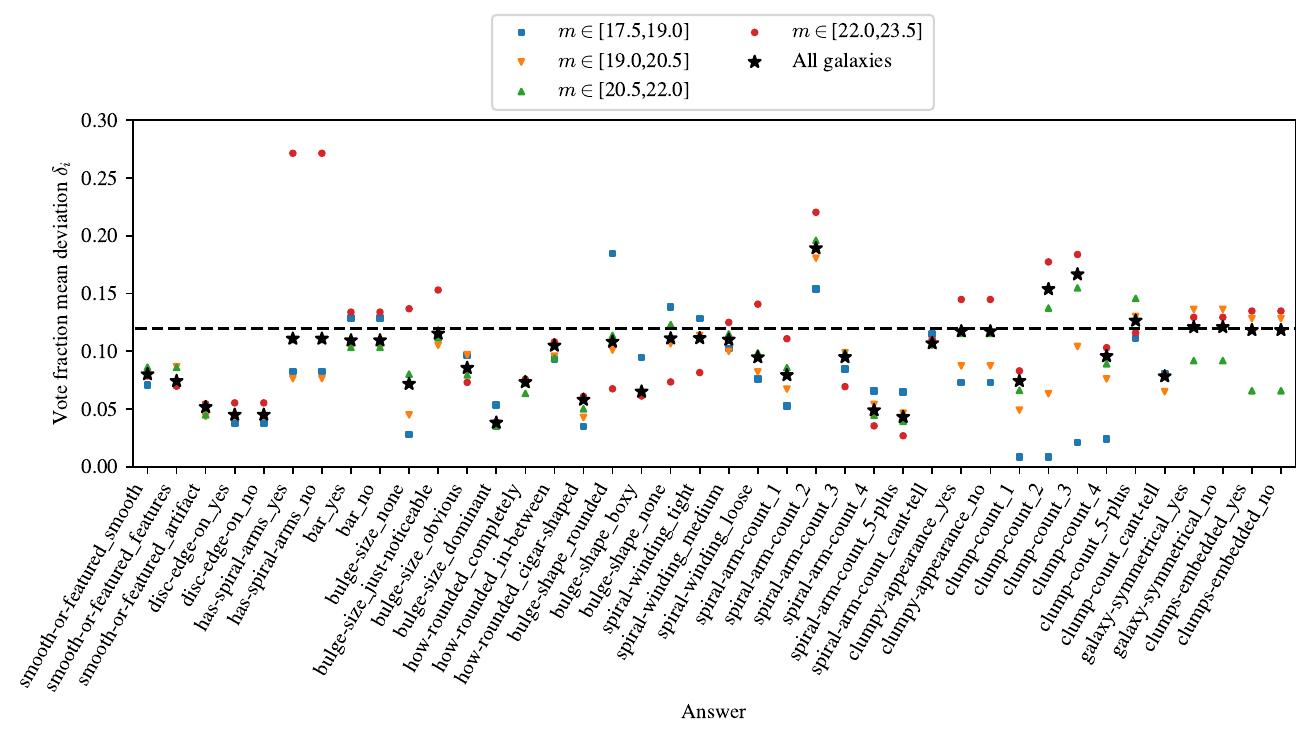}
\caption{Vote fraction mean deviations $\delta_i$ of the model predictions and the volunteer labels for the different morphology answers $i$ (see Eq.\,\ref{eq:vote_frac_mean_dev}). The model was trained with all galaxies from the complete set. The deviations are displayed for all galaxies of the test set and for galaxies within a magnitude interval with $m=m_{I814W}$. Lower $\delta_i$ indicates better performance. The black dashed line marks $12\,\%$ vote fraction mean deviation.}
\label{fig:deviations_euclid_mag}
\end{figure*}

\begin{figure}[htbp!]%
  \centering
  \includegraphics[width=\linewidth]{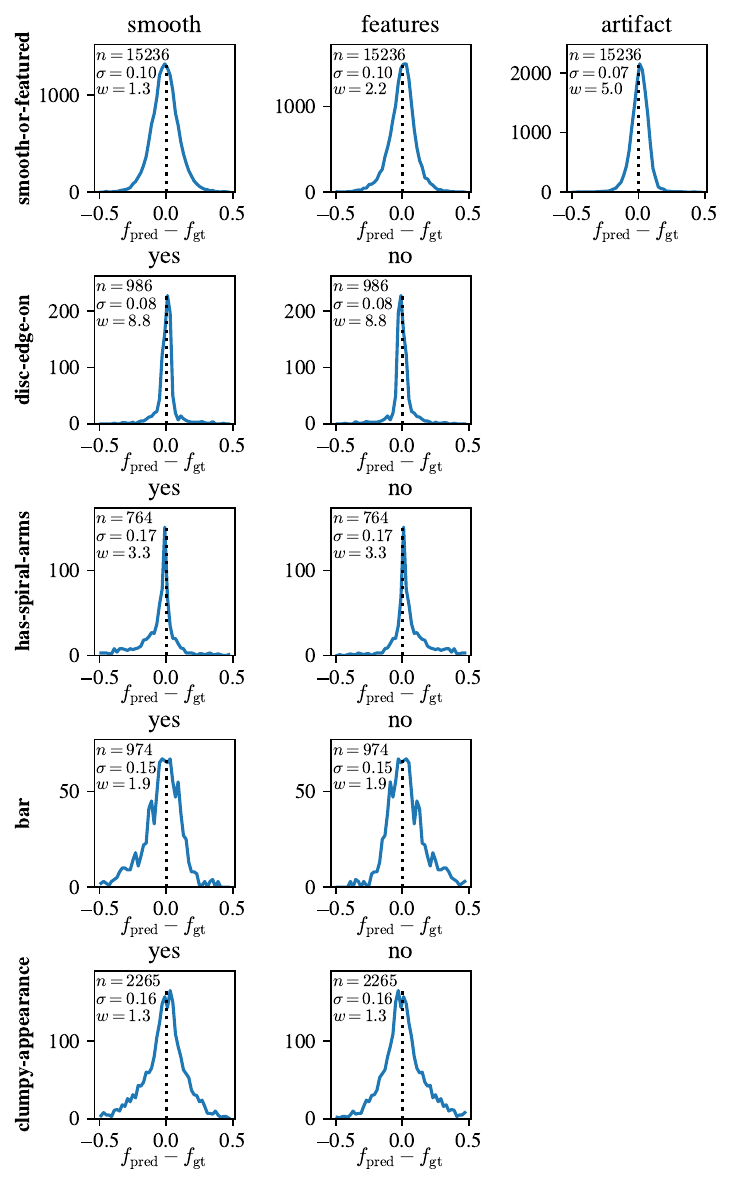}
  \caption{Histograms of the vote fraction deviations $f_{\textrm{pred}}-f_{\textrm{gt}}$ between the predicted $f_{\textrm{pred}}$ and volunteer vote fractions $f_{\textrm{gt}}$ for five selected questions. For each answer, we give the number of galaxies $n$, the standard deviation $\sigma,$ and the kurtosis $w$.}%
  \label{fig:histograms}
\end{figure}

\begin{figure*}[htbp!]
\centering
\includegraphics[width=\linewidth]{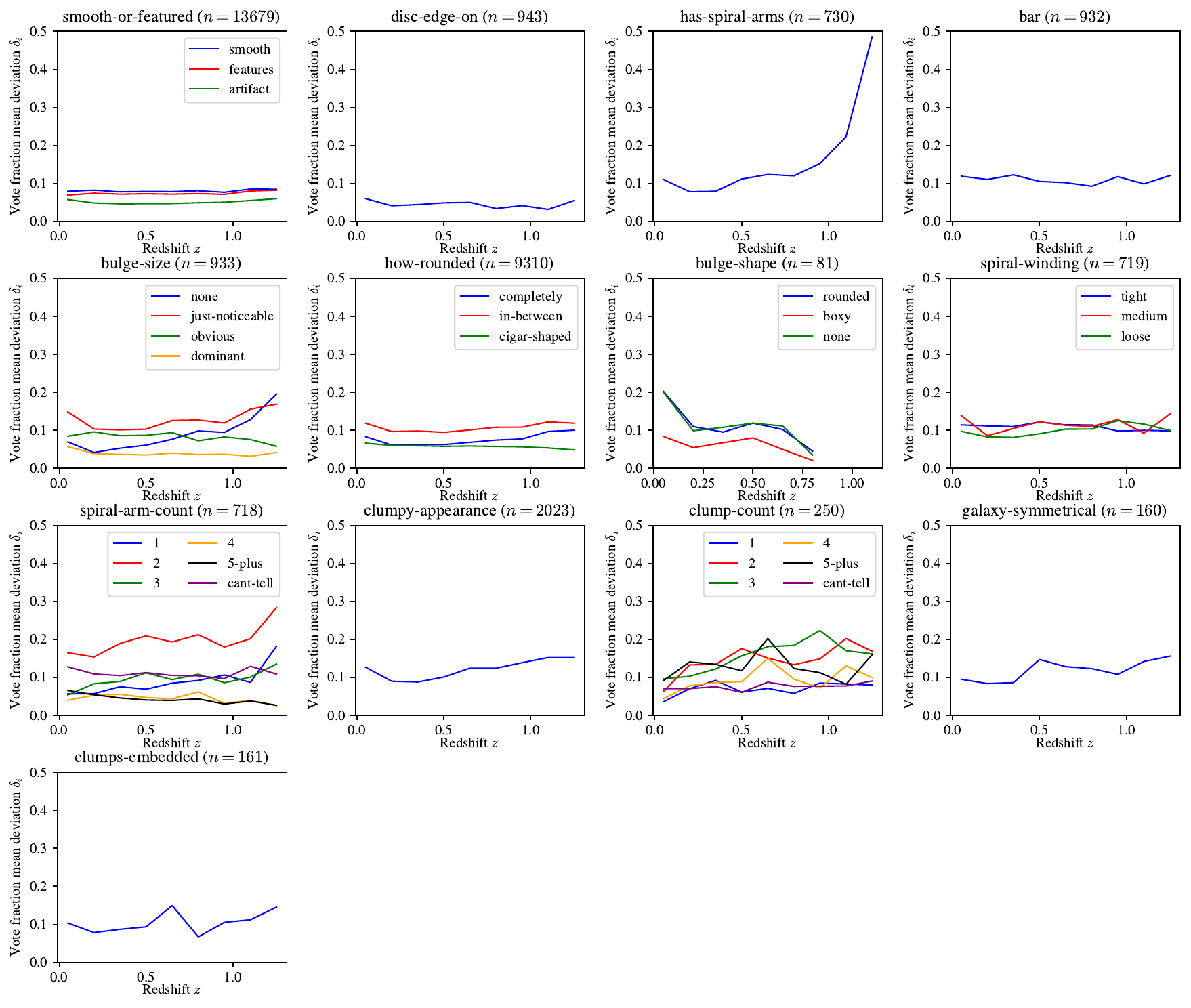}
\caption{Vote fraction mean deviations $\delta_i$ for all corresponding answers $i$ (different colours denoted in the legend) of the different GZH questions (see Table\,\ref{tab:questions}) as a function of redshift $z$ for the relevant galaxies of the complete test set (where at least half of the volunteers voted).}
\label{fig:redshift_fraction_questions}
\end{figure*}

\begin{figure*}[htbp!]
\centering
\includegraphics[width=\linewidth]{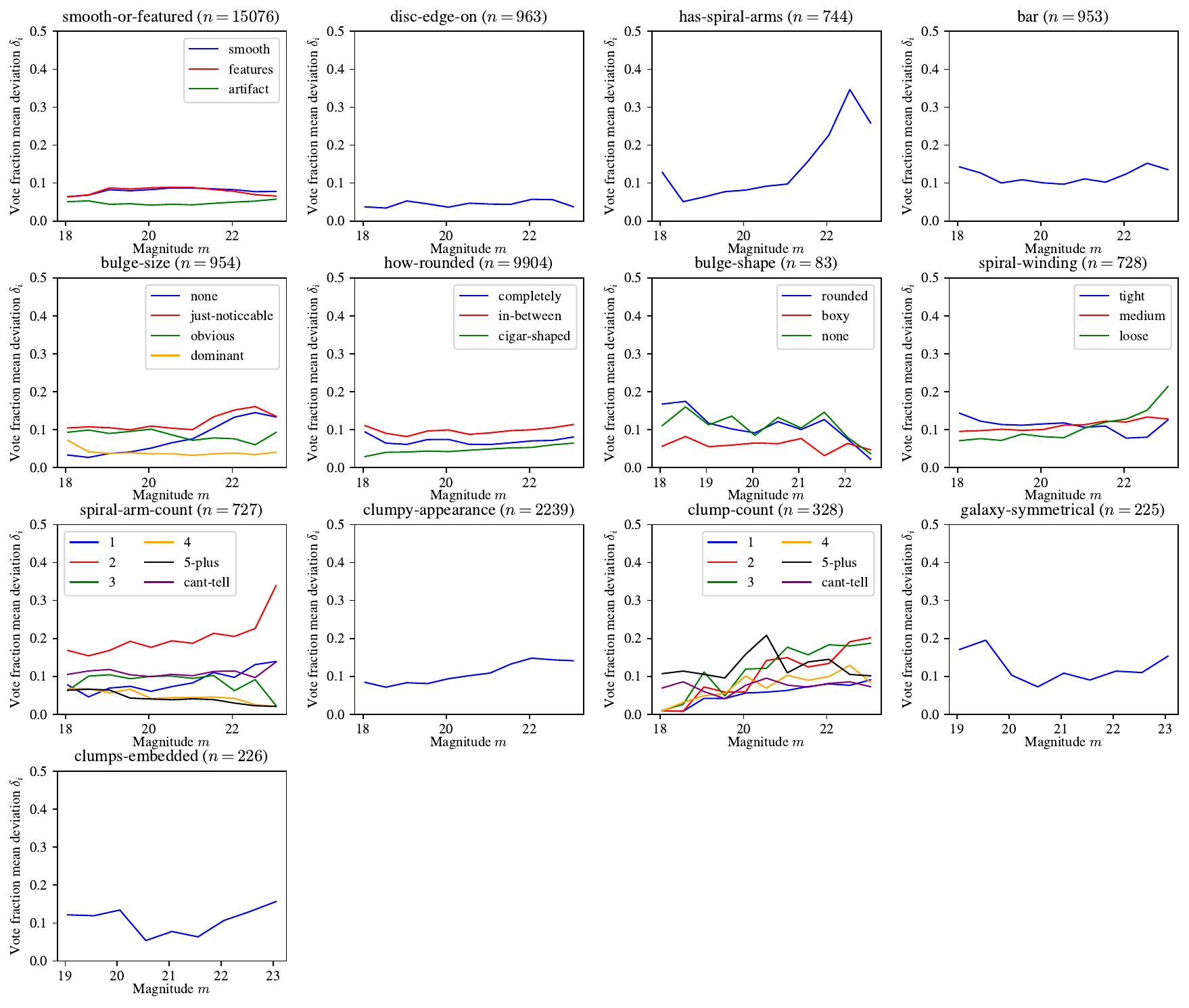}
\caption{Similar as Fig.\,\ref{fig:redshift_fraction_questions}, but with the vote fraction mean deviations $\delta_i$ depending on the magnitude $m$.}
\label{fig:magnitude_fraction_questions}
\end{figure*}

\begin{figure}[htbp!]%
  \centering
  \includegraphics[width=\linewidth]{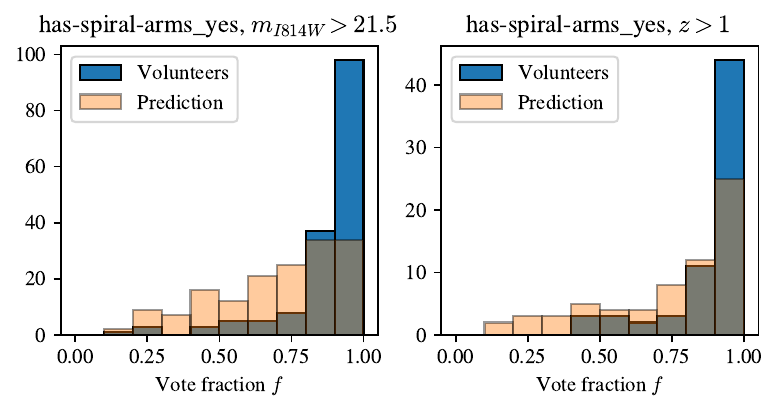}%
  \caption{Histograms of the predicted and volunteer vote fractions for the `has-spiral-arms' `yes' answer for faint and high-redshift galaxies.}%
  \label{fig:histograms_high_spiral}
\end{figure}

\begin{figure}[htbp!]%
  \centering
  \includegraphics[width=\linewidth]{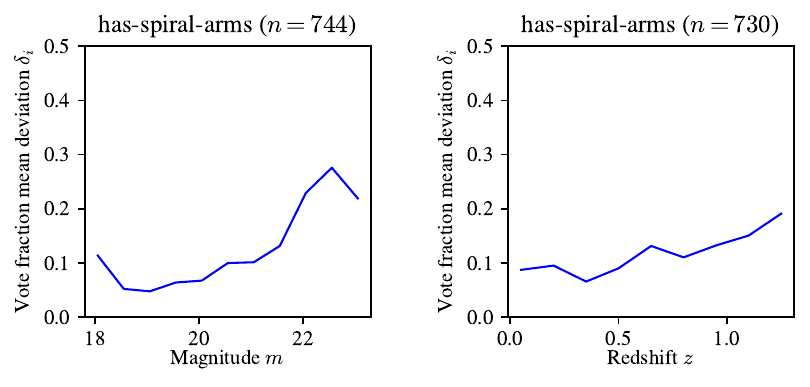}%
  \caption{Redshift and magnitude dependence of the `has-spiral-arms' vote fraction mean deviations $\delta_i$ for the model trained on the original HST COSMOS images. Compared to the \Euclid images (Figs.\,\ref{fig:redshift_fraction_questions} and \ref{fig:magnitude_fraction_questions}), the deviations for high-redshift and faint galaxies are substantially smaller.}%
  \label{fig:hubble_z_mag_spiral}
\end{figure}

We then evaluated the model performance by analysing the predicted vote fractions directly. We show the vote fraction mean deviations $\delta_i$ for all answers $i$ corresponding to different morphology types in Fig.\,\ref{fig:deviations_euclid_mag}. Moreover, we display how the performance varies with magnitude by selecting only galaxies from different magnitude intervals.

For almost all answers (36 of 40 answers), the vote fraction mean deviation is below $12\,\%$, while the performance varies between different answers. As before, the question with the lowest deviation for all answers is `disc-edge-on'. This can be attributed to the fact that it represents a less intricate feature, making it relatively easy to discern and learn. Conversely, questions related to spiral arms and clumps consistently yield the highest deviations. Once again, this can be explained with the inherent complexity of these questions, as they involve finer and more intricate structural details. We expect the quality of the morphological predictions to be better if more relevant labels for these morphology types were available, as indicated in Fig.\,\ref{fig:deviations_num_galaxies_questions}.

The dependence of the vote fraction mean deviation on the magnitude differs between answers. The mean deviation shows no substantial dependence on the magnitude for the `smooth-or-features', `disc-edge-on' and `how-rounded' questions. For the `has-spiral-arms' question, on the other hand, the differences between the deviations are the largest. While the model performs better for brighter galaxies ($m<20.5$) with a mean deviation below $10\,\%$, for faint galaxies ($m>22$) the deviations are the largest overall ($\sim\,27\,\%$). This indicates that identifying spiral arms in faint galaxies is a relatively difficult task. This is not surprising, as spiral arms are a finer structure. Once spiral arms are identified, the other tasks related to spiral arms, such as determining the winding of the spiral arms and counting them, do not show such a strong magnitude dependence. Finally, although clumps appear more frequently for faint galaxies, the model performs better in the case of brighter galaxies. This is not in contradiction to Section\,\ref{sec:zoobot_1000}, where the influence of a magnitude restriction for the galaxies used for training on the model performance on all galaxies of the complete test set was measured. Here, the performance of only one model, trained on all galaxies of the complete training set, is analysed for galaxies of different magnitudes from the complete test set.

\subsubsection{Histograms of the vote fraction deviations}\label{sec:histograms}

While we already investigated the mean of the (absolute) vote fraction deviations $f_{\textrm{pred}} - f_{\textrm{gt}}$, we show the corresponding histograms for five selected questions in Fig.\,\ref{fig:histograms}. Positive values indicate that \texttt{Zoobot} predicts a higher vote fraction than the volunteers, and negative values indicate that the volunteer vote fraction is higher. 

For most answers, the distributions are centred at 0, indicating that for most galaxies the vote fraction deviations are relatively small. The distributions are symmetrical around the centre, indicating that the model does not have a substantial bias. The widths of the distributions correspond to the mean vote fraction deviations (see Fig.\,\ref{fig:deviations_euclid_mag}), as expected.  

In contrast, for the `has-spiral-arms' answers, the distributions are not symmetrical. While the maximum of the distribution is at 0, indicating that most deviations are small, the volunteers' vote fractions for a galaxy to be spiral are higher than predicted from \texttt{Zoobot}. This can be explained by the high imbalance of the relevant `has-spiral-arms' answers (see Sect.\,\ref{sec:discrete}) with the extreme mean vote fraction for `yes' of $90.7\,\%$ in combination with the intrinsic difficulty of this task. Zoobot predicts for the most extreme volunteer vote fractions (close to 0 or 1) less extreme vote fractions \citep{Walmsley_2022_1}, leading to the asymmetry of the distribution. 

The `disc-edge-on' question also has an imbalance that is slightly less extreme (mean vote fraction for `no' is $83.9\,\%$). As it is easier to learn, the vote fraction mean deviation is much smaller ($\sim\,4\,\%$) than for `has-spiral-arms' ($\sim\,11\,\%$). Thus, the imbalance does not lead to a substantial asymmetry of the distribution.

\subsubsection{Magnitude and redshift dependence}\label{sec:magnitude_redshift}

Subsequently, we investigated the magnitude and redshift dependence of the mean deviations $\delta_i$ between the predicted and the volunteer vote fractions. These are shown in Figs.\,\ref{fig:redshift_fraction_questions} and \ref{fig:magnitude_fraction_questions} for the different questions. For some galaxies, there was no redshift information available. Thus, these galaxies were excluded. 

In general, the vote fraction mean deviation shows no substantial dependence on magnitude and redshift. For relatively easier morphology tasks (for example `disc-edge-on` and `smooth-or-featured') the deviations are smaller than for more complex ones, such as tasks related to spiral arms, bars, and clumps. 

For the `has-spiral-arms' question, the vote fraction mean deviation shows a strong increase for $z>1$ up to almost $50\,\%$. The same effect can be observed for fainter galaxies ($m_{I814W}>21.5$ in Fig.\,\ref{fig:magnitude_fraction_questions}), although the deviation is smaller. This indicates again that the difficulty of identifying spiral arms for high redshift and faint galaxies is substantially higher than other morphology tasks.

In Fig.\,\ref{fig:histograms_high_spiral}, we show the volunteer and model vote fraction for the `yes' answer of the `has-spiral-arms' question in a histogram for high-redshift and faint galaxies. For the majority of the galaxies, the volunteer vote fraction is above $90\,\%$, meaning that the volunteers are confidently classifying most galaxies to be spiral. \texttt{Zoobot}, on the other hand, shows a wider range of predicted vote fractions, with most being above $70\,\%$. While there are no galaxies (for high redshifts) or only one (for faint galaxies) galaxy to be confidently classified to have no spiral arms (vote fraction below $20\,\%$), \texttt{Zoobot} is confidently predicting that only two galaxies have no spiral arms. Therefore, \texttt{Zoobot} is not misclassifying galaxies, it is just not as confident as the volunteers. This could be explained with the lower resolution and the additional noise for the emulated \Euclid images compared to the original HST images. 

To check this, we show the redshift and magnitude dependence for \texttt{Zoobot} trained on the original HST images in Fig.\,\ref{fig:hubble_z_mag_spiral}. The vote fraction mean deviations are, although still present, substantially smaller than for \Euclid images supporting our interpretation. For a practical use, to identify spiral galaxies in high-redshift and high $m$ ranges, the selection cuts can be lowered when applying \texttt{Zoobot}.

\subsubsection{Comparing performance to original HST images}\label{sec:comparison_hubble}

\begin{figure*}[htbp!]
\centering
\includegraphics[width=\hsize]{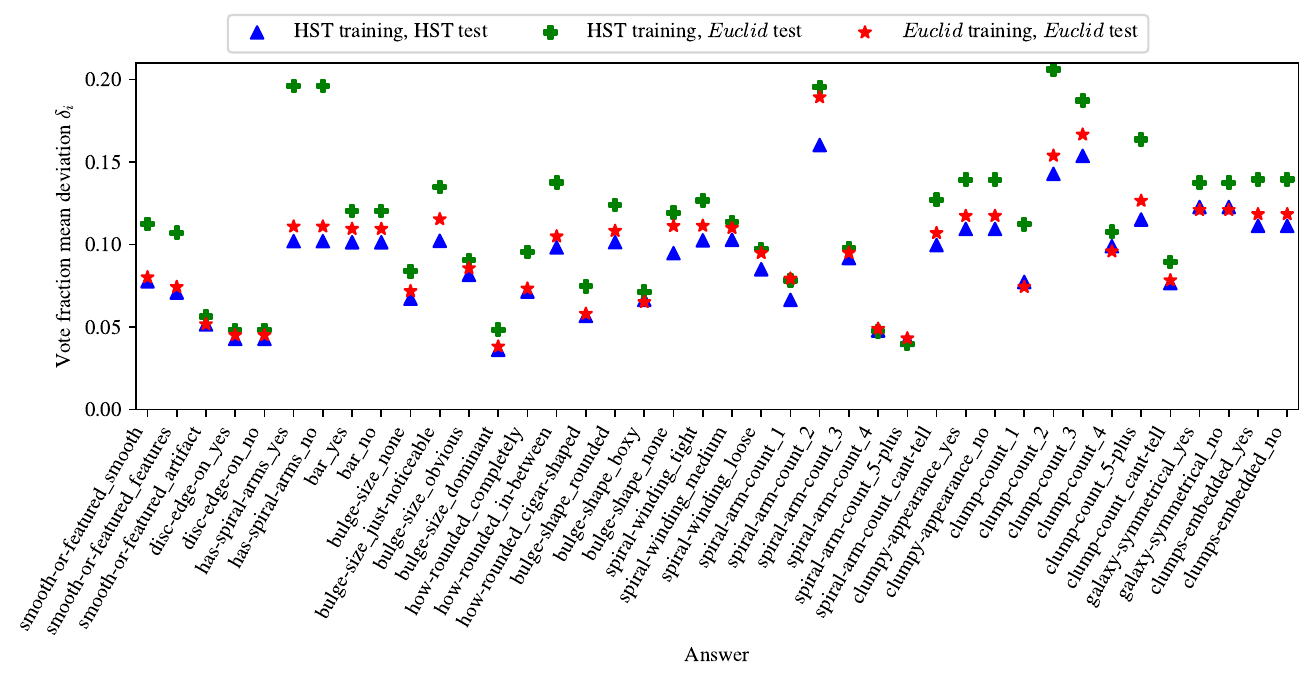}
\caption{Vote fraction mean deviations of the model predictions and the volunteer labels for the different answers of the decision tree for the model trained and tested on emulated \Euclid images and for the model trained on original HST images and tested on emulated \Euclid images and on HST images. In all cases, the models were trained and tested with the same galaxies from the complete set.}
\label{fig:deviations_hubble_euclid}
\end{figure*}

While we already investigated the influence of the lower resolution and the additional noise of the \Euclid images for identifying spiral arms, we show in Fig.\,\ref{fig:deviations_hubble_euclid} the performance of the model trained and tested on the emulated \Euclid images of the complete dataset and trained and tested on the original HST images for the same galaxies (see Sect.\,\ref{sec:Data}). The model trained on HST images was additionally tested on emulated \Euclid images.

As expected, the model trained and tested on HST images displays the lowest deviations. The deviations of the same model tested on emulated \Euclid images are for many answers substantially larger. This can be explained with the lower resolution (which is approximately two times poorer for \Euclid compared to HST) and additional noise of the emulated \Euclid images. The difference in the deviations varies with the different answers. For example for the ‘disc-edge-on’ question, the deviations are almost the same, supporting the previous discussion that this feature depends less on the resolution. On the other hand, for more complex features, such as spiral arms or clumps, the model performs substantially better for HST images than for \Euclid images. This is in agreement with the previous discussion that spiral arms and clumps are finer features and their detection depends on resolution. 

The deviations for emulated \Euclid images are substantially reduced when the model is trained directly on emulated \Euclid images, as shown in Fig.\,\ref{fig:deviations_hubble_euclid}. For many questions, such as `smooth-or-features', `disc-edge-on', or `how-rounded', the vote fraction mean deviation is almost the same. However, for questions related to spiral arms, bars and clumps, the performance of \texttt{Zoobot} trained and tested on HST images is still better than for \texttt{Zoobot} trained and tested on \Euclid images. This suggests that even when directly using \Euclid images in training, due to the lower resolution and noise of the \Euclid images, \texttt{Zoobot} performs worse for these finer features.

\section{Adapting \texttt{Zoobot} to a new morphology type}\label{sec:Peculiar}

We trained \texttt{Zoobot} to emulated \Euclid images with the GZH decision tree (see Table\,\ref{tab:questions}). Therefore, our model could be directly used for real \Euclid images, but is restricted to answering only the questions of the GZH decision tree. However, for \Euclid, there might be additional or other galaxy morphology tasks that are currently not included in our \texttt{Zoobot} model. We show that \texttt{Zoobot} can be easily adapted to a new morphology task that is not included in the GZH tasks with the example of peculiar galaxies. 

\subsection{Adaption procedure}

\begin{figure}[htbp!]
\centering
\includegraphics[width=\linewidth]{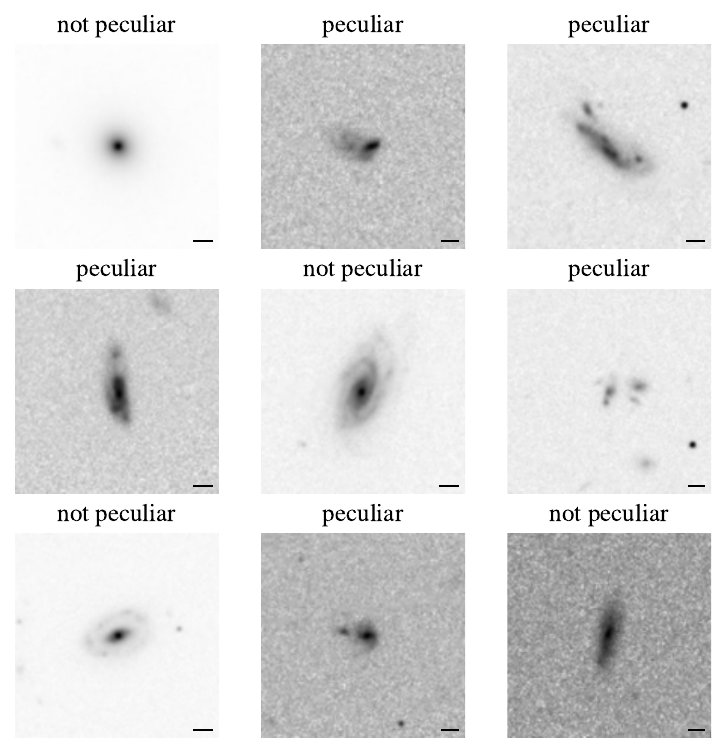}
\caption{Examples of galaxies and their expert labels as a peculiar or normal galaxy. The black bars represent a length of $\ang{;;1}$. There are no distinct morphological features that characterise peculiar galaxies, making this classification task rather difficult.}
\label{fig:peculiar_examples}
\end{figure} 

Peculiar galaxies are a type of irregular galaxy, with disorganised structure, often at high redshifts that do not typically fall into smooth, spheroid, or disc classes. The class of peculiar galaxies was not included in the GZH questions. Instead, we used labels for the same emulated \Euclid VIS images (see Sect.\,\ref{sec:Data}) from a different source, namely expert classification from the \Euclid Zoo project\footnote{\url{https://www.zooniverse.org/projects/sandorkruk/euclid-zoo/}}. \Euclid Zoo was an internal classification project in the \Euclid Consortium, with astronomers as classifiers. In total, 2006 galaxies were classified with $N=1$ to $N=3$ expert classifications per galaxy image. We selected a galaxy from the dataset to be classified as peculiar if the vote fraction for the peculiar class is larger than $50\,\%$, resulting in $231$ galaxy images for the peculiar class. In Fig.\,\ref{fig:peculiar_examples}, we show examples of galaxy images and their corresponding labels. We then applied a 70/10/20 percent train/validation/test split, leading to 1404 images for training, 200 for validation and 402 for testing. Next, we balanced our train and validation set by randomly dropping galaxy images that are not peculiar, leading to the same number of `not peculiar' and `peculiar' galaxy images. In total, 308 galaxy images were used for training and 54 for validation.

Similar to \citet{Walmsley_2022_2}, we used our best-performing model (trained on all images from the complete set) and replaced the final output layer by a new model `head', simply consisting of a final dense layer with a sigmoid activation function. We used the Adam optimizer \citep{Kingma_2014}. We then trained the new model with the dataset of peculiar galaxies while applying the same augmentation as before, namely random flips and rotations. To avoid overfitting, we stopped training as soon as the validation loss was not decreasing for 20 consecutive epochs. After training, \texttt{Zoobot} calculated predictions for the 402 images of the test set.

\subsection{Performance of the adapted \texttt{Zoobot}}

\begin{table}[h]
\begin{center}
    \caption{Classification metrics accuracy $A$, precision $P$, recall $R,$ and the unweighted and weighted $F_1$-scores $F_1$ and $F_1^\star$ for identifying peculiar galaxies in the test set with different confidence thresholds $c_{\textrm{th}}$. The corresponding confusion matrices are shown in Fig.\,\ref{fig:peculiar_confusion_matrix}.}
    \label{tab:metrics_peculiar}
    \tiny
    \begin{tabular}{lrlllll}
    \hline
    $c_{\textrm{th}}$ & $N_{\textrm{total}}$ & $A$ & $P$ & $R$ & $F_1$ & $F_1^\star$\\ \hline
    $0.5$ & 402 & $0.791$ & $0.670$ & $0.829$ & $0.689$ & $0.823$ \\
    $0.75$ & 402 & $0.915$ & $0.803$ & $0.823$ & $0.812$ & $0.917$
    \end{tabular}
\end{center}
\end{table}

\begin{figure}[htbp!]%
  \centering
  \includegraphics[width=\linewidth]{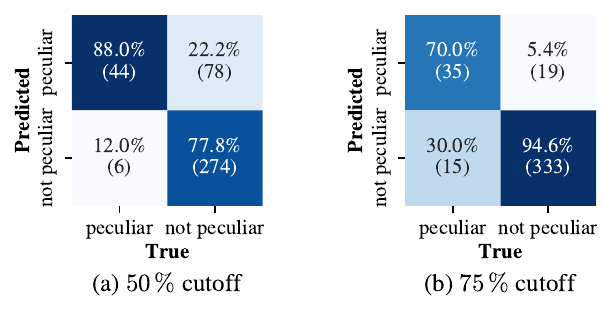}%
  \caption{Confusion matrices for finding peculiar galaxies for \texttt{Zoobot} pretrained on the emulated \Euclid images and the GZH tasks at (a) $50\,\%$ and (b) $75\,\%$ cutoff thresholds $c_{\textrm{th}}$ for selecting peculiar galaxies.}%
  \label{fig:peculiar_confusion_matrix}
\end{figure}

The performance of the model is listed in Table\,\ref{tab:metrics_peculiar} when evaluated with the classification metrics introduced in Sect.\,\ref{sec:discrete}. The corresponding confusion matrices are displayed in Fig.\,\ref{fig:peculiar_confusion_matrix}. The model achieves an accuracy of $79.1\,\%$ for a model confidence threshold of $c_{\textrm{th}}=0.5$ for selecting a galaxy to be peculiar. By applying $c_{\textrm{th}}=0.75$ for our model predictions, we obtain a higher accuracy of $91.5\,\%$ and a higher $F_1$-score of $81.2\,\%$ due to significantly higher precision. The values indicate that \texttt{Zoobot} performs well at the task of finding peculiar galaxies. 

The accurate identification of peculiar galaxies is particularly impressive, considering that it is a relatively challenging task even for an expert, due to the lack of clear morphological features. In addition to the inherent difficulty, there were only 231 examples of peculiar galaxies, of which 20\,\% were not included in the training dataset. This underscores our earlier discussion regarding the effectiveness of fine-tuning. Our \texttt{Zoobot} model was initially trained on all major GZ campaigns (as described in Sect.\,\ref{sec:Zoobot}) and subsequently on GZH using emulated \Euclid images, making it well-suited for adaptation to a new \Euclid morphology task.

This shows that \texttt{Zoobot} can easily be adapted to new problems, even if these are difficult and do not have many examples. For the application to \Euclid, our trained model can be used as a first step to predict detailed morphology for \Euclid with the GZH questions and can then be adapted to a new task in an effective way without requiring large labelled sets of galaxy images. Thus, in practice, if an astronomer is interested in finding all examples of a particular galaxy morphological type that is not included in the GZH questions for a given set of real \Euclid images, the following steps can be applied. First, a dataset needs to be labelled that is then used to fine-tune the trained  \texttt{Zoobot} model to the new galaxy morphology task. Once the model is fine-tuned, it can be used to classify all images of a given set of \Euclid images.

\section{Summary and conclusions}\label{sec:Conclusions}

This paper introduces automated and detailed predictions of galaxy morphology for emulated \Euclid images. These emulated images were generated by converting HST COSMOS images to \Euclid VIS images, considering the \Euclid PSF and adjusting them to match the expected noise level of \Euclid. The automated predictions were created using \texttt{Zoobot}, a Python package for creating deep learning models that classify galaxy morphology and for adapting (`fine-tuning') those models to new surveys and tasks. We fine-tuned a pre-existing \texttt{Zoobot} model (trained on 450\,000 non-Euclid galaxies from Galaxy Zoo) using emulated \Euclid images and labels derived from the Galaxy Zoo: Hubble (GZH) volunteer responses.

The model is able to accurately predict the detailed morphologies for emulated \Euclid galaxy images. It predicts various aspects, including the presence and quantity of clumps, detection, and counting of spiral arms, measurement of their winding, identification of disc galaxies, detection of bars, and determination of the presence, shape, and size of the central bulge, as well as measurement of the shape of featureless galaxies (refer to Table\,\ref{tab:questions}).

The \texttt{Zoobot} model fine-tuned on 60\,000 available emulated \Euclid images with GZH labels achieves a mean deviation of the predicted vote fraction from the volunteer classifications averaged over all answers of $9.5\,\%$ and below $12\,\%$ for nearly all answers individually (36 out of 40, as depicted in Fig.\,\ref{fig:deviations_euclid_mag}). Additionally, it achieves an accuracy of above $91\%$ for 12 of 13 questions when considering confident volunteer responses (refer to Table\,\ref{tab:metrics_confident}). However, the model's performance varies across different morphology classes.

For the top questions of the decision tree (global morphology type -- `smooth-or-features', disc orientation -- `disc-edge-on' or `bulge-size'), the model is able to predict within $10\,\%$ of the volunteers' vote fraction after being trained with only 1000 randomly selected galaxies. For other questions, such as `how-rounded', `spiral-arm-count', or `bulge-shape', 10\,000 training galaxies are needed, while for questions related to the more complex morphologies, such as `has-spiral-arms', `bar', `spiral-winding', or `clumpy-appearance', the full training set of 60\,000 galaxies is required to reach $12\,\%$ deviation from the volunteer classifications. This suggests that using a greater number of examples of complex morphology classes improves the performance of the model. Finally, our investigations of the effects of using the complete sample of available galaxies for training ($m_{I814W}<23.5$), or a subset of the brightest galaxies ($m_{I814W}<22.5$), suggest that the difference in performance is minimal; the number of galaxies with complex morphologies used for training has a higher impact. 

Our results have the following implications for \Euclid:
\begin{itemize}
\item \texttt{Zoobot}, trained on emulated \Euclid galaxies using volunteer labels from GZH, shows accurate predictions (within $10\%$ of human classifications) for global morphology (smooth versus featured), disc orientation (edge-on versus face-on), and bulge size.
\item To enhance the model's performance in predicting more complex detailed morphologies, such as bars, spiral arms, and clumps, additional labels are required. Based on Fig.\,\ref{fig:deviations_num_galaxies_questions}, approximately 60\,000 randomly selected galaxies would be needed to achieve a global vote fraction deviation of below $10\%$ and maintain deviations below $12\%$ for all labels. These additional labels could be obtained by initiating a Galaxy Zoo project for \Euclid using \Euclid Q1 data.
\item Our experiments indicate minimal performance differences when selecting galaxies with $m_{I814W}<23.5$ or brighter galaxies with $m_{I814W}<22.5$. Therefore, we suggest that the pool of explored galaxies for \Euclid be expanded with a restriction of $\IE<23.5$ (assuming VIS magnitudes are reasonably similar to I814W magnitudes). Fainter galaxies were not tested as no morphological labels were available for these galaxies. We expect the fraction of galaxies with features to decrease at higher magnitudes and smaller sizes, as observational effects cause these galaxies to appear smoother.
\item \texttt{Zoobot} can be adapted to new \Euclid morphology tasks using a few new labels. We demonstrate this adaptability by successfully training \texttt{Zoobot} for a new class of peculiar galaxies, consisting of only 261 examples, achieving an accuracy of $91.5\%$ (Fig.\,\ref{fig:peculiar_confusion_matrix}). Consequently, for new classes, it is feasible to set up a dedicated Galaxy Zoo-style workflow where volunteers are asked simple binary questions related to the morphology of the specific class of interest. The exact number of required labelled galaxies depends on the specific morphology class (Fig.\,\ref{fig:deviations_num_galaxies_questions}).
\item The proposed morphology classification scheme for \Euclid is outlined in a companion paper (Euclid Collaboration: OU-MER, in prep.).
\end{itemize}

Currently, the generation of structural parameters describing galaxy morphology with a morphology fitting code is included in the \Euclid data pipeline \citep{Bretonniere-EP26}. The algorithm generates morphological parameters for single- and double-S\'ersic components, and the measurements are reliable up to $\IE=23$ for one component and $\IE<21$ for two components. \citet{Bretonniere-EP26} conclude that robust structural parameters will be delivered for at least 400 million galaxies by the \Euclid Data Releases.

We estimated the number of galaxies for which detailed morphologies could be measured with our deep learning model. With $m_{I814W}<23.5$, there are approximately $70$\,000 galaxies in an area of $(1.2\times1.2)$\,deg$^2=1.44\,$deg$^2$ of the HST COSMOS survey. Scaling up to the total sky area measured by the \Euclid Wide Survey, of namely 15\,000\,deg$^2$, and assuming that VIS magnitudes are similar to the $I$814W magnitudes of HST, there would be approximately $800$ million galaxies with reliably measured morphologies up to $\IE=23.5$ (focusing on the brighter magnitudes, $\IE<22.5$, the estimated count would be approximately $300$ million galaxies). This is close to the 400 million galaxies estimated by \citet{Bretonniere-EP26} for $\IE<23$. Accounting for an average vote fraction of $29\,\%$ for galaxies to display features, we conclude that, of the 800 million measured galaxies from the \Euclid Wide Survey, approximately 230 million galaxies will display complex morphology. This closely matches the 250 million galaxies that are estimated to have complex structures by \citet{Bretonniere-EP13} for the Euclid Wide Survey and the Euclid Deep Survey.

In conclusion, we successfully showcase the feasibility of generating high-quality and detailed morphology predictions for \Euclid images. Our trained \texttt{Zoobot} model is now ready for deployment in the \Euclid pipeline to produce morphological catalogues for \Euclid images with Q1 data. As additional labels for more complex morphologies are obtained, the performance of \texttt{Zoobot} will improve for the upcoming Data Release 1 (DR1). Moreover, the model can be easily adapted to new morphology classes that are of interest to astronomers as new labels are gathered through crowd-sourcing projects.

\begin{acknowledgements}
  
\AckEC
The data in this paper are the result of the efforts of the Galaxy Zoo volunteers, without whom none of this work would be possible. Their efforts are individually acknowledged at \texttt{http://authors.galaxyzoo.org}.
This publication uses data generated via the Zooniverse.org platform, development of which is funded by generous support, including a Global Impact Award from Google, and by a grant from the Alfred P. Sloan Foundation.
\end{acknowledgements}


\bibliography{refs,Euclid}

\begin{thebibliography}{48}
\expandafter\ifx\csname natexlab\endcsname\relax\def\natexlab#1{#1}\fi

\bibitem[{Abadi {et~al.}(2016)Abadi, Barham, Chen, Chen, Davis, Dean, Devin, Ghemawat, Irving, Isard, Kudlur, Levenberg, Monga, Moore, Murray, Steiner, Tucker, Vasudevan, Warden, Wicke, Yu, \& Zheng}]{Abadi_2016}
Abadi, M., Barham, P., Chen, J., {et~al.} 2016, in Proceedings of the 12th USENIX Symposium on Operating Systems Design and Implementation, OSDI'16

\bibitem[{Baillard {et~al.}(2011)Baillard, Bertin, de~Lapparent, Fouqu{\'e}, Arnouts, Mellier, Pell{\'o}, Leborgne, Prugniel, Makarov, Makarova, McCracken, Bijaoui, \& Tasca}]{Baillard_2011}
Baillard, A., Bertin, E., de~Lapparent, V., {et~al.} 2011, \aap, 532, A74

\bibitem[{{Bait} {et~al.}(2017){Bait}, {Barway}, \& {Wadadekar}}]{Bait_2017}
{Bait}, O., {Barway}, S., \& {Wadadekar}, Y. 2017, \mnras, 471, 2687

\bibitem[{Cheng {et~al.}(2020)Cheng, Conselice, {Arag{\'o}n-Salamanca}, Li, Bluck, Hartley, Annis, Brooks, Doel, {Garc{\'i}a-Bellido}, James, Kuehn, Kuropatkin, Smith, Sobreira, \& Tarle}]{Cheng_2020}
Cheng, T.-Y., Conselice, C.~J., {Arag{\'o}n-Salamanca}, A., {et~al.} 2020, \mnras, 493, 4209

\bibitem[{{Conselice}(2003)}]{Conselice_2003}
{Conselice}, C.~J. 2003, \apjs, 147, 1

\bibitem[{Cropper {et~al.}(2016)Cropper, Pottinger, Niemi, Azzollini, Denniston, Szafraniec, Awan, Mellier, Berthe, Martignac, Cara, Di~Giorgio, Sciortino, Bozzo, Genolet, Cole, Philippon, Hailey, Hunt, Swindells, Holland, Gow, Murray, Hall, Skottfelt, Amiaux, Laureijs, Racca, Salvignol, Short, Lorenzo~Alvarez, Kitching, Hoekstra, Massey, \& Israel}]{Cropper_2016}
Cropper, M., Pottinger, S., Niemi, S., {et~al.} 2016, in Space Telescopes and Instrumentation 2016: Optical, Infrared, and Millimeter Wave, 99040Q

\bibitem[{{de Vaucouleurs}(1959)}]{deVaucouleurs_1959}
{de Vaucouleurs}, G. 1959, Handbuch der Physik, 53, 275

\bibitem[{{de Vaucouleurs} {et~al.}(1991){de Vaucouleurs}, {de Vaucouleurs}, Corwin, Buta, Paturel, \& Fouque}]{deVaucouleurs_1991}
{de Vaucouleurs}, G., {de Vaucouleurs}, A., Corwin, Jr., H.~G., {et~al.} 1991, Third {{Reference Catalogue}} of {{Bright Galaxies}} (Springer, New York)

\bibitem[{Dey {et~al.}(2019)Dey, Schlegel, Lang, Blum, Burleigh, Fan, Findlay, Finkbeiner, Herrera, Juneau, Landriau, Levi, McGreer, Meisner, Myers, Moustakas, Nugent, Patej, Schlafly, Walker, Valdes, Weaver, Y{\`e}che, Zou, Zhou, Abareshi, Abbott, Abolfathi, Aguilera, Alam, Allen, Alvarez, Annis, Ansarinejad, Aubert, Beechert, Bell, BenZvi, Beutler, Bielby, Bolton, Brice{\~n}o, {Buckley-Geer}, Butler, Calamida, Carlberg, Carter, Casas, Castander, Choi, Comparat, Cukanovaite, Delubac, DeVries, Dey, Dhungana, Dickinson, Ding, Donaldson, Duan, Duckworth, Eftekharzadeh, Eisenstein, Etourneau, Fagrelius, Farihi, Fitzpatrick, {Font-Ribera}, Fulmer, G{\"a}nsicke, Gaztanaga, George, Gerdes, Gontcho, Gorgoni, Green, Guy, Harmer, Hernandez, Honscheid, Huang, James, Jannuzi, Jiang, Joyce, Karcher, Karkar, Kehoe, {Jean-Paul}, {Kueter-Young}, Lan, Lauer, Guillou, Suu, Lee, Lesser, Levasseur, Li, Mann, Marshall, {Mart{\'i}nez-V{\'a}zquez}, Martini, des Bourboux, McManus, Meier, M{\'e}nard, Metcalfe,
  {Mu{\~n}oz-Guti{\'e}rrez}, Najita, Napier, Narayan, Newman, Nie, Nord, Norman, Olsen, Paat, {Palanque-Delabrouille}, Peng, Poppett, Poremba, Prakash, Rabinowitz, Raichoor, Rezaie, Robertson, Roe, Ross, Ross, Rudnick, Safonova, Saha, S{\'a}nchez, Savary, Schweiker, Scott, Seo, Shan, Silva, Slepian, Soto, Sprayberry, Staten, Stillman, Stupak, Summers, Tie, Tirado, {Vargas-Maga{\~n}a}, Vivas, Wechsler, Williams, Yang, Yang, Yapici, Zaritsky, Zenteno, Zhang, Zhang, Zhou, \& Zhou}]{Dey_2019}
Dey, A., Schlegel, D.~J., Lang, D., {et~al.} 2019, \aj, 157, 168

\bibitem[{Dieleman {et~al.}(2015)Dieleman, Willett, \& Dambre}]{Dieleman_2015}
Dieleman, S., Willett, K.~W., \& Dambre, J. 2015, \mnras, 450, 1441

\bibitem[{Dom{\'i}nguez~S{\'a}nchez {et~al.}(2019)Dom{\'i}nguez~S{\'a}nchez, {Huertas-Company}, Bernardi, Kaviraj, Fischer, Abbott, Abdalla, Annis, Avila, Brooks, {Buckley-Geer}, Carnero~Rosell, Carrasco~Kind, Carretero, Cunha, D'Andrea, {da~Costa}, Davis, De~Vicente, Doel, Evrard, Fosalba, Frieman, {Garc{\'i}a-Bellido}, Gaztanaga, Gerdes, Gruen, Gruendl, Gschwend, Gutierrez, Hartley, Hollowood, Honscheid, Hoyle, James, Kuehn, Kuropatkin, Lahav, Maia, March, Melchior, Menanteau, Miquel, Nord, Plazas, Sanchez, Scarpine, Schindler, Schubnell, Smith, Smith, {Soares-Santos}, Sobreira, Suchyta, Swanson, Tarle, Thomas, Walker, \& Zuntz}]{DominguezSanchez_2019}
Dom{\'i}nguez~S{\'a}nchez, H., {Huertas-Company}, M., Bernardi, M., {et~al.} 2019, \mnras, 484, 93

\bibitem[{Dom{\'i}nguez~S{\'a}nchez {et~al.}(2018)Dom{\'i}nguez~S{\'a}nchez, {Huertas-Company}, Bernardi, Tuccillo, \& Fischer}]{DominguezSanchez_2018}
Dom{\'i}nguez~S{\'a}nchez, H., {Huertas-Company}, M., Bernardi, M., Tuccillo, D., \& Fischer, J.~L. 2018, \mnras, 476, 3661

\bibitem[{{Euclid Collaboration: Bretonni{\`e}re} {et~al.}(2022){Euclid Collaboration: Bretonni{\`e}re}, {Huertas-Company}, {Boucaud}, {Lanusse}, {Jullo}, {Merlin}, {Tuccillo}, {Castellano}, {Brinchmann}, {Conselice}, {Dole}, {Cabanac}, {Courtois}, {Castander}, {Duc}, {Fosalba}, {Guinet}, {Kruk}, {Kuchner}, {Serrano}, {Soubrie}, {Tramacere}, {Wang}, {Amara}, {Auricchio}, {Bender}, {Bodendorf}, {Bonino}, {Branchini}, {Brau-Nogue}, {Brescia}, {Capobianco}, {Carbone}, {Carretero}, {Cavuoti}, {Cimatti}, {Cledassou}, {Congedo}, {Conversi}, {Copin}, {Corcione}, {Costille}, {Cropper}, {Da Silva}, {Degaudenzi}, {Douspis}, {Dubath}, {Duncan}, {Dupac}, {Dusini}, {Farrens}, {Ferriol}, {Frailis}, {Franceschi}, {Fumana}, {Garilli}, {Gillard}, {Gillis}, {Giocoli}, {Grazian}, {Grupp}, {Haugan}, {Holmes}, {Hormuth}, {Hudelot}, {Jahnke}, {Kermiche}, {Kiessling}, {Kilbinger}, {Kitching}, {Kohley}, {K{\"u}mmel}, {Kunz}, {Kurki-Suonio}, {Ligori}, {Lilje}, {Lloro}, {Maiorano}, {Mansutti}, {Marggraf}, {Markovic}, {Marulli},
  {Massey}, {Maurogordato}, {Melchior}, {Meneghetti}, {Meylan}, {Moresco}, {Morin}, {Moscardini}, {Munari}, {Nakajima}, {Niemi}, {Padilla}, {Paltani}, {Pasian}, {Pedersen}, {Pettorino}, {Pires}, {Poncet}, {Popa}, {Pozzetti}, {Raison}, {Rebolo}, {Rhodes}, {Roncarelli}, {Rossetti}, {Saglia}, {Schneider}, {Secroun}, {Seidel}, {Sirignano}, {Sirri}, {Stanco}, {Starck}, {Tallada-Cresp{\'\i}}, {Taylor}, {Tereno}, {Toledo-Moreo}, {Torradeflot}, {Valentijn}, {Valenziano}, {Wang}, {Welikala}, {Weller}, {Zamorani}, {Zoubian}, {Baldi}, {Bardelli}, {Camera}, {Farinelli}, {Medinaceli}, {Mei}, {Polenta}, {Romelli}, {Tenti}, {Vassallo}, {Zacchei}, {Zucca}, {Baccigalupi}, {Balaguera-Antol{\'\i}nez}, {Biviano}, {Borgani}, {Bozzo}, {Burigana}, {Cappi}, {Carvalho}, {Casas}, {Castignani}, {Colodro-Conde}, {Coupon}, {de la Torre}, {Fabricius}, {Farina}, {Ferreira}, {Flose-Reimberg}, {Fotopoulou}, {Galeotta}, {Ganga}, {Garcia-Bellido}, {Gaztanaga}, {Gozaliasl}, {Hook}, {Joachimi}, {Kansal}, {Kashlinsky}, {Keihanen}, {Kirkpatrick},
  {Lindholm}, {Mainetti}, {Maino}, {Maoli}, {Martinelli}, {Martinet}, {McCracken}, {Metcalf}, {Morgante}, {Morisset}, {Nightingale}, {Nucita}, {Patrizii}, {Potter}, {Renzi}, {Riccio}, {S{\'a}nchez}, {Sapone}, {Schirmer}, {Schultheis}, {Scottez}, {Sefusatti}, {Teyssier}, {Tutusaus}, {Valiviita}, {Viel}, {Whittaker}, \& {Knapen}}]{Bretonniere-EP13}
{Euclid Collaboration: Bretonni{\`e}re}, H., {Huertas-Company}, M., {Boucaud}, A., {et~al.} 2022, \aap, 657, A90

\bibitem[{{Euclid Collaboration: Bretonni{\`e}re} {et~al.}(2023){Euclid Collaboration: Bretonni{\`e}re}, {Kuchner}, {Huertas-Company}, {Merlin}, {Castellano}, {Tuccillo}, {Buitrago}, {Conselice}, {Boucaud}, {H{\"a}u{\ss}ler}, {K{\"u}mmel}, {Hartley}, {Alvarez Ayllon}, {Bertin}, {Ferrari}, {Ferreira}, {Gavazzi}, {Hern{\'a}ndez-Lang}, {Lucatelli}, {Robotham}, {Schefer}, {Wang}, {Cabanac}, {Dom{\'\i}nguez S{\'a}nchez}, {Duc}, {Fotopoulou}, {Kruk}, {La Marca}, {Margalef-Bentabol}, {Marleau}, {Tortora}, {Aghanim}, {Amara}, {Auricchio}, {Azzollini}, {Baldi}, {Bender}, {Bodendorf}, {Branchini}, {Brescia}, {Brinchmann}, {Camera}, {Capobianco}, {Carbone}, {Carretero}, {Castander}, {Cavuoti}, {Cimatti}, {Cledassou}, {Congedo}, {Conversi}, {Copin}, {Corcione}, {Courbin}, {Cropper}, {Da Silva}, {Degaudenzi}, {Dinis}, {Dubath}, {Duncan}, {Dupac}, {Dusini}, {Farrens}, {Ferriol}, {Frailis}, {Franceschi}, {Fumana}, {Galeotta}, {Garilli}, {Gillis}, {Giocoli}, {Grazian}, {Grupp}, {Haugan}, {Hoekstra}, {Holmes}, {Hormuth},
  {Hornstrup}, {Hudelot}, {Jahnke}, {Kermiche}, {Kiessling}, {Kohley}, {Kunz}, {Kurki-Suonio}, {Ligori}, {Lilje}, {Lloro}, {Mansutti}, {Marggraf}, {Markovic}, {Marulli}, {Massey}, {McCracken}, {Medinaceli}, {Melchior}, {Meneghetti}, {Meylan}, {Moresco}, {Moscardini}, {Munari}, {Niemi}, {Padilla}, {Paltani}, {Pasian}, {Pedersen}, {Percival}, {Pettorino}, {Polenta}, {Poncet}, {Pozzetti}, {Raison}, {Rebolo}, {Renzi}, {Rhodes}, {Riccio}, {Romelli}, {Rosset}, {Rossetti}, {Saglia}, {Sapone}, {Sartoris}, {Schneider}, {Secroun}, {Seidel}, {Sirignano}, {Sirri}, {Skottfelt}, {Starck}, {Tallada-Cresp{\'\i}}, {Taylor}, {Tereno}, {Toledo-Moreo}, {Tutusaus}, {Valentijn}, {Valenziano}, {Vassallo}, {Wang}, {Weller}, {Zamorani}, {Zoubian}, {Andreon}, {Bardelli}, {Colodro-Conde}, {Di Ferdinando}, {Graci{\'a}-Carpio}, {Lindholm}, {Mauri}, {Mei}, {Scottez}, {Zucca}, {Baccigalupi}, {Ballardini}, {Bernardeau}, {Biviano}, {Borgani}, {Borlaff}, {Burigana}, {Cappi}, {Carvalho}, {Casas}, {Castignani}, {Cooray}, {Coupon}, {Courtois},
  {Davini}, {De Lucia}, {Desprez}, {Escartin}, {Escoffier}, {Fabricius}, {Farina}, {Fontana}, {Ganga}, {Garcia-Bellido}, {George}, {Gozaliasl}, {Hildebrandt}, {Hook}, {Ilbert}, {Ili{\'c}}, {Joachimi}, {Kansal}, {Keihanen}, {Kirkpatrick}, {Loureiro}, {Macias-Perez}, {Magliocchetti}, {Maoli}, {Marcin}, {Martinelli}, {Martinet}, {Maturi}, {Monaco}, {Morgante}, {Nadathur}, {Nucita}, {Patrizii}, {Popa}, {Porciani}, {Potter}, {Pourtsidou}, {P{\"o}ntinen}, {Reimberg}, {S{\'a}nchez}, {Sakr}, {Schirmer}, {Sefusatti}, {Sereno}, {Stadel}, {Teyssier}, {Valiviita}, {van Mierlo}, {Veropalumbo}, {Viel}, {Weaver}, \& {Scott}}]{Bretonniere-EP26}
{Euclid Collaboration: Bretonni{\`e}re}, H., {Kuchner}, U., {Huertas-Company}, M., {et~al.} 2023, \aap, 671, A102

\bibitem[{{Euclid Collaboration: Scaramella} {et~al.}(2022){Euclid Collaboration: Scaramella}, {Amiaux}, {Mellier}, {Burigana}, {Carvalho}, {Cuillandre}, {Da Silva}, {Derosa}, {Dinis}, {Maiorano}, {Maris}, {Tereno}, {Laureijs}, {Boenke}, {Buenadicha}, {Dupac}, {Gaspar Venancio}, {G{\'o}mez-{\'A}lvarez}, {Hoar}, {Lorenzo Alvarez}, {Racca}, {Saavedra-Criado}, {Schwartz}, {Vavrek}, {Schirmer}, {Aussel}, {Azzollini}, {Cardone}, {Cropper}, {Ealet}, {Garilli}, {Gillard}, {Granett}, {Guzzo}, {Hoekstra}, {Jahnke}, {Kitching}, {Maciaszek}, {Meneghetti}, {Miller}, {Nakajima}, {Niemi}, {Pasian}, {Percival}, {Pottinger}, {Sauvage}, {Scodeggio}, {Wachter}, {Zacchei}, {Aghanim}, {Amara}, {Auphan}, {Auricchio}, {Awan}, {Balestra}, {Bender}, {Bodendorf}, {Bonino}, {Branchini}, {Brau-Nogue}, {Brescia}, {Candini}, {Capobianco}, {Carbone}, {Carlberg}, {Carretero}, {Casas}, {Castander}, {Castellano}, {Cavuoti}, {Cimatti}, {Cledassou}, {Congedo}, {Conselice}, {Conversi}, {Copin}, {Corcione}, {Costille}, {Courbin}, {Degaudenzi},
  {Douspis}, {Dubath}, {Duncan}, {Dusini}, {Farrens}, {Ferriol}, {Fosalba}, {Fourmanoit}, {Frailis}, {Franceschi}, {Franzetti}, {Fumana}, {Gillis}, {Giocoli}, {Grazian}, {Grupp}, {Haugan}, {Holmes}, {Hormuth}, {Hudelot}, {Kermiche}, {Kiessling}, {Kilbinger}, {Kohley}, {Kubik}, {K{\"u}mmel}, {Kunz}, {Kurki-Suonio}, {Lahav}, {Ligori}, {Lilje}, {Lloro}, {Mansutti}, {Marggraf}, {Markovic}, {Marulli}, {Massey}, {Maurogordato}, {Melchior}, {Merlin}, {Meylan}, {Mohr}, {Moresco}, {Morin}, {Moscardini}, {Munari}, {Nichol}, {Padilla}, {Paltani}, {Peacock}, {Pedersen}, {Pettorino}, {Pires}, {Poncet}, {Popa}, {Pozzetti}, {Raison}, {Rebolo}, {Rhodes}, {Rix}, {Roncarelli}, {Rossetti}, {Saglia}, {Schneider}, {Schrabback}, {Secroun}, {Seidel}, {Serrano}, {Sirignano}, {Sirri}, {Skottfelt}, {Stanco}, {Starck}, {Tallada-Cresp{\'\i}}, {Tavagnacco}, {Taylor}, {Teplitz}, {Toledo-Moreo}, {Torradeflot}, {Trifoglio}, {Valentijn}, {Valenziano}, {Verdoes Kleijn}, {Wang}, {Welikala}, {Weller}, {Wetzstein}, {Zamorani}, {Zoubian},
  {Andreon}, {Baldi}, {Bardelli}, {Boucaud}, {Camera}, {Di Ferdinando}, {Fabbian}, {Farinelli}, {Galeotta}, {Graci{\'a}-Carpio}, {Maino}, {Medinaceli}, {Mei}, {Neissner}, {Polenta}, {Renzi}, {Romelli}, {Rosset}, {Sureau}, {Tenti}, {Vassallo}, {Zucca}, {Baccigalupi}, {Balaguera-Antol{\'\i}nez}, {Battaglia}, {Biviano}, {Borgani}, {Bozzo}, {Cabanac}, {Cappi}, {Casas}, {Castignani}, {Colodro-Conde}, {Coupon}, {Courtois}, {Cuby}, {de la Torre}, {Desai}, {Dole}, {Fabricius}, {Farina}, {Ferreira}, {Finelli}, {Flose-Reimberg}, {Fotopoulou}, {Ganga}, {Gozaliasl}, {Hook}, {Keihanen}, {Kirkpatrick}, {Liebing}, {Lindholm}, {Mainetti}, {Martinelli}, {Martinet}, {Maturi}, {McCracken}, {Metcalf}, {Morgante}, {Nightingale}, {Nucita}, {Patrizii}, {Potter}, {Riccio}, {S{\'a}nchez}, {Sapone}, {Schewtschenko}, {Schultheis}, {Scottez}, {Teyssier}, {Tutusaus}, {Valiviita}, {Viel}, {Vriend}, \& {Whittaker}}]{Scaramella-EP1}
{Euclid Collaboration: Scaramella}, R., {Amiaux}, J., {Mellier}, Y., {et~al.} 2022, \aap, 662, A112

\bibitem[{Griffith {et~al.}(2012)Griffith, Cooper, Newman, Moustakas, Stern, Comerford, Davis, Lotz, Barden, Conselice, Capak, Faber, Kirkpatrick, Koekemoer, Koo, Noeske, Scoville, Sheth, Shopbell, Willmer, \& Weiner}]{Griffith_2012}
Griffith, R.~L., Cooper, M.~C., Newman, J.~A., {et~al.} 2012, \apjs, 200, 9

\bibitem[{Grogin {et~al.}(2011)Grogin, Kocevski, Faber, Ferguson, Koekemoer, Riess, Acquaviva, Alexander, Almaini, Ashby, Barden, Bell, Bournaud, Brown, Caputi, Casertano, Cassata, Castellano, Challis, Chary, Cheung, Cirasuolo, Conselice, Roshan~Cooray, Croton, Daddi, Dahlen, Dav{\'e}, {de Mello}, Dekel, Dickinson, Dolch, Donley, Dunlop, Dutton, Elbaz, Fazio, Filippenko, Finkelstein, Fontana, Gardner, Garnavich, Gawiser, Giavalisco, Grazian, Guo, Hathi, H{\"a}ussler, Hopkins, Huang, Huang, Jha, Kartaltepe, Kirshner, Koo, Lai, Lee, Li, Lotz, Lucas, Madau, McCarthy, McGrath, McIntosh, McLure, Mobasher, Moustakas, Mozena, Nandra, Newman, Niemi, Noeske, Papovich, Pentericci, Pope, Primack, Rajan, Ravindranath, Reddy, Renzini, Rix, Robaina, Rodney, Rosario, Rosati, Salimbeni, Scarlata, Siana, Simard, Smidt, Somerville, Spinrad, Straughn, Strolger, Telford, Teplitz, Trump, {van der Wel}, Villforth, Wechsler, Weiner, Wiklind, Wild, Wilson, Wuyts, Yan, \& Yun}]{Grogin_2011}
Grogin, N.~A., Kocevski, D.~D., Faber, S.~M., {et~al.} 2011, \apjs, 197, 35

\bibitem[{Hubble(1926)}]{Hubble_1926}
Hubble, E.~P. 1926, \apj, 64, 321

\bibitem[{{Huertas-Company} {et~al.}(2015){Huertas-Company}, Gravet, {Cabrera-Vives}, {P{\'e}rez-Gonz{\'a}lez}, Kartaltepe, Barro, Bernardi, Mei, Shankar, Dimauro, Bell, Kocevski, Koo, Faber, \& Mcintosh}]{Huertas-Company_2015}
{Huertas-Company}, M., Gravet, R., {Cabrera-Vives}, G., {et~al.} 2015, \apjs, 221, 8

\bibitem[{{Ivezi{\'c}} {et~al.}(2019){Ivezi{\'c}}, {Kahn}, {Tyson}, {Abel}, {Acosta}, {Allsman}, {Alonso}, {AlSayyad}, {Anderson}, {Andrew}, {Angel}, {Angeli}, {Ansari}, {Antilogus}, {Araujo}, {Armstrong}, {Arndt}, {Astier}, {Aubourg}, {Auza}, {Axelrod}, {Bard}, {Barr}, {Barrau}, {Bartlett}, {Bauer}, {Bauman}, {Baumont}, {Bechtol}, {Bechtol}, {Becker}, {Becla}, {Beldica}, {Bellavia}, {Bianco}, {Biswas}, {Blanc}, {Blazek}, {Blandford}, {Bloom}, {Bogart}, {Bond}, {Booth}, {Borgland}, {Borne}, {Bosch}, {Boutigny}, {Brackett}, {Bradshaw}, {Brandt}, {Brown}, {Bullock}, {Burchat}, {Burke}, {Cagnoli}, {Calabrese}, {Callahan}, {Callen}, {Carlin}, {Carlson}, {Chandrasekharan}, {Charles-Emerson}, {Chesley}, {Cheu}, {Chiang}, {Chiang}, {Chirino}, {Chow}, {Ciardi}, {Claver}, {Cohen-Tanugi}, {Cockrum}, {Coles}, {Connolly}, {Cook}, {Cooray}, {Covey}, {Cribbs}, {Cui}, {Cutri}, {Daly}, {Daniel}, {Daruich}, {Daubard}, {Daues}, {Dawson}, {Delgado}, {Dellapenna}, {de Peyster}, {de Val-Borro}, {Digel}, {Doherty}, {Dubois},
  {Dubois-Felsmann}, {Durech}, {Economou}, {Eifler}, {Eracleous}, {Emmons}, {Fausti Neto}, {Ferguson}, {Figueroa}, {Fisher-Levine}, {Focke}, {Foss}, {Frank}, {Freemon}, {Gangler}, {Gawiser}, {Geary}, {Gee}, {Geha}, {Gessner}, {Gibson}, {Gilmore}, {Glanzman}, {Glick}, {Goldina}, {Goldstein}, {Goodenow}, {Graham}, {Gressler}, {Gris}, {Guy}, {Guyonnet}, {Haller}, {Harris}, {Hascall}, {Haupt}, {Hernandez}, {Herrmann}, {Hileman}, {Hoblitt}, {Hodgson}, {Hogan}, {Howard}, {Huang}, {Huffer}, {Ingraham}, {Innes}, {Jacoby}, {Jain}, {Jammes}, {Jee}, {Jenness}, {Jernigan}, {Jevremovi{\'c}}, {Johns}, {Johnson}, {Johnson}, {Jones}, {Juramy-Gilles}, {Juri{\'c}}, {Kalirai}, {Kallivayalil}, {Kalmbach}, {Kantor}, {Karst}, {Kasliwal}, {Kelly}, {Kessler}, {Kinnison}, {Kirkby}, {Knox}, {Kotov}, {Krabbendam}, {Krughoff}, {Kub{\'a}nek}, {Kuczewski}, {Kulkarni}, {Ku}, {Kurita}, {Lage}, {Lambert}, {Lange}, {Langton}, {Le Guillou}, {Levine}, {Liang}, {Lim}, {Lintott}, {Long}, {Lopez}, {Lotz}, {Lupton}, {Lust}, {MacArthur}, {Mahabal},
  {Mandelbaum}, {Markiewicz}, {Marsh}, {Marshall}, {Marshall}, {May}, {McKercher}, {McQueen}, {Meyers}, {Migliore}, {Miller}, {Mills}, {Miraval}, {Moeyens}, {Moolekamp}, {Monet}, {Moniez}, {Monkewitz}, {Montgomery}, {Morrison}, {Mueller}, {Muller}, {Mu{\~n}oz Arancibia}, {Neill}, {Newbry}, {Nief}, {Nomerotski}, {Nordby}, {O'Connor}, {Oliver}, {Olivier}, {Olsen}, {O'Mullane}, {Ortiz}, {Osier}, {Owen}, {Pain}, {Palecek}, {Parejko}, {Parsons}, {Pease}, {Peterson}, {Peterson}, {Petravick}, {Libby Petrick}, {Petry}, {Pierfederici}, {Pietrowicz}, {Pike}, {Pinto}, {Plante}, {Plate}, {Plutchak}, {Price}, {Prouza}, {Radeka}, {Rajagopal}, {Rasmussen}, {Regnault}, {Reil}, {Reiss}, {Reuter}, {Ridgway}, {Riot}, {Ritz}, {Robinson}, {Roby}, {Roodman}, {Rosing}, {Roucelle}, {Rumore}, {Russo}, {Saha}, {Sassolas}, {Schalk}, {Schellart}, {Schindler}, {Schmidt}, {Schneider}, {Schneider}, {Schoening}, {Schumacher}, {Schwamb}, {Sebag}, {Selvy}, {Sembroski}, {Seppala}, {Serio}, {Serrano}, {Shaw}, {Shipsey}, {Sick}, {Silvestri},
  {Slater}, {Smith}, {Smith}, {Sobhani}, {Soldahl}, {Storrie-Lombardi}, {Stover}, {Strauss}, {Street}, {Stubbs}, {Sullivan}, {Sweeney}, {Swinbank}, {Szalay}, {Takacs}, {Tether}, {Thaler}, {Thayer}, {Thomas}, {Thornton}, {Thukral}, {Tice}, {Trilling}, {Turri}, {Van Berg}, {Vanden Berk}, {Vetter}, {Virieux}, {Vucina}, {Wahl}, {Walkowicz}, {Walsh}, {Walter}, {Wang}, {Wang}, {Warner}, {Wiecha}, {Willman}, {Winters}, {Wittman}, {Wolff}, {Wood-Vasey}, {Wu}, {Xin}, {Yoachim}, \& {Zhan}}]{Ivezic_2019}
{Ivezi{\'c}}, {\v{Z}}., {Kahn}, S.~M., {Tyson}, J.~A., {et~al.} 2019, \apj, 873, 111

\bibitem[{{Kaifu} {et~al.}(2000){Kaifu}, {Usuda}, {Hayashi}, {Itoh}, {Akiyama}, {Yamashita}, {Nakajima}, {Tamura}, {Inutsuka}, {Hayashi}, {Maihara}, {Iwamuro}, {Motohara}, {Iwai}, {Tanabe}, {Taguchi}, {Hata}, {Terada}, {Goto}, {Ando}, {Aoki}, {Chikada}, {Doi}, {Ebizuka}, {Fukuda}, {Hamabe}, {Hasegawa}, {Horaguchi}, {Ichikawa}, {Ichikawa}, {Imanishi}, {Imi}, {Inata}, {Isobe}, {Iye}, {Kamata}, {Kanzawa}, {Karoji}, {Kashikawa}, {Kataza}, {Kato}, {Kobayashi}, {Kobayashi}, {Kodaira}, {Kosugi}, {Kurakami}, {Mikami}, {Miyama}, {Miyashita}, {Miyata}, {Miyazaki}, {Mizumoto}, {Nakagiri}, {Nakajima}, {Nakamura}, {Nariai}, {Nishihara}, {Nishikawa}, {Nishimura}, {Nishimura}, {Nishino}, {Noguchi}, {Noguchi}, {Noumaru}, {Ogasawara}, {Okada}, {Okita}, {Omata}, {Oshima}, {Otsubo}, {Sasaki}, {Sasaki}, {Sekiguchi}, {Sekiguchi}, {Shelton}, {Simpson}, {Suto}, {Takami}, {Takata}, {Takato}, {Tanaka}, {Tanaka}, {Tomono}, {Torii}, {Waseda}, {Watanabe}, {Watanabe}, {Yagi}, {Yamashita}, {Yasuda}, {Yoshida}, {Yoshida}, \&
  {Yutani}}]{Kaifu_2000}
{Kaifu}, N., {Usuda}, T., {Hayashi}, S.~S., {et~al.} 2000, \pasj, 52, 1

\bibitem[{Kingma \& Ba(2015)}]{Kingma_2014}
Kingma, D.~P. \& Ba, J. 2015, in 3rd International Conference on Learning Representations, ICLR 2015 - Conference Track Proceedings

\bibitem[{Koekemoer {et~al.}(2007)Koekemoer, Aussel, Calzetti, Capak, Giavalisco, Kneib, Leauthaud, Le~Fevre, McCracken, Massey, Mobasher, Rhodes, Scoville, \& Shopbell}]{Koekemoer_2007}
Koekemoer, A.~M., Aussel, H., Calzetti, D., {et~al.} 2007, \apjs, 172, 196

\bibitem[{Kruk {et~al.}(2018)Kruk, Lintott, Bamford, Masters, Simmons, H{\"a}u{\ss}ler, Cardamone, Hart, Kelvin, Schawinski, Smethurst, \& Vika}]{Kruk_2018}
Kruk, S.~J., Lintott, C.~J., Bamford, S.~P., {et~al.} 2018, \mnras, 473, 4731

\bibitem[{Laureijs {et~al.}(2011)Laureijs, Amiaux, Arduini, Augu{\`e}res, Brinchmann, Cole, Cropper, Dabin, Duvet, Ealet, Garilli, Gondoin, Guzzo, Hoar, Hoekstra, Holmes, Kitching, Maciaszek, Mellier, Pasian, Percival, Rhodes, Saavedra~Criado, Sauvage, Scaramella, Valenziano, Warren, Bender, Castander, Cimatti, Le~F{\`e}vre, {Kurki-Suonio}, Levi, Lilje, Meylan, Nichol, Pedersen, Popa, Rebolo~Lopez, Rix, Rottgering, Zeilinger, Grupp, Hudelot, Massey, Meneghetti, Miller, Paltani, {Paulin-Henriksson}, Pires, Saxton, Schrabback, Seidel, Walsh, Aghanim, Amendola, Bartlett, Baccigalupi, Beaulieu, Benabed, Cuby, Elbaz, Fosalba, Gavazzi, Helmi, Hook, Irwin, Kneib, Kunz, Mannucci, Moscardini, Tao, Teyssier, Weller, Zamorani, Zapatero~Osorio, Boulade, Foumond, Di~Giorgio, Guttridge, James, Kemp, Martignac, Spencer, Walton, Bl{\"u}mchen, Bonoli, Bortoletto, Cerna, Corcione, Fabron, Jahnke, Ligori, Madrid, Martin, Morgante, Pamplona, Prieto, Riva, Toledo, Trifoglio, Zerbi, Abdalla, Douspis, Grenet, Borgani, Bouwens,
  Courbin, Delouis, Dubath, Fontana, Frailis, Grazian, Koppenh{\"o}fer, Mansutti, Melchior, Mignoli, Mohr, Neissner, Noddle, Poncet, Scodeggio, Serrano, Shane, Starck, Surace, Taylor, {Verdoes-Kleijn}, Vuerli, Williams, Zacchei, Altieri, Escudero~Sanz, Kohley, Oosterbroek, Astier, Bacon, Bardelli, Baugh, Bellagamba, Benoist, Bianchi, Biviano, Branchini, Carbone, Cardone, Clements, Colombi, Conselice, Cresci, Deacon, Dunlop, Fedeli, Fontanot, Franzetti, Giocoli, {Garcia-Bellido}, Gow, Heavens, Hewett, Heymans, Holland, Huang, Ilbert, Joachimi, Jennins, Kerins, Kiessling, Kirk, Kotak, Krause, Lahav, {van Leeuwen}, Lesgourgues, Lombardi, Magliocchetti, Maguire, Majerotto, Maoli, Marulli, Maurogordato, McCracken, McLure, Melchiorri, Merson, Moresco, Nonino, Norberg, Peacock, Pello, Penny, Pettorino, Di~Porto, Pozzetti, Quercellini, Radovich, Rassat, Roche, Ronayette, Rossetti, Sartoris, Schneider, Semboloni, Serjeant, Simpson, Skordis, Smadja, Smartt, Spano, Spiro, Sullivan, Tilquin, Trotta, Verde, Wang,
  Williger, Zhao, Zoubian, \& Zucca}]{laureijs_2011}
Laureijs, R., Amiaux, J., Arduini, S., {et~al.} 2011 [\eprint[arxiv]{1110.3193}]

\bibitem[{Lintott {et~al.}(2008)Lintott, Schawinski, Slosar, Land, Bamford, Thomas, Raddick, Nichol, Szalay, Andreescu, Murray, \& Vandenberg}]{Lintott_2008}
Lintott, C.~J., Schawinski, K., Slosar, A., {et~al.} 2008, \mnras, 389, 1179

\bibitem[{{Lotz} {et~al.}(2004){Lotz}, {Primack}, \& {Madau}}]{Lotz_2004}
{Lotz}, J.~M., {Primack}, J., \& {Madau}, P. 2004, \aj, 128, 163

\bibitem[{Lu {et~al.}(2015)Lu, Behbood, Hao, Zuo, Xue, \& Zhang}]{Lu_2015}
Lu, J., Behbood, V., Hao, P., {et~al.} 2015, Knowledge-Based Systems, 80, 14

\bibitem[{Masters(2019)}]{Masters_2019}
Masters, K.~L. 2019, Proceedings of the International Astronomical Union, 14, 205

\bibitem[{Masters {et~al.}(2010)Masters, Mosleh, Romer, Nichol, Bamford, Schawinski, Lintott, Andreescu, Campbell, Crowcroft, Doyle, Edmondson, Murray, Raddick, Slosar, Szalay, \& Vandenberg}]{Masters_2010}
Masters, K.~L., Mosleh, M., Romer, A.~K., {et~al.} 2010, \mnras, 405, 783

\bibitem[{{Peng} {et~al.}(2002){Peng}, {Ho}, {Impey}, \& {Rix}}]{Peng_2002}
{Peng}, C.~Y., {Ho}, L.~C., {Impey}, C.~D., \& {Rix}, H.-W. 2002, \aj, 124, 266

\bibitem[{Sakamoto {et~al.}(1999)Sakamoto, Okumura, Ishizuki, \& Scoville}]{Sakamoto_1999}
Sakamoto, K., Okumura, S.~K., Ishizuki, S., \& Scoville, N.~Z. 1999, \apj, 525, 691

\bibitem[{Sandage(1961)}]{Sandage_1961}
Sandage, A. 1961, The {{Hubble Atlas}} of {{Galaxies}} (Washington: Carnegie Institution)

\bibitem[{Scoville {et~al.}(2007{\natexlab{a}})Scoville, Abraham, Aussel, Barnes, Benson, Blain, Calzetti, Comastri, Capak, Carilli, Carlstrom, Carollo, Colbert, Daddi, Ellis, Elvis, Ewald, Fall, Franceschini, Giavalisco, Green, Griffiths, Guzzo, Hasinger, Impey, Kneib, Koda, Koekemoer, Lefevre, Lilly, Liu, McCracken, Massey, Mellier, Miyazaki, Mobasher, Mould, Norman, Refregier, Renzini, Rhodes, Rich, Sanders, Schiminovich, Schinnerer, Scodeggio, Sheth, Shopbell, Taniguchi, Tyson, Urry, Van~Waerbeke, Vettolani, White, \& Yan}]{Scoville_2007a}
Scoville, N., Abraham, R.~G., Aussel, H., {et~al.} 2007{\natexlab{a}}, \apjs, 172, 38

\bibitem[{Scoville {et~al.}(2007{\natexlab{b}})Scoville, Aussel, Brusa, Capak, Carollo, Elvis, Giavalisco, Guzzo, Hasinger, Impey, Kneib, LeFevre, Lilly, Mobasher, Renzini, Rich, Sanders, Schinnerer, Schminovich, Shopbell, Taniguchi, \& Tyson}]{Scoville_2007}
Scoville, N., Aussel, H., Brusa, M., {et~al.} 2007{\natexlab{b}}, \apjs, 172, 1

\bibitem[{{S\'ersic}(1968)}]{Sersic_1968}
{S\'ersic}, J.~L. 1968, {Atlas de galaxias australes} (Cordoba, Argentina: Observatorio Astronomico, 1968)

\bibitem[{Simmons {et~al.}(2017)Simmons, Lintott, Willett, Masters, Kartaltepe, H{\"a}u{\ss}ler, Kaviraj, Krawczyk, Kruk, McIntosh, Smethurst, Nichol, Scarlata, Schawinski, Conselice, Almaini, Ferguson, Fortson, Hartley, Kocevski, Koekemoer, Mortlock, Newman, Bamford, Grogin, Lucas, Hathi, McGrath, Peth, Pforr, Rizer, Wuyts, Barro, Bell, Castellano, Dahlen, Dekel, Ownsworth, Faber, Finkelstein, Fontana, Galametz, Gr{\"u}tzbauch, Koo, Lotz, Mobasher, Mozena, Salvato, \& Wiklind}]{Simmons_2017}
Simmons, B.~D., Lintott, C., Willett, K.~W., {et~al.} 2017, \mnras, 464, 4420

\bibitem[{Tan \& Le(2019)}]{Tan_2019}
Tan, M. \& Le, Q.~V. 2019, 36th International Conference on Machine Learning, ICML 2019 [\eprint[arxiv]{1905.11946}]

\bibitem[{Taniguchi {et~al.}(2007)Taniguchi, Scoville, Murayama, Sanders, Mobasher, Aussel, Capak, Ajiki, Miyazaki, Komiyama, Shioya, Nagao, Sasaki, Koda, Carilli, Giavalisco, Guzzo, Hasinger, Impey, LeFevre, Lilly, Renzini, Rich, Schinnerer, Shopbell, Kaifu, Karoji, Arimoto, Okamura, \& Ohta}]{Taniguchi_2007}
Taniguchi, Y., Scoville, N., Murayama, T., {et~al.} 2007, \apjs, 172, 9

\bibitem[{{van den Bergh}(1976)}]{vandenBergh_1976}
{van den Bergh}, S. 1976, \apj, 206, 883

\bibitem[{{Vega-Ferrero} {et~al.}(2021){Vega-Ferrero}, Dom{\'i}nguez~S{\'a}nchez, Bernardi, {Huertas-Company}, Morgan, Margalef, Aguena, Allam, Annis, Avila, Bacon, Bertin, Brooks, Carnero~Rosell, Carrasco~Kind, Carretero, Choi, Conselice, Costanzi, {da Costa}, Pereira, De~Vicente, Desai, Ferrero, Fosalba, Frieman, {Garc{\'i}a-Bellido}, Gruen, Gruendl, Gschwend, Gutierrez, Hartley, Hinton, Hollowood, Honscheid, Hoyle, Jarvis, Kim, Kuehn, Kuropatkin, Lima, Maia, Menanteau, Miquel, Ogando, Palmese, {Paz-Chinch{\'o}n}, Plazas, Romer, Sanchez, Scarpine, Schubnell, Serrano, {Sevilla-Noarbe}, Smith, Suchyta, Swanson, Tarle, Tarsitano, To, Tucker, Varga, \& Wilkinson}]{Vega-Ferrero_2021}
{Vega-Ferrero}, J., Dom{\'i}nguez~S{\'a}nchez, H., Bernardi, M., {et~al.} 2021, \mnras, 506, 1927

\bibitem[{Walmsley {et~al.}(2023{\natexlab{a}})Walmsley, Allen, Aussel, Bowles, Gregorowicz, Slijepcevic, Lintott, Scaife, Jab{\l}o{\'n}ska, Karchev, Lanzieri, Mohan, O'Ryan, Saiguhan, Su{\'a}rez, {Guerra-Varas}, \& Velu}]{Walmsley_2023}
Walmsley, M., Allen, C., Aussel, B., {et~al.} 2023{\natexlab{a}}, Journal of Open Source Software, 8, 5312

\bibitem[{Walmsley {et~al.}(2023{\natexlab{b}})Walmsley, Géron, Kruk, Scaife, Lintott, Masters, Dawson, Dickinson, Fortson, Garland, Mantha, O’Ryan, Popp, Simmons, Baeten, \& Macmillan}]{Walmsley_2023_2}
Walmsley, M., Géron, T., Kruk, S., {et~al.} 2023{\natexlab{b}}, \mnras, 526, 4768

\bibitem[{Walmsley {et~al.}(2022{\natexlab{a}})Walmsley, Lintott, G{\'e}ron, Kruk, Krawczyk, Willett, Bamford, Kelvin, Fortson, Gal, Keel, Masters, Mehta, Simmons, Smethurst, Smith, Baeten, \& Macmillan}]{Walmsley_2022_1}
Walmsley, M., Lintott, C., G{\'e}ron, T., {et~al.} 2022{\natexlab{a}}, \mnras, 509, 3966

\bibitem[{Walmsley {et~al.}(2022{\natexlab{b}})Walmsley, Scaife, Lintott, Lochner, Etsebeth, G{\'e}ron, Dickinson, Fortson, Kruk, Masters, Mantha, \& Simmons}]{Walmsley_2022_2}
Walmsley, M., Scaife, A. M.~M., Lintott, C., {et~al.} 2022{\natexlab{b}}, \mnras, 513, 1581

\bibitem[{Walmsley {et~al.}(2022{\natexlab{c}})Walmsley, Slijepcevic, Bowles, \& Scaife}]{Walmsley_2022_3}
Walmsley, M., Slijepcevic, I.~V., Bowles, M., \& Scaife, A. M.~M. 2022{\natexlab{c}}, Machine Learning for Astrophysics Workshop at the 39th International Conference on Machine Learning, ICML 2022 [\eprint[arxiv]{2206.11927}]

\bibitem[{Willett {et~al.}(2017)Willett, Galloway, Bamford, Lintott, Masters, Scarlata, Simmons, Beck, Cardamone, Cheung, Edmondson, Fortson, Griffith, H{\"a}u{\ss}ler, Han, Hart, Melvin, Parrish, Schawinski, Smethurst, \& Smith}]{Willett_2017}
Willett, K.~W., Galloway, M.~A., Bamford, S.~P., {et~al.} 2017, \mnras, 464, 4176

\bibitem[{Willett {et~al.}(2013)Willett, Lintott, Bamford, Masters, Simmons, Casteels, Edmondson, Fortson, Kaviraj, Keel, Melvin, Nichol, Raddick, Schawinski, Simpson, Skibba, Smith, \& Thomas}]{Willett_2013}
Willett, K.~W., Lintott, C.~J., Bamford, S.~P., {et~al.} 2013, \mnras, 435, 2835

\end{thebibliography}

\begin{appendix}
\onecolumn

\section{The GZH decision tree}\label{sec:appendix_a}

\begin{figure*}[htbp!]
\centering
\includegraphics[width=0.9\textwidth]{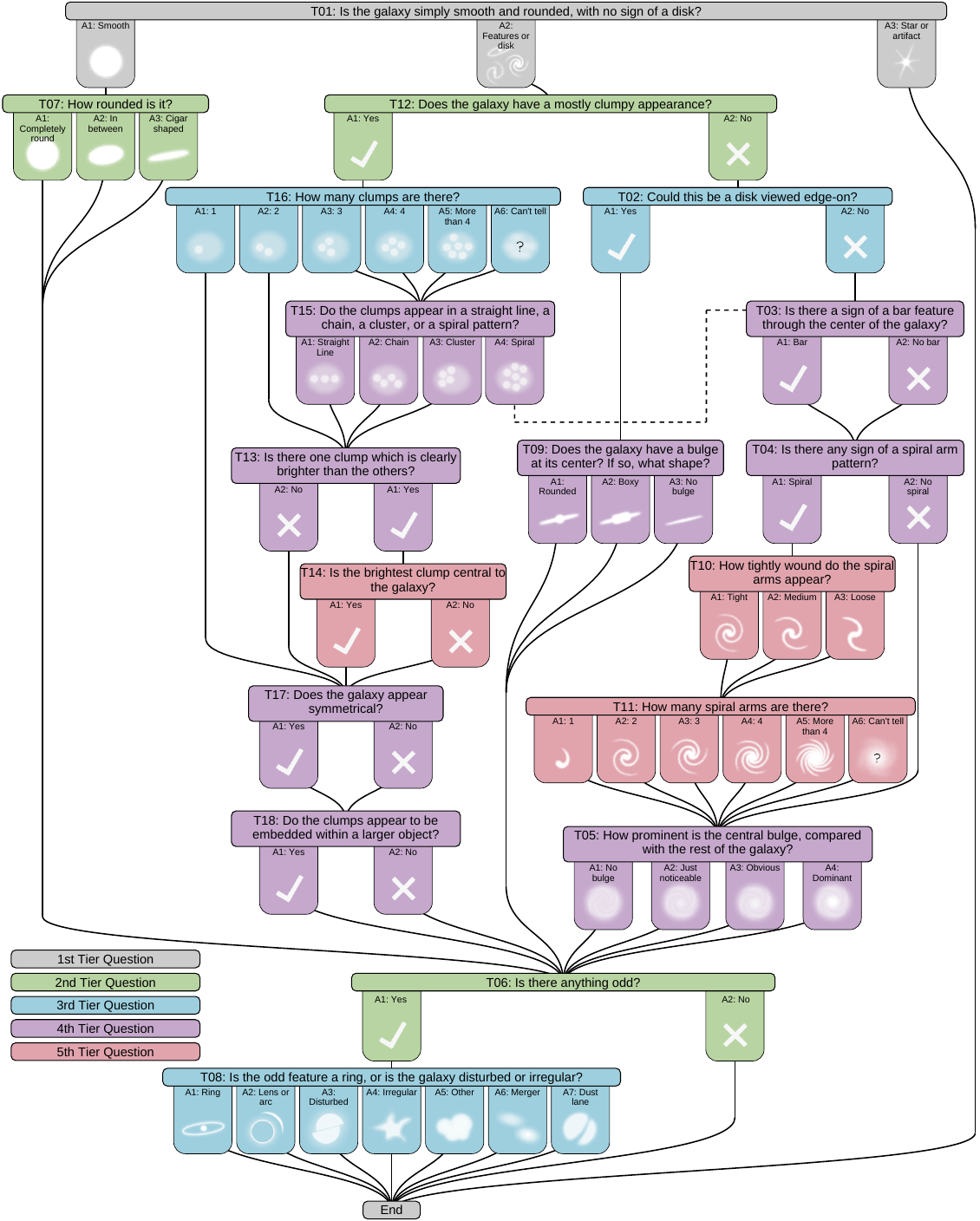}
\caption{The original GZH decision tree \citep{Willett_2017}. The questions T06, T08, T13, T14 and T15 were not included in this study (see Sect.\,\ref{sec:Training}).}
\label{fig:gz_decision_tree}
\end{figure*}
\FloatBarrier

\twocolumn

\section{Experimenting with different initial weights}\label{sec:appendix_weights}

\begin{figure}[htbp!]
\centering
\includegraphics[width=\hsize]{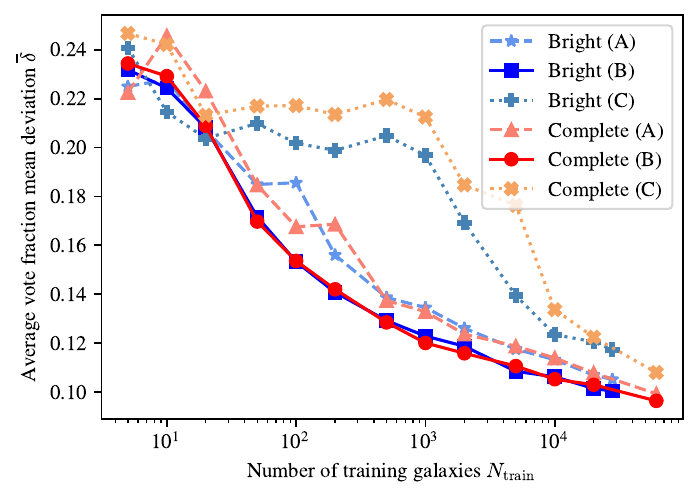}
\caption{The vote fraction mean deviation averaged over all answers depending on the number of galaxies used for training for initial \texttt{weights A} (pretraining on GZD-5), \texttt{B} (pretraining on all major GZ campaigns except GZH) and \texttt{C} (no pretraining) and training on the bright and complete training set. In all cases, the predictions were done on the complete test set. Lower values indicate better performance.
}
\label{fig:num_galaxies_deviations}
\end{figure}

\begin{figure*}[htbp!]
\centering
\includegraphics[width=\hsize]{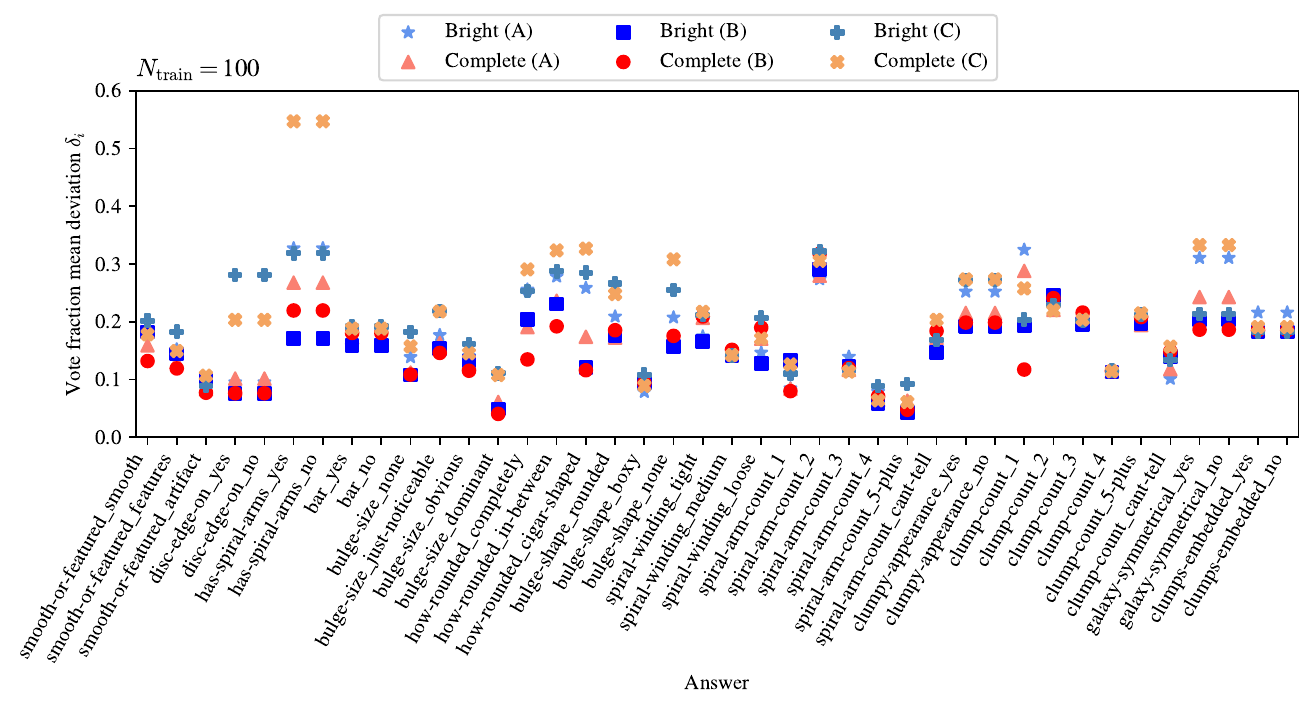}
\caption{Vote fraction mean deviations of the model predictions and the volunteer labels for the different answers of the decision tree for the models trained on 100 galaxies of the different datasets (bright and complete) and with different initial weights (\texttt{weights A}, \texttt{B} and \texttt{C}). Lower $\delta_i$ indicates better performance.}
\label{fig:deviations_datasets_weights_100}
\end{figure*}

\begin{figure*}[htbp!]
\centering
\includegraphics[width=\hsize]{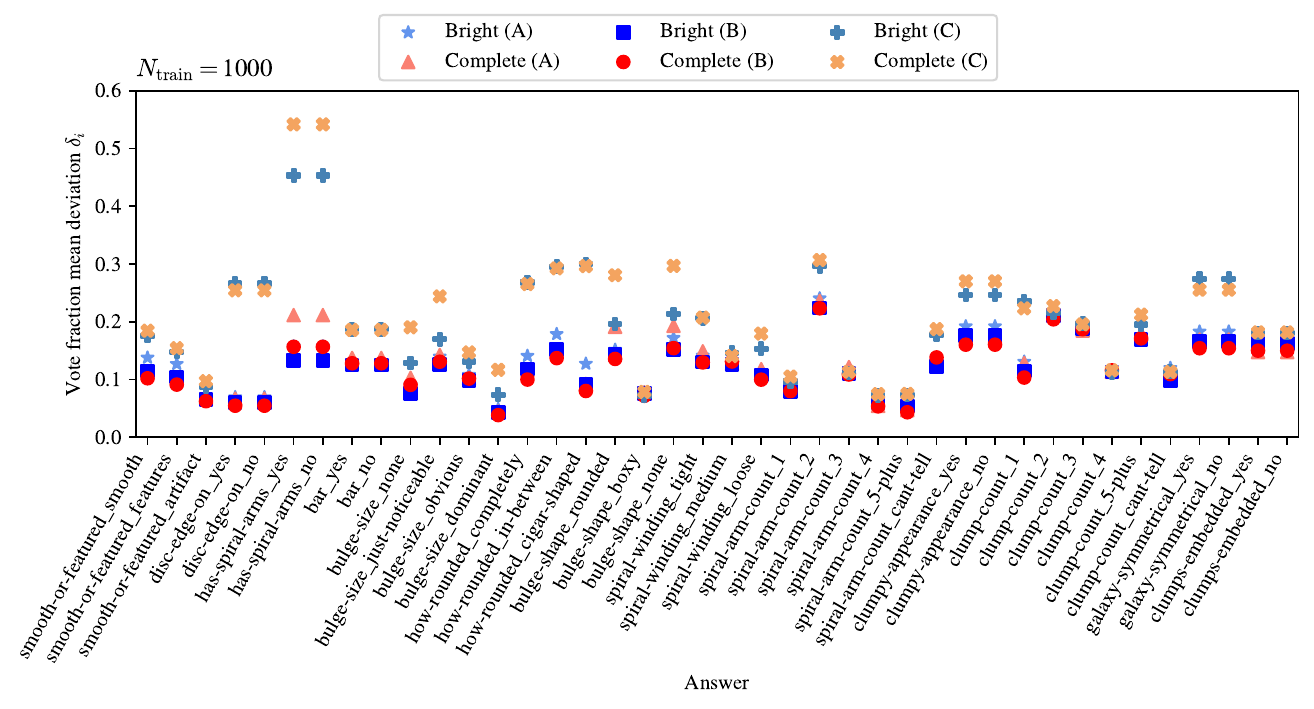}
\caption{Similar to Fig.\,\ref{fig:deviations_datasets_weights_100} with $N_{\textrm{train}}=1000$ galaxies.}
\label{fig:deviations_datasets_weights_1000}
\end{figure*}

\begin{figure*}[htbp!]
\centering
\includegraphics[width=\hsize]{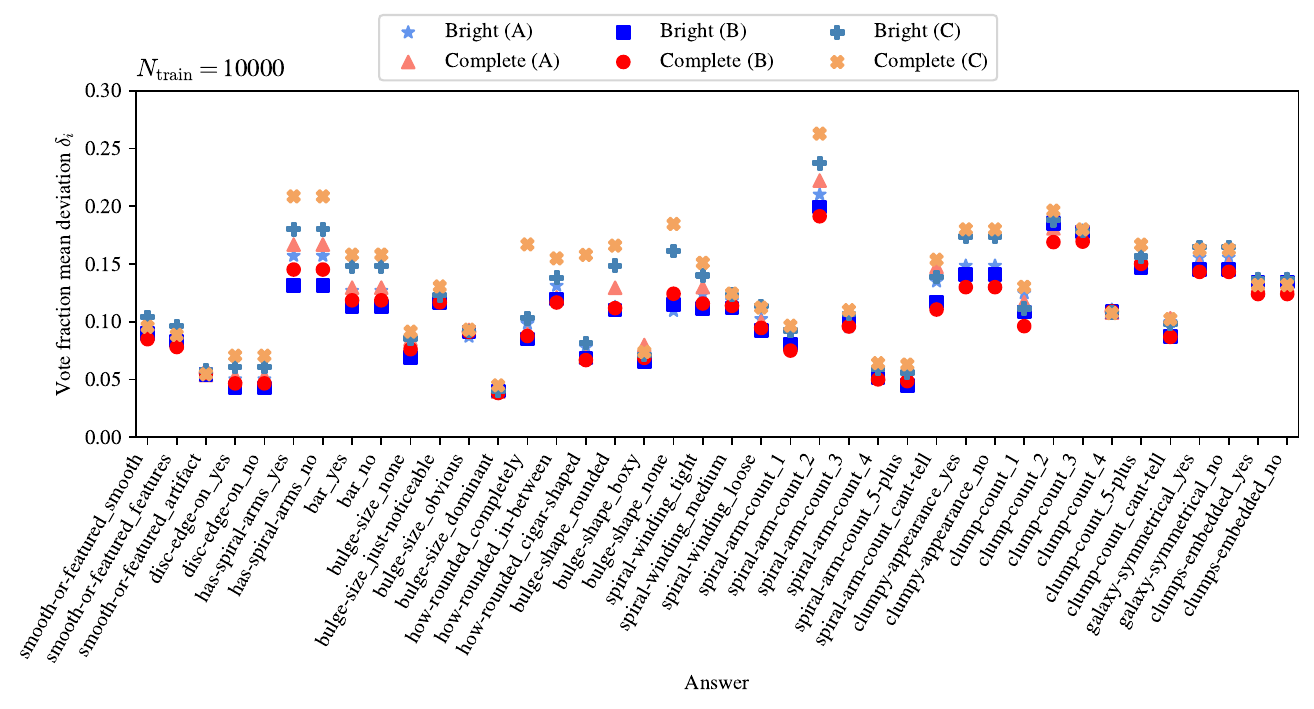}
\caption{Similar to Fig.\,\ref{fig:deviations_datasets_weights_100} with $N_{\textrm{train}}=10\,000$ galaxies.}
\label{fig:deviations_datasets_weights_10000}
\end{figure*}

We also conducted the experiment described in the main paper with different initial weights, meaning the utilization of \texttt{Zoobot} pretrained on different data. The weights described in the main text are denoted as \texttt{weights B}. Additionally, we tested with \texttt{Zoobot} pretrained on only GZD-5 (\citealt{Walmsley_2022_1}, \texttt{weights A}) and \texttt{Zoobot} without pretraining (random weights, \texttt{weights C}). We show the dependence of the averaged vote fraction mean deviation $\Bar{\delta}$ on the number of galaxies $N_{\textrm{train}}$ in Fig.\,\ref{fig:num_galaxies_deviations}. To compare the models, we use in all cases the predictions on the same complete test set of 15\,236 images (see Table\,\ref{tab:datasets}). Additionally, we show for all answers the deviations for all models trained on 100, 1000 and 10\,000 galaxies in Figs.\,\ref{fig:deviations_datasets_weights_100}, \ref{fig:deviations_datasets_weights_1000} and \ref{fig:deviations_datasets_weights_10000}.

With increasing number of training galaxies, the average mean deviation $\Bar\delta$ is decreasing: The more galaxy examples (of different types) are used for training, the better the model predictions get for all answers. In the regime with $N_{\textrm{train}}<20$, the models perform similarly. With more training galaxies, the performance of the model is for all numbers of training galaxies best with initial \texttt{weights B}, followed by \texttt{weights A} and then \texttt{weights C}. The model pretrained on all Galaxy Zoo campaigns except GZH leads, for the same number of galaxies, to a better performance than for a pretraining with only GZD labels. This is due to the better generalization of the model in the pretraining. The model without pretraining (\texttt{weights C}) shows the worst performance of the three, as expected.

Especially, for 100 to 10\,000 galaxies, the difference between pretrained models and models used from scratch is most evident. Thus, for a limited number of labelled galaxies, transfer learning is substantially more effective for training to a new problem than training from scratch. In comparison to \texttt{weights A} and \texttt{B}, for \texttt{weights C}, there is a difference between training with bright and with random (complete) galaxies, namely bright galaxies lead to a better model performance especially between 100 and 10\,000 galaxies. This difference could be explained with the pretraining for the models with \texttt{weights A} and \texttt{B}. While these have seen many types of galaxies with different magnitudes, they are more reliable and the magnitude cut does not have a significant impact. For \texttt{weights C}, the model was not trained before and thus, learns the galaxies morphologies for the first time. Bright galaxies with morphology that is easier accessible in general seem to be more effective when training from scratch. 

More details of the differences between models are shown in Figs.\,\ref{fig:deviations_datasets_weights_100}, \ref{fig:deviations_datasets_weights_1000} and \ref{fig:deviations_datasets_weights_10000}. The impact of pretraining can be best seen for the `disc-edge-on' question. For the pretrained models (\texttt{weights A} and \texttt{B}), 100 galaxy images are enough in training to get below a deviation of $10\,\%$. This deviation is reached for the model from scratch (\texttt{weights C}) at 10\,000 galaxies. In contrast, for the questions related to clumps, the differences between the models for 100, 1000, and 10\,000 training galaxies are relatively similar, indicating that the influence of pretraining is smaller for these questions. Only for \texttt{weights B} questions regarding clumps were included in the pretraining, supporting this interpretation.

\FloatBarrier

\onecolumn

\section{Additional confusion matrices}\label{sec:appendix_conf}

\begin{figure}[h]%
  \centering
  \includegraphics[width=0.55\linewidth]{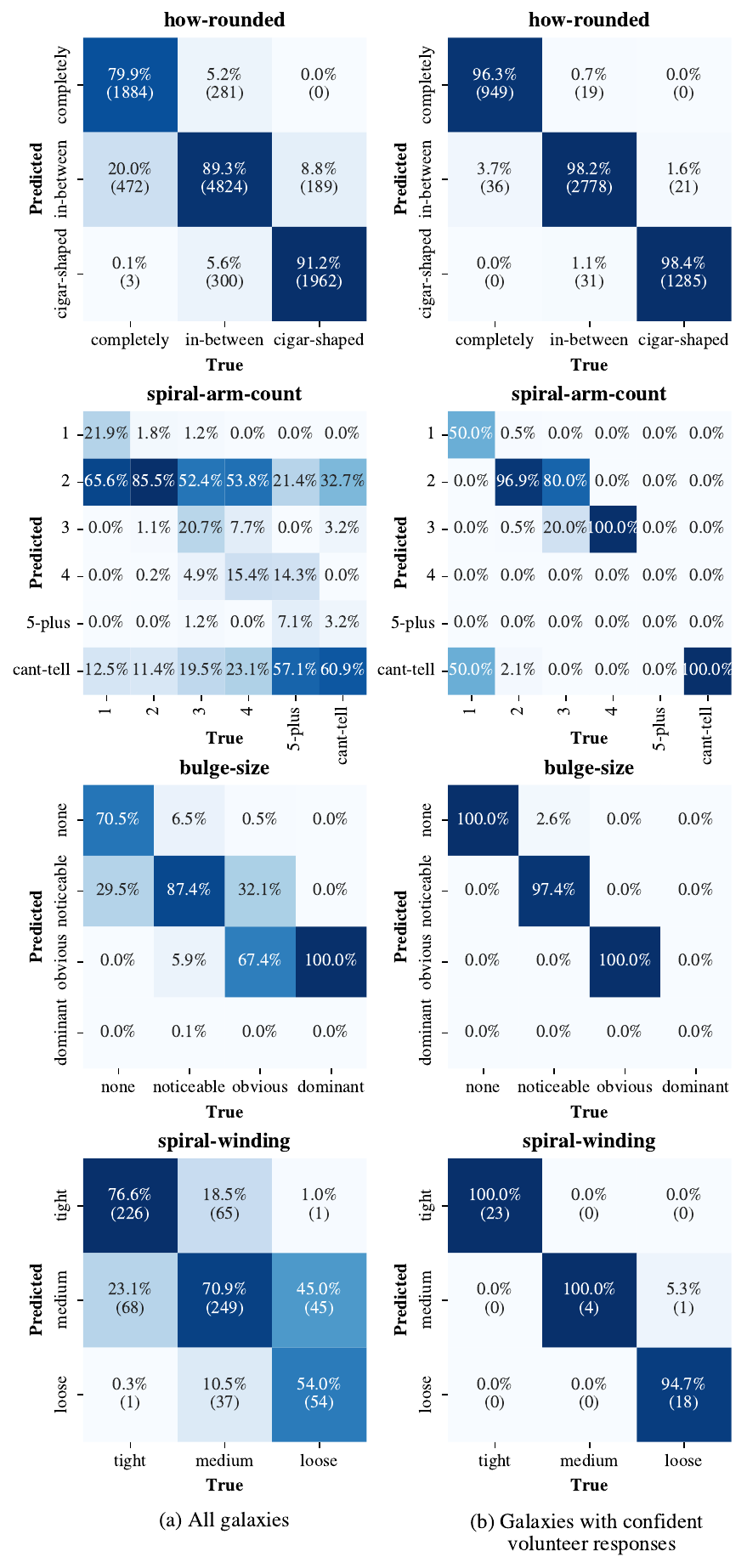}%
  \caption{Confusion matrices continued from Fig.\,\ref{fig:confusion_matrices} after binning to the class with the highest predicted vote fraction. The colour map corresponds to the fraction of the ground truth values for the different classes. To improve the readability, for the tasks with more than three answers, only the percentage is stated.}%
  \label{fig:confusion_matrices2}
\end{figure}

\begin{figure}[htbp!]%
  \centering
  \includegraphics[width=0.6\linewidth]{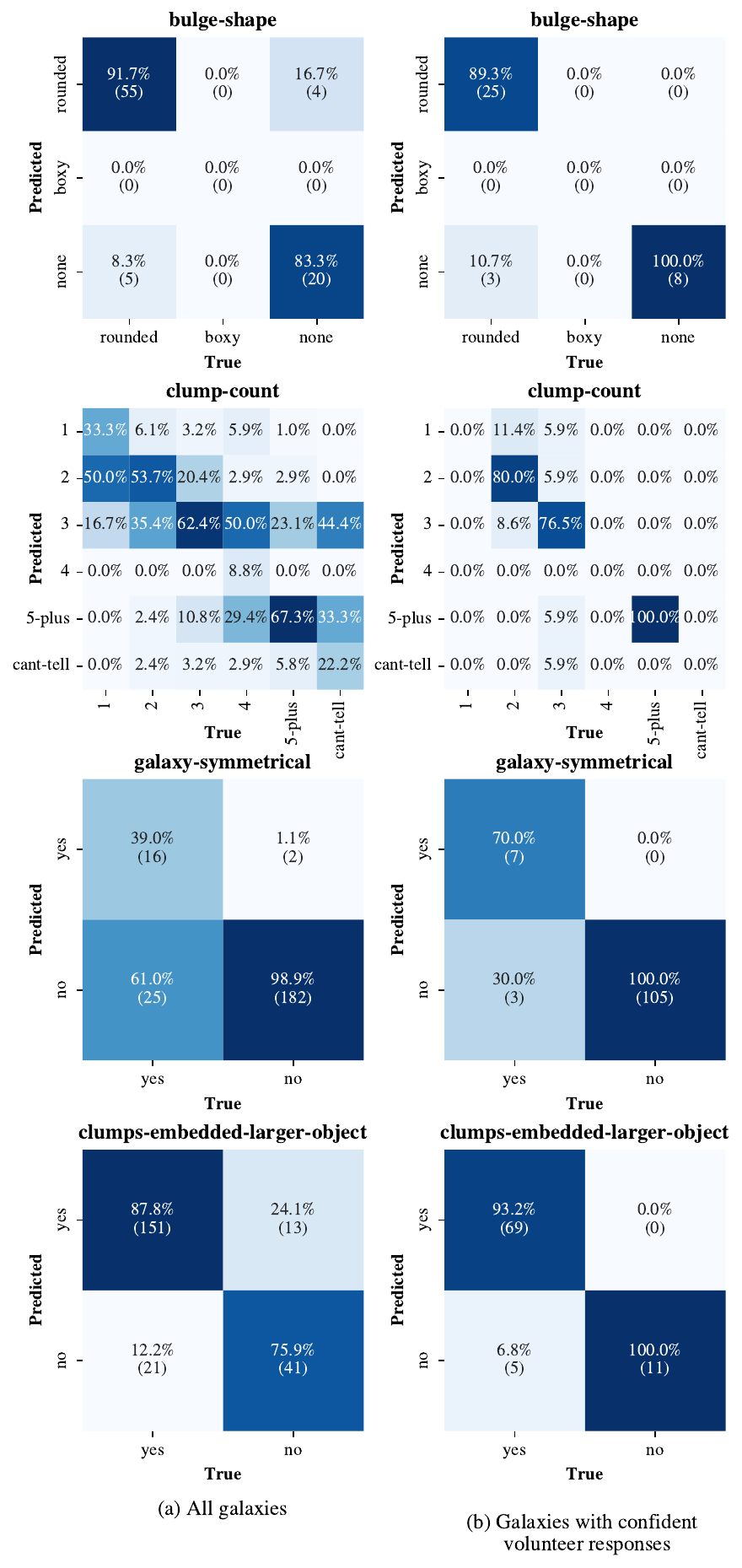}%
  \caption{Fig.\,\ref{fig:confusion_matrices2} continued.}%
  \label{fig:confusion_matrices3}
\end{figure}


\twocolumn

\section{Volunteer uncertainty}

\begin{figure}[htbp!]
\centering
\includegraphics[width=\linewidth]{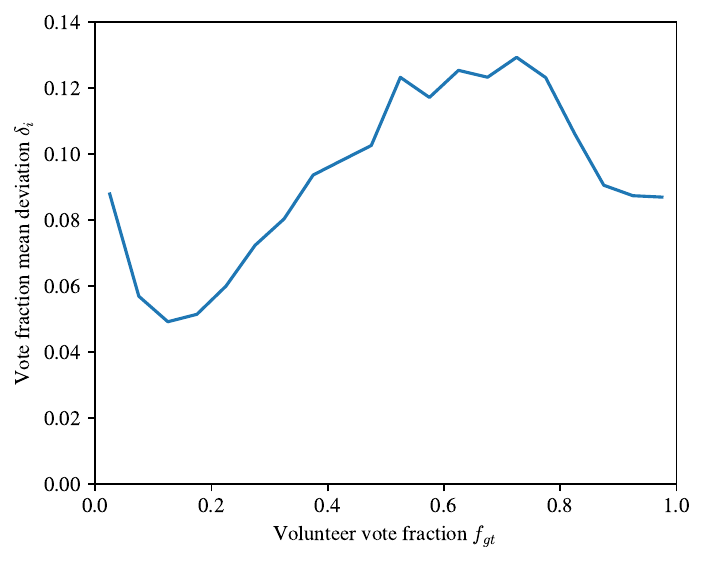}
\caption{Vote fraction mean deviations $\delta_i$ of the model predictions and the volunteer labels for the `smooth-or-features\_features' answer depending on the volunteer vote fraction $f_{gt}$.}
\label{fig:vote_fraction_deviations_comp_sf}
\end{figure}

We show in Fig.\,\ref{fig:vote_fraction_deviations_comp_sf} the mean vote fraction deviation $\delta_i$ for the `features' answer of the `smooth-or-features' question, depending on the volunteer vote fraction $f_{gt}$. In general, the deviations are smaller for confident volunteer responses (vote fraction lower than $0.2$ or greater than $0.8$) compared to more uncertain volunteer responses (vote fraction between $0.2$ and $0.8$). As expected, the model performs better for confident volunteer responses and with increasing uncertainty in the volunteer responses the deviations also increase. Moreover, an asymmetry of the deviations can be observed, as deviations are substantially smaller for vote fractions below $0.2$ compared to vote fractions above $0.8$. This could be explained with the fact that most galaxies of the dataset do not display features. Additionally, the deviations for the lowest volunteer vote fractions are higher than for vote fractions of $\sim0.1$. This is due to the characteristic of Zoobot to predict for the most extreme volunteer vote fractions (close to 0 or 1) less extreme vote fractions \citep{Walmsley_2022_1}, which should not affect practical use.

\section{Reproducibility}

The \texttt{Zoobot} CNN is publicly available\footnote{\url{https://github.com/mwalmsley/zoobot}}. The code for the creation of the images, the training of the \texttt{Zoobot} CNN and for the analysis of the results is also publicly available\footnote{\url{https://github.com/baussel/ZoobotEuclid}}.

\end{appendix}

\end{document}